\documentclass[final,3p,times]{elsarticle}

\usepackage[english]{babel}
\usepackage[utf8]{inputenc}
\usepackage{amsmath}
\usepackage{amsthm}
\usepackage{graphicx}
\usepackage{amssymb}
\usepackage{enumerate}
\usepackage{algorithm2e}
\usepackage{bm}
\usepackage{natbib}
\usepackage{color}
\newtheorem{prop}{Proposition}[section]
\newtheorem{thm}{Theorem}[section]
\newtheorem{lemma}{Lemma}[section]

\newtheorem{ex}{Example}[section]
\newtheorem{rem}{Remark}[section]

\newcommand{\btau}{{\bm{\tau}}}
\newcommand{\eps}{\varepsilon}
\newcommand{\E}{\mathbb{E}}
\newcommand{\R}{\mathbb{R}}
\newcommand{\Z}{\mathbb{Z}}
\newcommand{\N}{\mathbb{N}}

\newcommand{\cum}{\mathrm{cum}}
\newcommand{\Ind}[1]{\mathbb{I}{\{ #1 \}}}
\newcommand{\Cov}{\mathrm{Cov}}
\newcommand{\Op}{O_{\mathbb{P}}}
\newcommand{\op}{o_{\mathbb{P}}}

\newcommand{\p}{\mathbb{P}}

\newcommand{\vb}{{\bm{v}}}
\newcommand{\ub}{{\bm{u}}}

\newcommand{\Dkonv}{\stackrel{\mathcal{D}}{\longrightarrow}}%Pfeil f?r  Verteilungskonvergenz
\newcommand{\Rb}{{\mathbb{R}}}
\newcommand{\Zb}{{\mathbb{Z}}}
\newcommand{\Nb}{{\mathbb{N}}}
\newcommand{\taus}{\tau_j}

\journal{}

\begin{document}

\begin{frontmatter}



\title{Model assessment for time series dynamics using copula spectral densities:\\ a graphical tool}


\author[adr1]{Stefan Birr}
\address[adr1]{Ruhr Universit\"at Bochum. Research partially supported by project C1 of SFB 823 of the DFG}
\author[adr2]{Tobias Kley}
\address[adr2]{University of Bristol. Research partially supported by the Engineering and Physical Sciences Research Council grant no. EP/L014246/1}
\author[adr3]{Stanislav Volgushev}
\address[adr3]{University of Toronto. Research partially supported by a discovery grant from Natural Sciences and Engineering Research Council of Canada.}

\begin{abstract}
Finding parametric models that accurately describe the dependence structure of observed data is a central task in the analysis of time series. Classical frequency domain methods provide a popular set of tools for fitting and diagnostics of time series models, but their applicability is seriously impacted by the limitations of covariances as a measure of dependence. Motivated by recent developments of frequency domain methods that are based on copulas instead of covariances, we propose a novel graphical tool that allows to access the quality of time series models for describing dependencies that go beyond linearity. We provide a thorough theoretical justification of our approach and show in simulations that it can successfully distinguish between subtle differences of time series dynamics, including non-linear dynamics which result from GARCH and EGARCH models. We also demonstrate the utility of the proposed tools through an application to modeling returns of the S\&P~500 stock market index. 
\end{abstract}


\end{frontmatter}


\section{Introduction}
Non-parametric methods provide valuable tools for dependence modeling. If a parametric candidate model is available, we can compare the corresponding estimate with a non-parametric one to evaluate how well the chosen model describes the data. If no candidate model is available, non-parametric techniques can be used to get a first impression of the underlying dependence and inform about potentially suitable parametric models.

In time series analysis, methods that are based on spectral densities and periodograms have a long and successful history. Priestley \cite{priestley1981} suggests to use spectral densities as a graphical tool for model validation by comparing the \textit{spectral shape} of a dataset with standard ones from well known parametric models. Tools based on spectral distributions were considered, among others, by Bartlett \cite{bartlett1954fr, bartlett1978introduction}, who proposed to use the normalized cumulative periodogram to asses whether a process is uncorrelated and to detect hidden periodicities. Rigorous tests for the hypothesis $H_0: f = f_0$ for a fixed $f_0$ were derived by Anderson \cite{anderson1993goodness}, while the more general testing problem $H_0: f \in \mathcal{F}_\theta,$ where $\mathcal{F}_\theta$ is a parametric class of spectral densities, was considered by Paparoditis \cite{paparoditis2000spectral}. Fan and Zhang \cite{FanZhang2004} consider generalised likelihood ratio tests for the same hypothesis. There also is a rich literature on non-parametric comparison of the (multivariate) spectra of two time series, here some recent references include \cite{diggle1991nonparametric, dette2009bootstrapping, McElroyHolan2009, JentschPauly2015, ChauEtAl2017}, but this list is by no means complete.

\begin{figure}[tbh] \label{fig:DAX}
\begin{center}
\includegraphics[width = 0.4\textwidth]{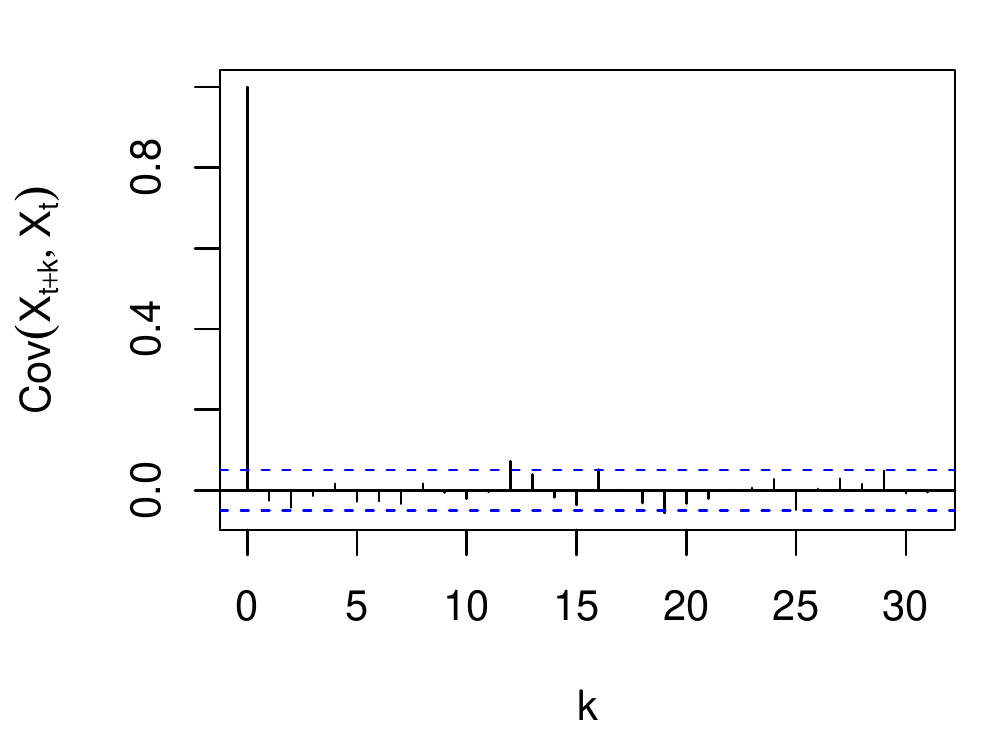} \hspace{0.5cm}
\includegraphics[width = 0.4\textwidth]{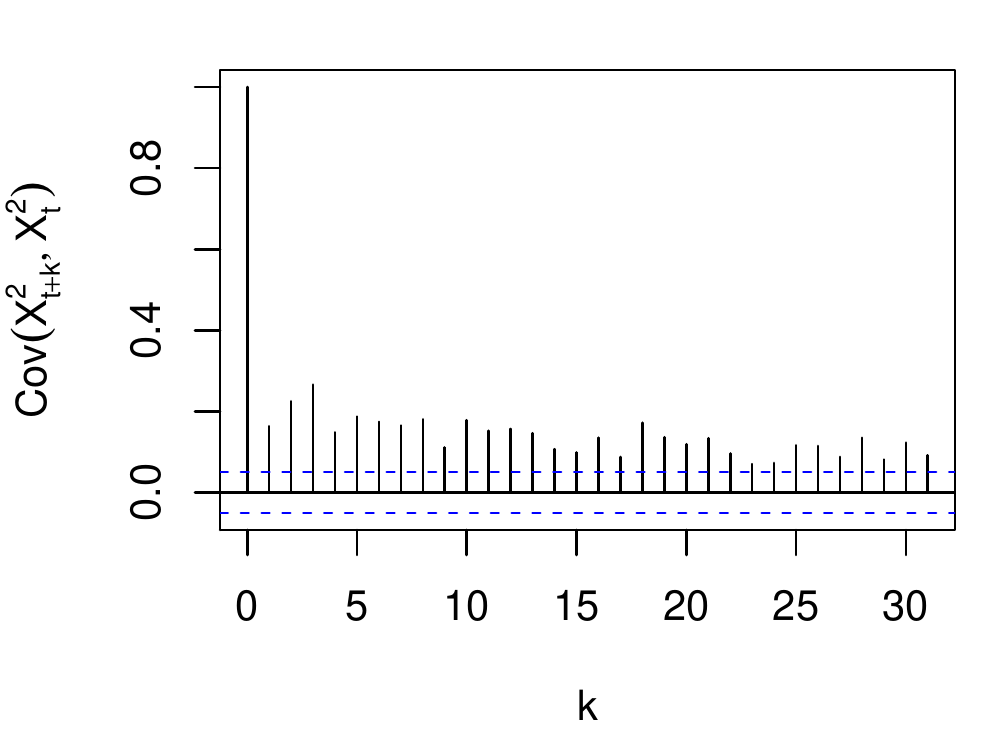}
\end{center}
\caption{Autocorrelation functions of (daily) log-returns and squared log-returns of the S\&P~500 between 2000 and 2005.}
\end{figure}

All of the references cited above deal with classical spectral analysis which is based on the autocovariance function and therefore restricted to the aspects of time series dynamics that can be captured by second-order moments. The autocovariance function does provide a complete description of the dependence of Gaussian processes, but it can completely miss dependencies in a non-Gaussian setting. One such example arises in Economics when first order differences of stock market data (more precisely, of log prices) are analyzed. For illustration, Figure \ref{fig:DAX} shows the autocorrelations of the log-returns $X_t$ and the squared log-returns $X_t^2$ calculated from the S\&P~500 between 2005 and 2010. While the observations $X_t$ appear to be uncorrelated, we can clearly see positive correlation in the squared observations $X_t^2.$ This shows that $X_t$ in fact exhibits strong dependence, which however can not be described though the autocovariance function and therefore also completely escapes classical spectral analysis.

This limitation has recently attracted much attention, and new frequency domain tools that can capture non-linear dynamics have been proposed. Pioneering contributions in that direction were made by Hong \cite{hong1999, hong2000} who introduced \textit{generalized spectra} that are based on joint distributions and joint characteristic functions, respectively. Generalized spectra were later utilized by Hong and Lee \cite{HongLee2003, honglee2005} and Escanciano \cite{Escanciano2006} to test for the validity of various forms of parametric time series models.  

More recently, related approaches were taken by Li \cite{li2008,li2012}, who coined the names of {\it Laplace spectrum} and {\it Laplace periodogram}. Those ideas were further developed by Hagemann \cite{hagemann2011} and extended to {\it cross-spectrum} and spectral {\it kernel} concepts by Dette et. al \cite{dhkv2014, kvdh2014}, who introduced the notion of {\it copula spectral densities}, Barun{\'i}k and Kley \cite{BarunikKley2015}, who introduced {\it quantile coherency} to measure dependence between economic variables and Birr et. al~\cite{birr2014quantile} who consider copula spectra for strictly locally stationary time series.  

In the present paper, we utilize copula spectral densities to develop a graphical tool for determining suitable parametric models for time series. We would like to emphasize that our main goal is not to construct yet another test for the hypothesis that a time series is generated by a certain parametric model. Rather, we provide a graphical tool that can indicate whether a chosen model accurately reflects the dependence present in the observed data. By providing useful information about which aspects of the dependence are not captured (if the model is not appropriate) our procedure goes beyond goodness-of-fit tests that merely aim to reject a class of candidate models. 

The remaining part of the paper is organized as follows. Section~\ref{sec:interpr} contains a summary of basic properties of \emph{copula spectral densities} and provides guidelines for their interpretation. In Section~\ref{sec:tool} we provide details on the proposed algorithm. Section~\ref{chap:asymptotics} gives a theoretical justification for the graphical approach in the form of a central limit theorem for triangular arrays of processes and Section~\ref{chap:simandex} contains a simulation study and a real data example. All proofs are deferred to the online supplement.

\section{Description of Method} \label{chap:method}

\subsection{Copula spectral densities: definition and interpretation}\label{sec:interpr}

We begin by briefly recalling the definition of copula spectral densities. Consider a strictly stationary process $(X_{t})_{t \in \mathbb{Z}},$ denote by $F$ its marginal distribution function (which we assume to be continuous), by $F_{h}$ denote the bivariate distribution functions of $(X_{t+h}, X_{t})$ and by $C_h$ the corresponding copulas. Then, the copula spectral density for the process $X_{t}$ is given by 
\begin{equation*}
f_{\btau}(\omega) := \frac{1}{2\pi} \sum_{k \in \Zb} \Cov(\Ind{F(X_{k}) \leq \tau_1},\Ind{F(X_{0}) \leq \tau_2}) {\rm e}^{-{\rm i}k\omega}
= \frac{1}{2\pi} \sum_{k \in \Zb} (C_k(\tau_1,\tau_2) - \tau_1\tau_2) {\rm e}^{-{\rm i}k\omega},
\end{equation*}
where $\btau = (\tau_1,\tau_2) \in (0,1)^2$, $\Ind{A}$ denotes the indicator function of $A$, and we assume that the terms in the series are absolutely summable. Estimation of copula spectral densities is discussed in the next paragraph, and we begin by providing more insights about their interpretation. Some of the properties mentioned below have been described in \cite{hagemann2011, dhkv2014, kvdh2014}, while other parts are new. 

We begin by noting that, being based on copulas, the copula spectral density is invariant under strictly increasing marginal transformations of the time series and is thus truly providing information about the temporal dependence structure of the process under consideration. This also implies that the copula spectra of a pair-wise independent time series takes the particularly simple form $f_{(\tau,\eta)}(\omega) \equiv (\tau\wedge\eta - \tau\eta)/2\pi$ which is independent of the marginal distribution.     

\begin{figure}
\hspace*{0.2\textwidth}(a) \hspace*{0.51\textwidth} (b)

\includegraphics[width = 0.43\textwidth]{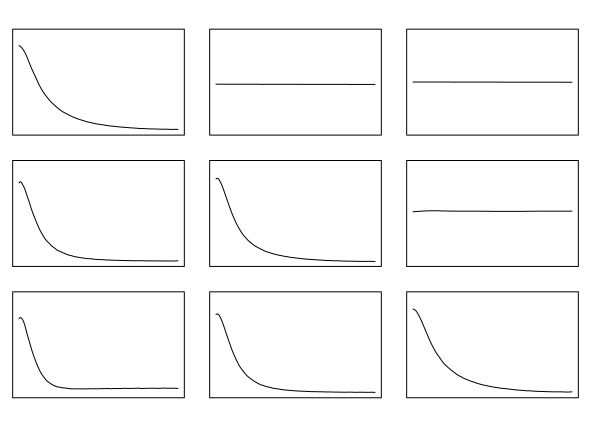} \hfill
\includegraphics[width = 0.43\textwidth]{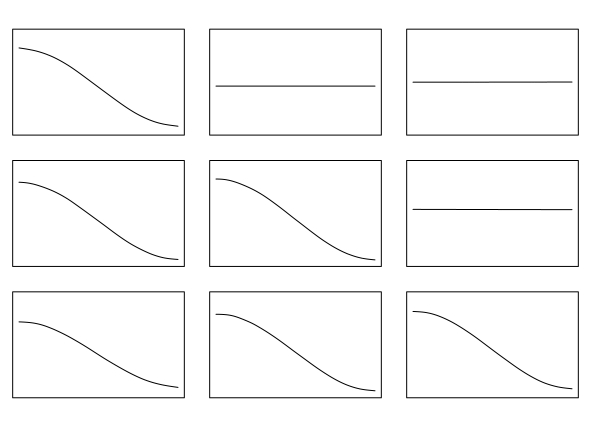}

\hspace*{0.2\textwidth}(c) \hspace*{0.51\textwidth} (d)

\includegraphics[width = 0.43\textwidth]{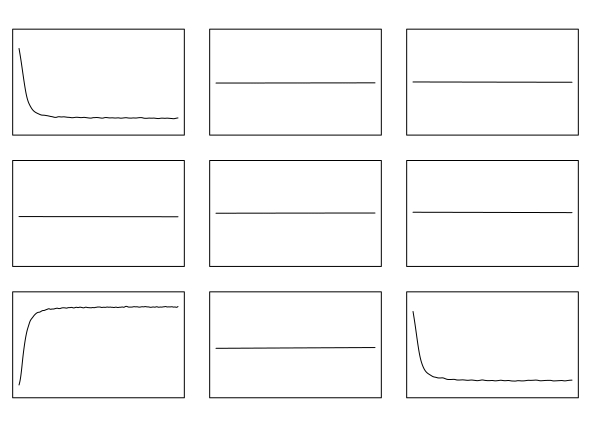} 
\hfill
\includegraphics[width = 0.43\textwidth]{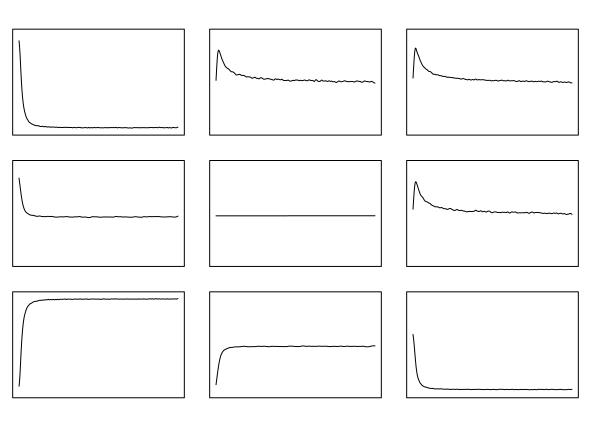} 
\caption{Copula Spectral Densities for $\btau \in \{0.1,0.5,0.9\}^2$ of an (a) AR(1), (b) MA(1), (c) GARCH(1,1) and (d) EGARCH(1,1) process. } \label{fig.example_ts}
\end{figure}

Next we note that $\omega \mapsto f_\btau(\omega)$ is $2\pi$-periodic for arbitrary $\btau \in [0,1]^2$ and that $f_\btau$ satisfies
\[
f_{(\tau_1,\tau_2)}(\omega) = \overline{f_{(\tau_1,\tau_2)}(-\omega)} = \overline{f_{(\tau_2,\tau_1)}(\omega)},
\]
where $\overline{a}$ denotes the complex conjugate of $a$. Those properties imply that the values of $\{f_\btau(\omega): \btau \in [0,1]^2,\omega \in [0,\pi]\}$ contain the complete information about the copula spectral density. Even given those restrictions, $f_\btau(\omega)$ is still a complex-valued function of three arguments with each argument taking values in an interval and thus difficult to visualize. One option to get a quick impression of the most important features of the copula spectral density of a given process is to consider all values of $\btau \in \{0.1,0.5,0.9\}^2$ and plot the functions $f_{\btau}(\omega)$ against $\omega \in [0,\pi]$. This requires nine plots which are organized as follows  
\begin{center}
\begin{tabular}{c|c|c}
$f_{(0.1,0.1)}(\omega)$ & ${\Im {f}_{(0.5,0.1)}({\omega})}$ & ${\Im { f}_{(0.9,0.1)}({\omega})}
$\\\noalign{\smallskip} \hline \noalign{\smallskip}
${\Re{f}_{(0.1,0.5)}({\omega})}$ & ${{f}_{(0.5,0.5)}({\omega})}$  &  ${\Im{f}_{(0.9,0.5)}({\omega})} 
$\\\noalign{\smallskip} \hline \noalign{\smallskip}
 ${\Re{f}_{(0.1,0.9)}({\omega})}$  &  ${\Re f_{(0.5,0.9)}({\omega})}$  &  ${{f}_{(0.9,0.9)}({\omega})}.$
\end{tabular}
\end{center}
In Figure~\ref{fig.example_ts}, examples of plots of copula spectral densities corresponding to different parametric models are shown. Those plots will be used to illustrate various properties of copula spectral densities given below. %\change{An alternative way to compare model-based copula spectra with spectra from data that is suitable for considering many quantiles simultaneously will be provided below.}

Begin by observing that $f_{(\tau,\tau)}(\omega)$ is real-valued (for any $\tau \in [0,1]$ and $\omega \in \R$). It corresponds to the `classical' $L^2$ spectral density of the clipped process $(I\{F(X_t) \leq \tau\})_{t\in \Z}$ and hence contains information about dynamics of the level-crossing behavior of the process $(X_t)_{t\in \Z}$. A closer look at Figure~\ref{fig.example_ts} reveals several interesting features. First, for linear Gaussian processes (i.\,e., AR(1) and MA(1) with i.i.d. normal innovations) in (a) and (b), the shape of $f_{(\tau,\tau)}(\omega)$ is similar for all values of $\tau$ and also similar to the corresponding shape of their $L^2$-spectral density. In contrast, the two non-linear models in (c), (d) have copula spectral densities with shape varying across quantile levels. Both models show no dependence at $\tau = 0.5$, which corresponds to an absence of `central dependence'. Yet, both processes show a strong dependence (as indicated by sharp peaks for small values of $\omega$) for more extreme quantiles (corresponding to $\tau=0.1,0.9$). Note also that the EGARCH model shows an asymmetric dependence with a higher peak at $\tau = 0.1$ compared to $\tau = 0.9$ indicating a stronger serial dependence in the lower tail. In contrast, the dependence in the GARCH model is completely symmetric.

For $\tau \neq \eta$, $f_{(\tau,\eta)}(\omega)$ can be complex-valued. To interpret the real part of $f_{(\tau,\eta)}(\omega)$, note that after a simple computation we obtain for $\tau < \eta$ 
\begin{align*}
\Re f_{(\tau,\eta)}(\omega) &= - \Re \sum_{k \in \Z} {\rm e}^{-{\rm i}k\omega}\Big( P(X_k \leq q_\tau, X_0 \geq q_\eta) - \tau(1-\eta) \Big)
\\
& = \tau(1-\eta) - \sum_{k\geq 1} \cos(k \omega)\Big( P(X_k \leq q_\tau, X_0 \geq q_\eta) - \tau(1-\eta) \Big)
 - \sum_{k\geq 1} \cos(k \omega)\Big( P(X_k \geq q_\eta, X_0 \leq q_\tau) - \tau(1-\eta) \Big).
\end{align*}
Hence, the function $\omega \mapsto f_{(\tau,\eta)}(\omega)$ contains information about $X_t$ switching between being below $q_\tau$ to above $q_\eta$ and vice versa. In particular, for $\tau$ `small' and $\eta$ `large' it can be interpreted as describing the dynamics of the process switching between two `extreme' states. Here, the negative peak of $\Re f_{(0.1,0.9)}$ at small values of $\omega$ in (c) indicates that the corresponding GARCH process is likely to switch from a high to a low value (or vice versa), which is exactly what happens in periods of high volatility. Similarly, the positive peak in the same function for (a), (b) corresponds to processes that are unlikely to switch from high to low states immediately, which corresponds to AR(1) or MA(1) dynamics with positive coefficients. It is also interesting to observe that for the two linear processes in (a) and (b) the shapes of $\Re f_{(\tau,\eta)}$ are similar to $_{(\tau,\tau)}$ for all combinations of $\tau, \eta$.

The imaginary part of $f_{(\tau,\eta)}(\omega)$ for $\tau < \eta$ takes the form
\[
\Im f_{(\tau,\eta)}(\omega) = - \sum_{k \geq 1} \sin(\omega k)\Big( P(X_k \leq q_\tau, X_0 \geq q_\eta) - P(X_k \geq q_\eta, X_{0} \leq q_\tau) \Big). 
\] 
Note that $\Im f_{(\tau,\eta)} \equiv 0 \Leftrightarrow P(X_k \leq q_\tau, X_0 \geq q_\eta) = P(X_k \geq q_\eta, X_{0} \leq q_\tau) \forall k \in\Z$, which shows that $\Im f_{(\tau,\eta)}$ contains information about asymmetry in going from above $q_\tau$ to below $q_\eta$ and vice versa. Non-zero imaginary parts thus indicate time-irreversibility of the dynamics in the observed time series. In particular, if $\Im f_{(\tau,\eta)} \equiv 0$ for all $\tau,\eta$ then this indicates that the process under consideration is pairwise time-reversible, i.e. $C_k(\tau,\eta) = C_{-k}(\tau,\eta)$ for all $k,\tau,\eta$. The Gaussian linear processes in (a), (b) of Figure~\ref{fig.example_ts} are time reversible, which is confirmed by the flat imaginary parts of their copula spectra. It is also noteworthy that the imaginary parts of the processes in (c) and (d) show very different behavior, with clear time-irreversibility for the EGARCH process in (d) and no immediately visible evidence of the same for the GARCH process in (c).

\subsection{Graphical tools for model validation} \label{sec:tool}

We begin by briefly reviewing estimation of copula spectral densities as discussed in \cite{kvdh2014} (see also \cite{hagemann2011} who considered the case $\tau_1 = \tau_2$ and \cite{dhkv2014} for alternative estimators based on ranks and quantile regression). Given observations $X_{0},\dots,X_{n-1}$ we calculate their empirical distribution function $\hat F_n(x) := n^{-1} \sum_{t=0}^{n-1} \Ind{X_t \leq x}$ and the copula periodogram
\[
I_{\btau,n}(\omega) = \frac{1}{2 \pi n} d_{\tau_1,n}(\omega) d_{\tau_2,n}(-\omega),
\]
where $\btau = (\tau_1, \tau_2)$ and
\[
d_{\tau,n}(\omega) = \sum_{t=0}^{n-1} \Ind{\hat F_n(X_{t}) \leq \tau} {\rm e}^{-{\rm i} \omega t}.
\]
To obtain a consistent estimator we smooth the copula periodograms over frequencies
\begin{equation} \label{eq:deffn}
\hat f_{\btau}(\omega) =  \frac{1}{2\pi n}\sum_{s=1}^n W_n(\omega - 2\pi s/n) I_{\btau,n}(2\pi s /n),
\end{equation}
where $W_n$ denotes a sequence of weighting functions which are specified in assumption (W). Kley et. al \cite{kvdh2014} proved asymptotic normality (uniformly in~$\btau$) of this estimator and computation is possible via the R package \textbf{quantspec} (see \cite{kley2016}). 

Now, given observations $X_1,\dots,X_n$ we want to decide if that data could have been produced by a parametric model $P^{\theta_0}$ where $\{P^{\theta}: \theta\in\Theta\}$ is a collection of candidate models and $\theta_0 \in \Theta$ is an unknown parameter. To this end we propose to apply parametric bootstrap ideas in the form of Algorithms~\ref{alg1} and~\ref{alg2} given below and on the subsequent page. 

\begin{algorithm}%[H]
\KwData{Observations $X_1,\dots,X_n$}
\KwIn{Class of parametric models $(P^{\theta})_{\theta\in\Theta}$, an estimator $\hat\theta$, a collection of frequencies $\omega_1,\dots,\omega_K \in [0,\pi]$, and a quantile level $\btau = (\tau_1,\tau_2)$}
\KwOut{Plot comparing copula spectral density estimated from data with `typical regions' created by a parametric bootstrap}
\Begin{
Estimate $\hat{\theta}$ from $X_1,\dots,X_n$\\
\For(\tcc*[f]{parametric bootstrap}){r in 1:R }{
$X_1^{\hat\theta,r},\dots,X_n^{\hat\theta,r} = $ simulate from the model $P^{\hat{\theta}}$ \\
$\hat{f}^{\hat\theta,r}_{\btau}(\omega_k) = $ estimated copula spectral density from $X_1^{\hat\theta,r},\dots,X_n^{\hat\theta,r}$
}\noindent
\tcc{Calculate lower and upper bounds, separately for real and imaginary parts:}
$l^\Re_{\btau,R}(\omega_k) = \alpha/2-\text{quantile}(\Re \hat{f}^{\hat\theta,1}_{\btau}(\omega_k),\dots,\Re \hat{f}^{\hat\theta,R}_{\btau}(\omega_k))$\\
$l^\Im_{\btau,R}(\omega_k) = \alpha/2-\text{quantile}(\Im \hat{f}^{\hat\theta,1}_{\btau}(\omega_k),\dots,\Im \hat{f}^{\hat\theta,R}_{\btau}(\omega_k))$\\
$u^\Re_{\btau,R}(\omega_k) = (1-\alpha/2)-\text{quantile}(\Re \hat{f}^{\hat\theta,1}_{\btau}(\omega_k),\dots,\Re \hat{f}^{\hat\theta,R}_{\btau}(\omega_k))$\\
$u^\Im_{\btau,R}(\omega_k) = (1-\alpha/2)-\text{quantile}(\Im \hat{f}^{\hat\theta,1}_{\btau}(\omega_k),\dots,\Im \hat{f}^{\hat\theta,R}_{\btau}(\omega_k))$\\
\tcc{Estimate the Copula Spectral Density for the data:}
$\hat f_{\btau}(\omega) =$ estimate the copula spectral density from $X_1,\dots,X_n$\\
\tcc{Plot the result}
plot($\{\hat f_{\btau}(\omega_k)\}_{k=1,...,K}$)\\
plot(Intervals computed from $(l_{\btau,R}(\omega_k), u_{\btau,R}(\omega_k))_{k=1,...,K}$ (separately for real and imaginary parts))
}
\label{alg1}
\caption{Graphical representation of `typical regions' from a parametric model (with parameter estimated from the data) together with the estimator based on observations.}
\end{algorithm}

Algorithm~\ref{alg1} provides a graphical summary of the copula spectral density estimated from data (blue lines) for a few distinct combinations of quantile levels (in the present paper, $(\tau_1, \tau_2) \in \{0.1, 0.5, 0.9\}^2$) as a function of $\omega$ together with typical regions (grey areas) that would contain this estimator with probability $1-\alpha$ if the corresponding class of parametric candidate models was specified correctly (see Proposition~\ref{prop:just} for a formal statement). One potential concern with Algorithm~\ref{alg1} is that the graphics can become overwhelming if many different quantile levels need to be considered simultaneously. Algorithm~\ref{alg2} below provides a useful supplement to Algorithm~\ref{alg1} which allows to consider many quantile levels at the same time. This necessitates a different graphical representation. The results from Algorithm~\ref{alg2} can be displayed in two different ways. The first provides a summary over all quantile levels $(\tau_1,\tau_2) \in M$ (in the present paper, we choose $M = \{0.05,...,0.95\}^2$) indicating whether the candidate model class produces spectral densities which are compatible with the data for a given frequency but uniformly over quantile levels. If a deviation is detected for a given frequency, a second plot for that particular frequency can be used to determine at which quantile levels the mismatch between the data and the parametric candidate model appears.

\begin{algorithm}%[H]
\KwData{Observations $X_1,\dots,X_n$}
\KwIn{Class of parametric models $(P^{\theta})_{\theta\in\Theta}$, an estimator $\hat\theta$, a frequency $\omega \in [0,\pi]$, and a number $K$ of how many equally spaced quantile levels should be used, quantile level $\beta$}
\KwOut{Heat-plot of signed p-values indicating whether estimated copula spectral density is within `typical regions' created by a parametric bootstrap}
\Begin{
Estimate $\hat{\theta}$ from $X_1,\dots,X_n$\\
\For(\tcc*[f]{parametric bootstrap}){r in 1:R }{
$X_1^{\hat\theta,r},\dots,X_n^{\hat\theta,r} = $ simulate from the model $P^{\hat{\theta}}$ \\
$\hat{f}^{\hat\theta,r}_{\btau}(\omega) = $ estimated copula spectral density from $X_1^{\hat\theta,r},\dots,X_n^{\hat\theta,r}$, {$\forall \btau \in M := \{1/(K+1), \ldots, K/(K+1)\}^2$}
}
\tcc{Calculate scaling factors, separately for real \& imaginary parts:}
Let $l_{\btau,R}^\Re(\omega), u_{\btau,R}^\Re(\omega)$ denote $\beta/2$ and $1-\beta/2$ quantile of $\Re \hat{f}^{\hat\theta,1}_{\btau}(\omega),\dots,\Re \hat{f}^{\hat\theta,R}_{\btau}(\omega)$, respectively (same for $\Im$)\\ 
%$l^\Re_{\btau,R}^\Re(\omega) = \beta/2-\text{quantile}(\Re \hat{f}^{\hat\theta,1}_{\btau}(\omega),\dots,\Re \hat{f}^{\hat\theta,R}_{\btau}(\omega))$\\
%$u^\Re_{\btau,R}(\omega) = (1-\beta/2)-\text{quantile}(\Re \hat{f}^{\hat\theta,1}_{\btau}(\omega),\dots,\Re \hat{f}^{\hat\theta,R}_{\btau}(\omega))$ and
%similar for imaginary parts.\\
Define
\[
c^\Re_{\btau,R}(\omega) = (u^\Re_{\btau,R}(\omega) + l^\Re_{\btau,R}(\omega))/2, \quad
c^\Im_{\btau,R}(\omega) = (u^\Im_{\btau,R}(\omega) + l^\Im_{\btau,R}(\omega))/2
\]\\
and
\[
\Delta^\Re_{\btau,R}(\omega) = (u^\Re_{\btau,R}(\omega) - l^\Re_{\btau,R}(\omega))/2, \quad  \Delta^\Im_{\btau,R}(\omega) = (u^\Im_{\btau,R}(\omega) - l^\Im_{\btau,R}(\omega))/2 + 10^{-6} I\{u^\Im_{\btau,R}(\omega) = l^\Im_{\btau,R}(\omega)\}
\]\\
%$\Delta^\Re_{\btau,R}(\omega) = (u^\Re_{\btau,R}(\omega)-l^\Re_{\btau,R}(\omega))/2$,
%$\Delta^\Im_{\btau,R}(\omega) = (u^\Im_{\btau,R}(\omega)-l^\Im_{\btau,R}(\omega))/2 + 0.0000001 I\{u^\Im_{\btau,R}(\omega) = l^\Im_{\btau,R}(\omega)\}$, 
%\\ 
%$c^\Re_{\btau,R}(\omega) = (u^\Re_{\btau,R}(\omega)+l^\Re_{\btau,R}(\omega))/2$, and
%$c^\Im_{\btau,R}(\omega) = (u^\Im_{\btau,R}(\omega)+l^\Im_{\btau,R}(\omega))/2$.
%\\
The scaled and centred bootstrap replicate is
\[
A_{r}^\Re(\omega) := \max_{\btau = (\tau_1,\tau_2) \in M} |\Re \hat{f}^{\hat\theta,R}_{\btau}(\omega) - c^\Re_{\btau,R}(\omega)| / \Delta^\Re_{\btau,R}(\omega), \quad
A_{r}^\Im(\omega) := \max_{\btau = (\tau_1,\tau_2) \in M} |\Im \hat{f}^{\hat\theta,R}_{\btau}(\omega) - c^\Im_{\btau,R}(\omega)| / \Delta^\Im_{\btau,R}(\omega).
\]\\
\tcc{Estimate the Copula Spectral Density for the data:}
$\hat f_{\btau}(\omega) =$ estimate the copula spectral density from $X_1,\dots,X_n$. Define\\
\[
E_{\btau}^\Re(\omega) := |\Re \hat f_{\btau}(\omega) - c^\Re_{\btau,R}(\omega)| / \Delta^\Re_{\btau,R}(\omega), \quad
E_{\btau}^\Im(\omega) := |\Im \hat f_{\btau}(\omega) - c^\Im_{\btau,R}(\omega)| / \Delta^\Im_{\btau,R}(\omega)
\]\\
%$S_{\btau}^\Re(\omega) := 2 I\{|\Re \hat f_{\btau}(\omega) - c^\Re_{\btau,R}(\omega)| \geq 0\} - 1$, and
%$S_{\btau}^\Im(\omega) := 2 I\{|\Im \hat f_{\btau}(\omega) - c^\Im_{\btau,R}(\omega)| \geq 0\} - 1$.\\
\tcc{Calculate p-values, separately for real and imaginary parts:}
Define $\hat F_R$ as the empirical cdf of $\max\{A_{1}^\Re(\omega),  A_{1}^\Im(\omega)\},...,\max\{A_{R}^\Re(\omega),  A_{R}^\Im(\omega)\}$ and compute
\[
p_{\btau,R}^\Re(\omega):= 1 - \hat F_R( E_{\btau}^\Re(\omega) -), \quad p_{\btau,R}^\Im(\omega):= 1 - \hat F_R( E_{\btau}^\Im(\omega) -), \quad p_{\min,R}(\omega) := \min_{\btau \in M} \min\{p_{\btau,R}^{\Re}(\omega), p_{\btau,R}^{\Im}(\omega)\}.
\]\\
\tcc{Plot the result}
plot 1: $\omega \mapsto  p_{\min,R}(\omega)$; x-axis from $1/R$ to 1, in log-scale. $p_{\min,R}(\omega) = 0$ is indicated by a red circle on the x-axis.)\\
plot 2: $K \times K$ panels for each $\omega$. The position within each panel corresponds to $\btau \in M$, the symbols used correspond to the magnitude of $p_{\btau,R}^\Re(\omega),p_{\btau,R}^\Im(\omega)$ ($1, 2$ and $3$ triangles correspond to $p_{\btau,R}^\cdot(\omega) < 0.05, < 0.01$ and $< 0.001$, respectively), and sign of $\Re \hat f_{\btau}(\omega) - c^\Re_{\btau,R}(\omega), \Im \hat f_{\btau}(\omega) - c^\Im_{\btau,R}(\omega)$ (red triangles facing up indicate a positive and blue triangles facing down indicate a negative value). Information corresponding to $p_{(\tau_i,\tau_j),R}^\Re(\omega)$ ($i \geq j$) is in row $i$ column $j$ and information on $p_{(\tau_i,\tau_j),R}^\Im(\omega)$ ($i < $) in row $i$ column $j$.
}
\label{alg2}
\caption{Graphical representation of `critical $\btau$s' from a parametric model (with parameter estimated from the data) together with the estimator based on observations.}
\end{algorithm}

%\change{We now provide an informal motivation for both algorithms (which are similar in spirit to parametric bootstrap approaches, see e.g. \textbf{Add literature}), while a more formal justification is given in Section~\ref{chap:asymptotics} below. For motivating both algorithms, observe that} if the data are generated from a model $P^{\theta_0}$, repeatedly simulating data and computing the corresponding, estimated copula spectral densities from $P^{\theta_0}$ will provide a precise range for the (random) variation in estimators for this particular choice of sample size and smoothing parameters. 

%Thus $\hat f_{\btau}(\omega_k) \in (l_{\btau,R}(\omega_k), u_{\btau,R}(\omega_k))$ for all $k$, where $l_{\btau,R}(\omega_k)$ and $u_{\btau,R}(\omega_k)$ denote the bounds obtained from the simulated copula spectral density estimates, indicates that the estimated copula spectrum observed from the data could have been produced by a model in the parametric candidate class. If $\hat f_{\btau}(\omega_k) \notin (l_{\btau,R}(\omega_k), u_{\btau,R}(\omega_k))$ this means that a model in the parametric class can not produce this type of copula spectrum (even when taking into account variation in the estimator due to randomness in the data), indicating that the model class is misspecified. Taking a closer look at deviations of the estimator $\hat f_{\btau}(\omega_k)$ from the intervals $(l_{\btau,R}(\omega_k), u_{\btau,R}(\omega_k))$ also provides useful information about potential alternative models that could be considered instead of the current parametric class.  

\begin{figure}[h!]
\includegraphics[width = 0.5\textwidth]{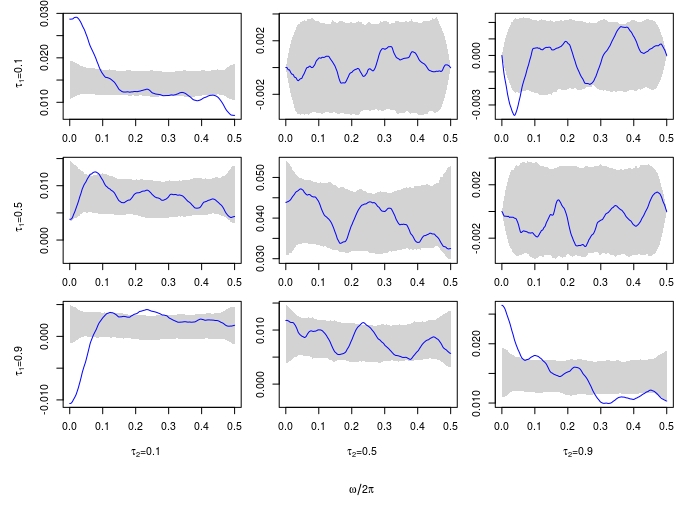}  
\includegraphics[width = 0.5\textwidth]{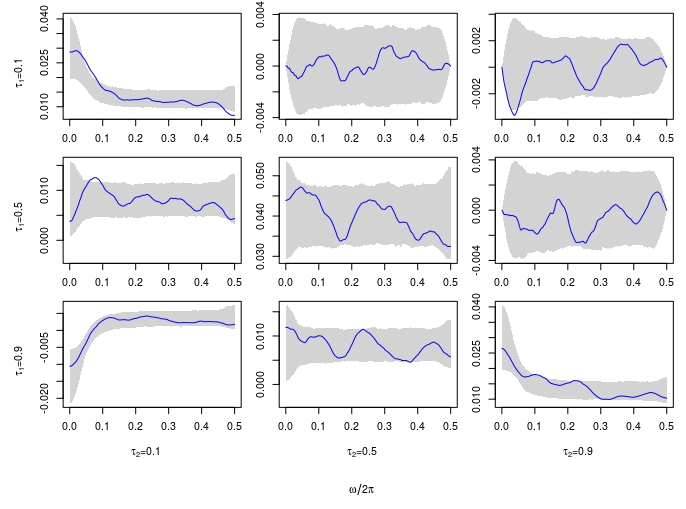}
\vspace*{-.8cm}
\caption{Example using our graphical tool (Algorithm~\ref{alg1}) on data generated from a GARCH(1,1) model with $n = 1024$ observations. We are fitting an AR(3) model (left) and a GARCH(1,1) model (right) with $\alpha = 0.05$.}
\label{fig.example_tool}
\end{figure}

Figure~\ref{fig.example_tool} illustrates Algorithm~\ref{alg1} in a simple example. Here the data is a single path simulated from a GARCH(1,1) model, and we considered two classes of parametric models: AR(3) and GARCH(1,1) (the true model is in the latter class, but the parameter was not specified). The blue line shows the estimated copula spectral density $\hat f_\btau$ (the plot is organized as discussed in Section~\ref{sec:interpr}; $\Re\hat f_{\btau}$ on/below and $\Im\hat f_{\btau}$ above the diagonal, respectively) and the grey area corresponds to the typical regions for $\alpha = 0.05$ (see Algorithm~\ref{alg1} for details). We clearly see that an AR(3) model is unable to describe the dynamics of a GARCH model, as it fails to capture the dependency in the extreme quantiles ($\btau = (0.1,0.1), (0.9,0.9), (0.1,0.9)$), especially at low frequencies. Considering the true model class (right panel) on the other hand leads to typical regions that almost completely contain the estimated spectrum (note that typical regions are computed pointwise in $\btau, \omega$, so the estimator can occasionally be just outside of the boundary of typical regions). 

\begin{figure}[h!]
\includegraphics[width = 0.5\textwidth]{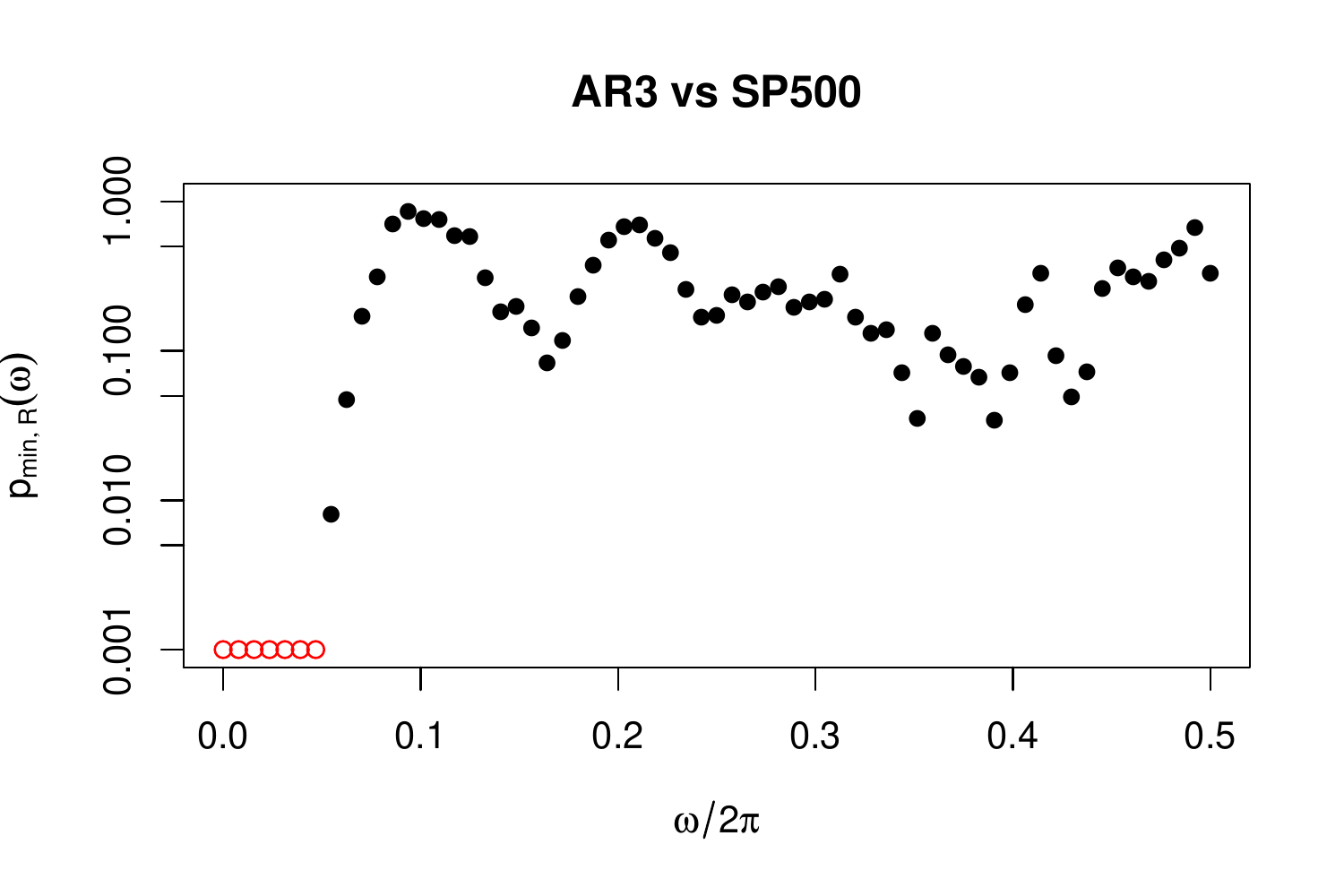}  
\includegraphics[width = 0.5\textwidth]{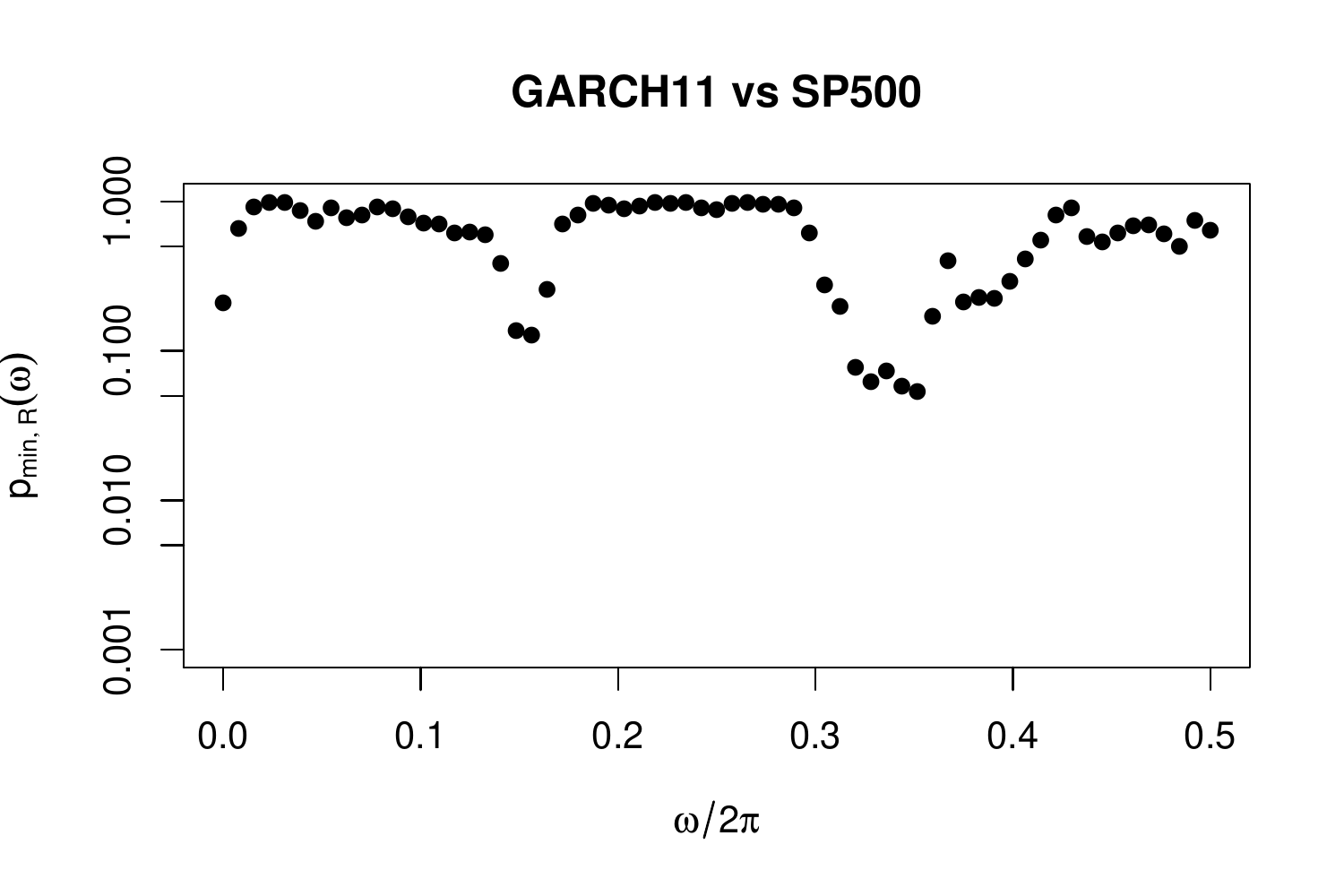}
\vspace*{-1cm}
\caption{Summary plot produced by Algorithm~\ref{alg2} on data generated from a GARCH(1,1) model with $n = 1024$ observations. We are fitting an AR(3) model (left) and a GARCH(1,1) model (right).}
\label{fig.example_tool2}
\end{figure}

\begin{figure}[h!]
\includegraphics[width = 0.5\textwidth]{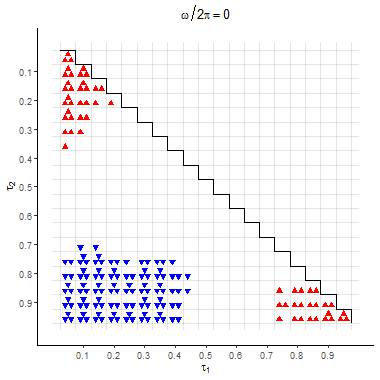}
\includegraphics[width = 0.5\textwidth]{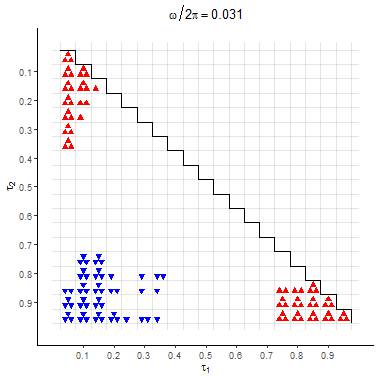}
\caption{Detailed plots for two frequencies produced by Algorithm~\ref{alg2} on data generated from a GARCH(1,1) model with $n = 1024$ observations with fitted AR(3) model (compare left panel of Figure~\ref{fig.example_tool2}).}
\label{fig.example_tool3}
\end{figure}

The output of Algorithm~\ref{alg2} for the same data set and models is illustrated in Figures~\ref{fig.example_tool2} and~\ref{fig.example_tool3}. First, consider the summary plots in Figure~\ref{fig.example_tool2} with frequencies on the x-axis and $p_{\min,R}(\omega)$ (see Algorithm~\ref{alg2}) on the y-axis; for better visibility of very low values the y-axis is in log-scale. By Proposition~\ref{prop:just}, proved below, the values on the y-axis can be interpreted as p-values (uniform in $\btau$ and pointwise in $\omega$) of a test for the null hypothesis that the data was generated from a model in the given parametric class against a non-parametric alternative. The left panel of Figure~\ref{fig.example_tool2} shows the plot corresponding to an AR(3) model class. This plot clearly indicates that the candidate model class does not match the data; this is particularly visible at the lowest frequencies where several p-values in a row are below $0.001$. In contrast to that, the right panel which uses the true model class does not show evidence of a miss-specified model.

The plots in Figure~\ref{fig.example_tool3} provide more detailed information about the quantile levels at which a mismatch between the data-based spectrum and a spectrum from the candidate parametric model is detected for the frequencies $\omega = 0, \omega = 4\pi/64$. Here, blue triangles facing down indicate that the data-based spectrum is smaller compared to the candidate model spectrum (with $1, 2$ and $3$ triangles indicating significance at the $5\%, 1\%$ and $0.1\%$ level, respectively) while red triangles facing up indicate a data-based spectrum which is larger compared to the candidate model spectrum. The corresponding plots reveal that most of the disagreement between data and model dynamics happens in the real parts of spectra corresponding to quantile levels $(\tau_1,\tau_2)$ where both $\tau_1$ and $\tau_2$ are either unexpectedly small or unexpectedly large. This confirms the first impression obtained from the $3 \times 3$ plots in Figure~\ref{fig.example_tool} and provides a more detailed view of the quantiles where data and model spectra disagree.

\section{Formal justification of graphical tools} \label{chap:asymptotics}
In this section we present a formal justification for the graphical approaches introduced in Section~\ref{sec:tool}. Denote by $\Theta \subset \R^d$ a candidate parameter space. For any $\theta \in \Theta$ let $(X_t^\theta)_{t \in \mathbb{Z}}$ be a strictly stationary process distributed according to $P^{\theta}$. Furthermore, let $F^{\theta}$ denote the cumulative (marginal) distribution function of $X_{t}^\theta$ and denote by $F_h^{\theta}$ the bivariate distribution function of $(X_{t+h}^\theta, X_t^\theta)$. Let $C_h^{\theta}$ denote the copula of $(X_{t+h}^\theta, X_t^\theta)$. We denote the copula spectral density of the process $X_{t}^\theta$ by 
\[
f_{\btau}^{\theta}(\omega) : = \frac{1}{2\pi} \sum_{k \in \Zb} (C_k^{\theta}(\tau_1,\tau_2) - \tau_1\tau_2) {\rm e}^{-{\rm i}k\omega}.
\]
The corresponding estimator $\hat f^{\theta}$, which is computed from $X_{1}^\theta,...,X_{n}^\theta$, is denoted by $\hat f_{\btau}^{\theta}$. We make the following technical assumptions

\begin{description}
\item[(LC)] The copulas $C_h^{\theta}$ are Lipschitz continuous with respect to the parameter $\theta$ in a neighborhood of $\theta_0$ uniformly in $\mathcal{T} \subseteq [0,1]^2$, i.e. there exist constants $\eps > 0, L < \infty$ such that $\|\theta - \theta_0\| \leq \eps$ 
\[
\sup_{h \in \Z} \sup_{\btau \in \mathcal{T}}|C_h^{\theta}(\btau) - C_h^{\theta_0}(\btau)| \leq L\|\theta - \theta_0\|.
\]
\item[(C)] For any $p \in \mathbf{N}$ there exist constants $\rho_p \in (0,1)$ and $K_p < \infty$ such that, for arbitrary intervals $A_1,\dots,A_p \subset \Rb$ and arbitrary $t_1,\dots,t_p \in \Zb,$
\[
\sup_{\|\theta - \theta_0\| \leq \eps}|\cum(\Ind{X_{t_1}^\theta \in A_1},\dots,\Ind{X_{t_p}^\theta \in A_p})| \leq K_p \rho_p^{\max_{i,j}|t_i-t_j|}.
\]

\item[(W)] The weight function $W$ is real-valued and even with support $[-\pi,\pi];$ moreover it has bounded variation and satisfies $\int W(u) {\rm d}u = 1.$ We denote by $b_n$ a sequence of scaling parameters such that $b_n \rightarrow 0$ and $nb_n \rightarrow \infty$, and assume that $W_n$ in~\eqref{eq:deffn} takes the form
\[
W_n(u) := \sum_{j \in \Zb} b_n^{-1}W[b_n^{-1} (u +2\pi j) ].  
\]
\end{description}

\begin{rem}
Assumption (C) is fulfilled under certain mixing assumptions (see Propositions~3.1 and~3.2 in \cite{kvdh2014}) and (W) places restriction on the smoothing function which are standard in time series analysis (see for instance page 147 of \cite{brill}). (LC) assures that if $\theta_n$ is close to $\theta_0$ we also have that the corresponding copula spectral densities are close. Below we show that this assumption is satisfied for ARMA(p,q) processes with normal innovations.   
\end{rem}

\begin{ex}\label{lem:arma}
{\rm Let $(X_t^{\theta})_{t \in \mathbb{Z}}$ be a strictly stationary ARMA(p,q) process where $\theta = (a_1,\dots,a_p,b_1,\dots,b_q)$ denotes the AR and MA coefficients, that means $X_t^{\theta}$ solves
\begin{equation}\label{WeWillShowThis}
X_t^{\theta} - \sum_{j=1}^p a_j X_{t-j}^{\theta} = \epsilon_t  + \sum_{i=1}^q b_i \epsilon_{t-i},
\end{equation}
where and $\epsilon_t$ are centered i.i.d $\mathcal{N}(0,1)$ random variables. Using the backshift operator $B$ we can write this as
$
P^{\theta}(B)X_t^{\theta} = Q_\theta(B)\epsilon_t,
$
where $P^{\theta}$ and $Q^{\theta}$ are the polynomials,
\[
P^{\theta}(z) := 1-a_1z - \dots - a_pz^p, \quad Q^{\theta}(z) := 1 + b_1z + \dots + b_q z^q, \quad z \in \mathbb{C}.
\]
To guarantee the existence of a unique strictly stationary and causal solution (see \cite{BrockLind}) we assume that $\Theta$ is a set such that for all $\theta \in \Theta$ the polynomials $P^{\theta}$ and $Q^{\theta}$ have no common roots and $P^{\theta}(z)$ only has roots outside roots the unit circle $\{z \in \mathbb{C}: |z| \leq 1\}$. Under these conditions (LC) holds for any open $\mathcal{T} \subset [0,1]^2$ and any $\theta_0$ in the interior of $\Theta$. This statement will be proved in Section~\ref{sec:proofex}}
\end{ex}

The main result in this section is Proposition~\ref{prop:just}. It implies that, if the parametric model is specified correctly, the intervals $[l^\Re_{\btau,R}(\omega), u^\Re_{\btau,R}(\omega)]$ and $[l^\Im_{\btau,R}(\omega), u^\Im_{\btau,R}(\omega)]$ will (asymptotically) contain the real and imaginary parts of the estimator $\hat f_{\btau}(\omega)$ with given probability $\alpha$. This provides a formal justification for the graphical approach introduced in Algorithm~\ref{alg1}. The second part of Proposition shows that the output of Algorithm~\ref{alg2} can indeed be interpreted as p-values for the null that the class of candidate models contains the true model.

\begin{prop}\label{prop:just}
Assume that the data $X_1,\dots,X_n$ are generated from the model $P^{\theta_0}$ and let $\hat \theta_n$ be a $\sqrt{n}$-consistent estimator of $\theta_0.$ Let assumptions $(LC),(C),(W)$ hold, assume that $R = R_n \to \infty$ as $n \to \infty$ and that there exist constants $k \in \Nb$ and $\kappa > 0$ with
\[ 
b_n = o(n^{-1/(2k+1)}) \qquad b_n n^{1-\kappa} \rightarrow \infty. 
\]
Then, for $l_{\btau,R_n}(\omega), u_{\btau,R_n}(\omega)$ defined in Algorithm~\ref{alg1} we have, as $n \rightarrow \infty$, for any $\btau \in \mathcal{T}, \omega \in \R$ with $Var(\Re H_0(\btau,\omega)) \neq 0$,
\begin{equation}\label{eq:main}
P\Big( l_{\btau,R_n}^\Re(\omega) \leq \Re\hat f_{\btau}(\omega) \leq u_{\btau,R_n}^\Re (\omega) \Big) \to 1 - \alpha. 
\end{equation}
The same holds for the imaginary parts. If additionally $\min_{\btau \in M} Var(\Re H_0(\btau,\omega)) > 0$ and $\min_{\btau \in M} Var(\Im H_0(\btau,\omega)) > 0$ then, for $p_{\btau,R}^\Re(\omega),p_{\btau,R}^\Im(\omega)$ defined in Algorithm~\ref{alg2} we have, as $n \rightarrow \infty$ and any $\omega \in \R$,
\begin{equation}\label{eq:main:unif}
P\Big(\min_{\btau \in M} \min\{p_{\btau,R}^\Re(\omega),p_{\btau,R}^\Im(\omega)\} < \alpha \Big) \to \alpha.
\end{equation}
\end{prop}

%\begin{prop} \label{prop:uniftau}
%Assume that the data $X_1,...,X_n$ are generated from the model $P^{\theta_0}$ and let $\hat \theta_n$ be a $\sqrt{n}$ consistent estimator for $\theta_0$. Let (LC), (C), (W) hold and assume that $R = R_n \to \infty$ as $n \to \infty$ and assume that there exist constants $k \in \N, \kappa> 0$ such that \sv{add this assumptions to Prp 3.1 as well}
%\[
%b_n = o(n^{-1/(2k+1)}), \quad  b_n n^{1-\kappa} \to \infty.
%\] 
%Then for $p_{2,\btau,R}^\Re(\omega), p_{2,\btau,R}^\Im(\omega)$ defined in Algorithm~\ref{alg2} we have  
%\[
%P\Big(\min_{\btau \in M} \min\{|p_{2,\btau,R}^\Re(\omega)|,|p_{2,\btau,R}^\Im(\omega)|\} < \alpha \Big) \to \alpha.
%\]
%\end{prop}

The key technical ingredient for proving Proposition~\ref{prop:just} is given by the following theorem. It provides a generalization of Theorem 3.6 in \cite{kvdh2014} to a particular kind of triangular array asymptotics. This result is of independent interest, and hence we chose to state it separately.

\begin{thm} \label{thm:main}
Let assumptions $(LC),(C),(W)$ hold, and assume that there exist constants $k \in \Nb$ and $\kappa > 0$ with
\[ 
b_n = o(n^{-1/(2k+1)}) \qquad b_n n^{1-\kappa} \rightarrow \infty. 
\]
If $\theta_n = \theta_0 + O(n^{-1/2})$ then, for $\mathcal{T}$ from assumption (LC),
\begin{equation*}
\sqrt{nb_n}(\hat f_{\btau}^{\theta_n}(\omega) - f_{\btau}^{\theta_0}(\omega) - B_n^{(k)}(\btau,\omega) )_{\btau \in \mathcal{T}} \leadsto H_0(\cdot;\omega)
\end{equation*}
in $\ell^\infty(\mathcal{T})$ where
\[
B_n^{(k)}(\btau,\omega) := \begin{cases}
\sum_{j=2}^k \frac{b_n^j}{j!} \int_{-\pi}^\pi u^jW(u) {\rm d}u (f^{\theta_0}_{\btau})^{(j)}(\omega) & \omega \neq 0\mod 2\pi \\
\frac{n}{2\pi}\tau_1\tau_2 & \omega = 0\mod 2\pi
\end{cases} 
\]
and $H_0(\cdot;\omega)$ is a complex-valued, centered Gaussian process characterized by 
\begin{align*}
& \Cov(H_0((u_1,v_1);\omega),H_0((u_2,v_2);\omega))
= 2\pi \int_{-\pi}^\pi W^2(u) {\rm d}u \\
& \quad \times \Big[ (f_{(u_1,u_2)}^{\theta_0}(\omega)f_{(v_1,v_2)}^{\theta_0}(-\omega))
+ (f_{(u_1,v_2)}^{\theta_0}(\omega)f_{(v_1,u_2)}^{\theta_0}(-\omega))\Ind{\omega = 0 \mod \pi}\Big].
\end{align*}
\end{thm}

%\newpage

\section{Simulation study and data example} \label{chap:simandex}

In this section we present a simulation study and an application to the returns of the S\&P~500 stock index between 2000 and 2005 and 1966 and 1970. 

\subsection{Real data example: S\&P~500 returns}

In this section we demonstrate how our graphical tools can be utilized to find an appropriate parametric model for a given time series and further provide an example where none of the standard models seem to work. To this end we consider the daily log-returns of the S\&P~500 between 03.01.2000 and 30.12.2005 (corresponding to $n = 1508$ observations) as well as between 03.01.1966 and 31.12.1970 (corresponding to $n = 1233$ observations). Throughout this section we use the Epanechnikov kernel for $W_n,$ a moderate bandwidth $b_n = 0.1$ and set $\alpha = 0.05$ in Algorithm~\ref{alg1}.

\begin{figure}[h!]
\hspace{0.22\textwidth}(a) \hspace{0.5\textwidth} (b)

\hspace*{-0.8cm}\includegraphics[width = 0.55\textwidth]{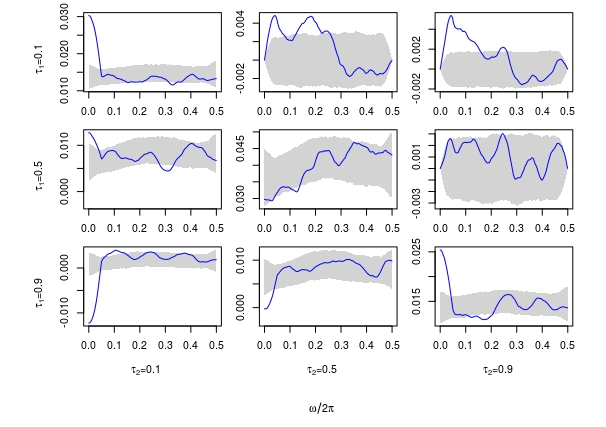}
\includegraphics[width = 0.55\textwidth]{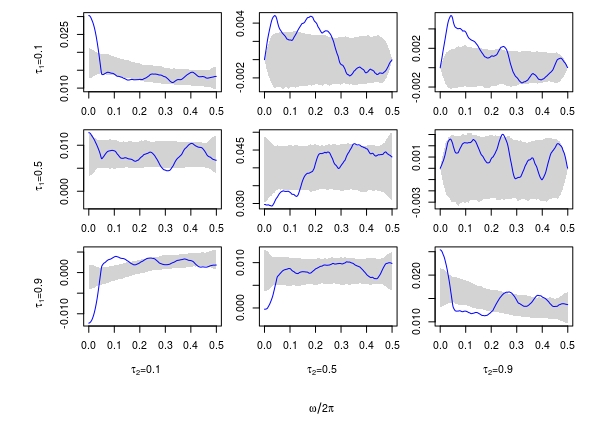}

\hspace{0.22\textwidth}(c) \hspace{0.5\textwidth} (d)

\hspace*{-0.8cm}\includegraphics[width = 0.55\textwidth]{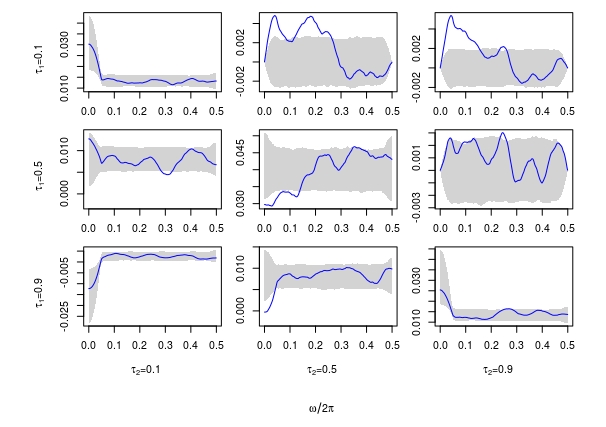}
\includegraphics[width = 0.55\textwidth]{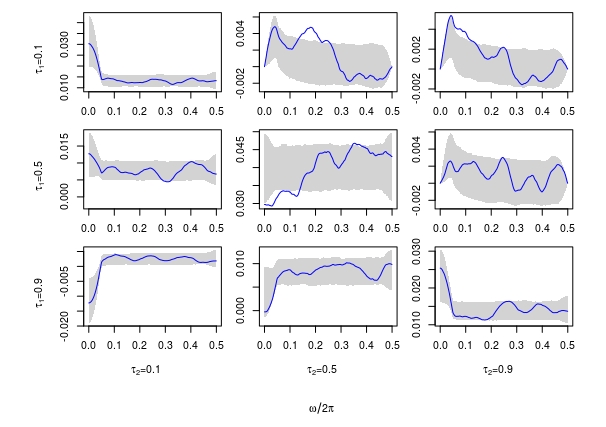}
\vspace*{-1cm}
\caption{Estimated copula spectral densities based on the daily log-returns of the S\&P~500 between 2000 and 2005. Figure displays the plots produced by Algorithm~\ref{alg1} for the following model classes (a) AR(3), (b) ARCH(1), (c) GARCH(1,1) and (d) EGARCH(1,1).} \label{fig.data_example}

\end{figure}

\begin{figure}[h!]
\begin{center}
\hspace{0.03\textwidth} (a) \hspace{0.42\textwidth} (b)

\vspace{-.2cm}
\includegraphics[width = 0.45\textwidth]{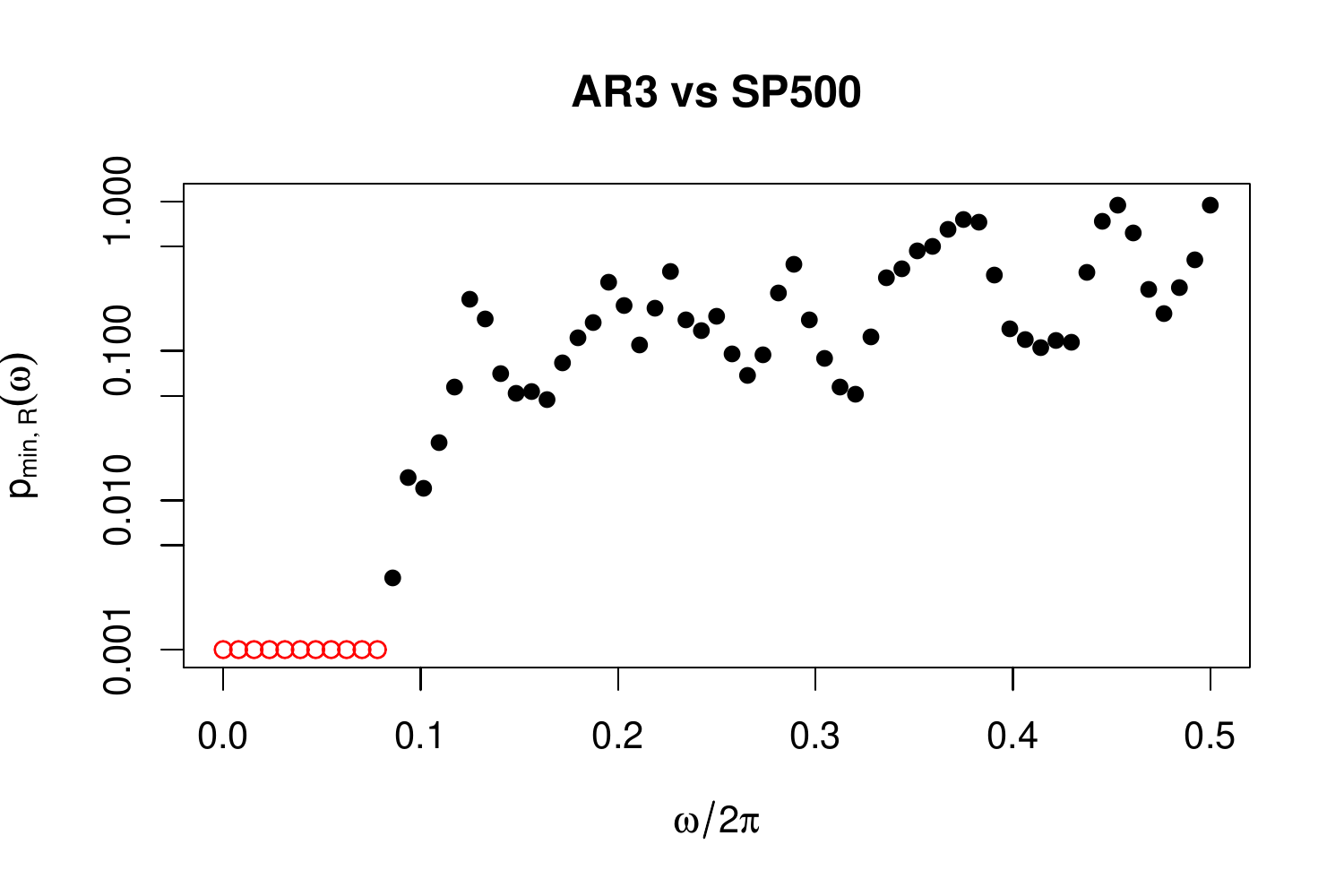}
\includegraphics[width = 0.45\textwidth]{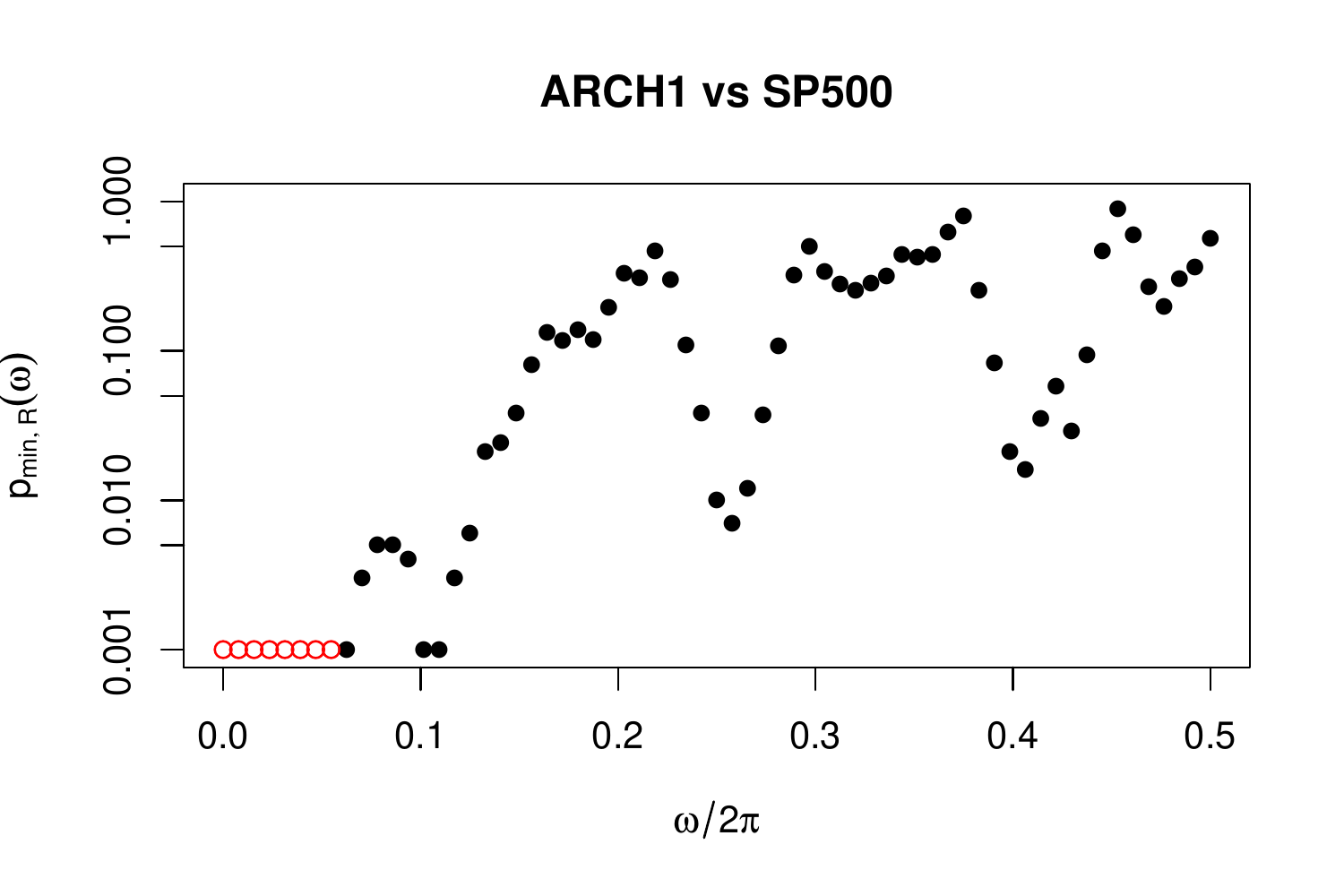}

\hspace{0.03\textwidth} (c) \hspace{0.42\textwidth} (d)

\vspace{-.2cm}
\includegraphics[width = 0.45\textwidth]{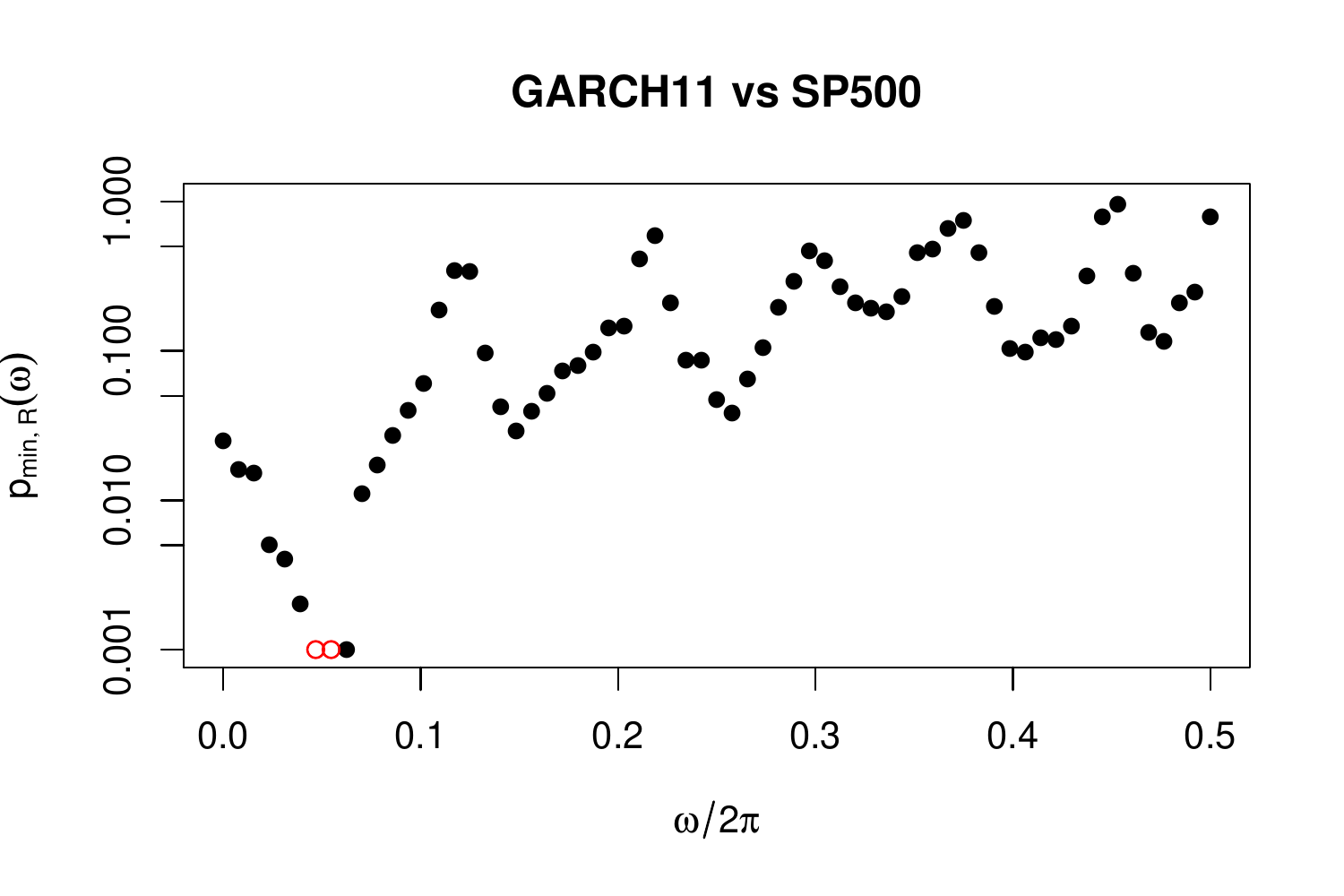}
\includegraphics[width = 0.45\textwidth]{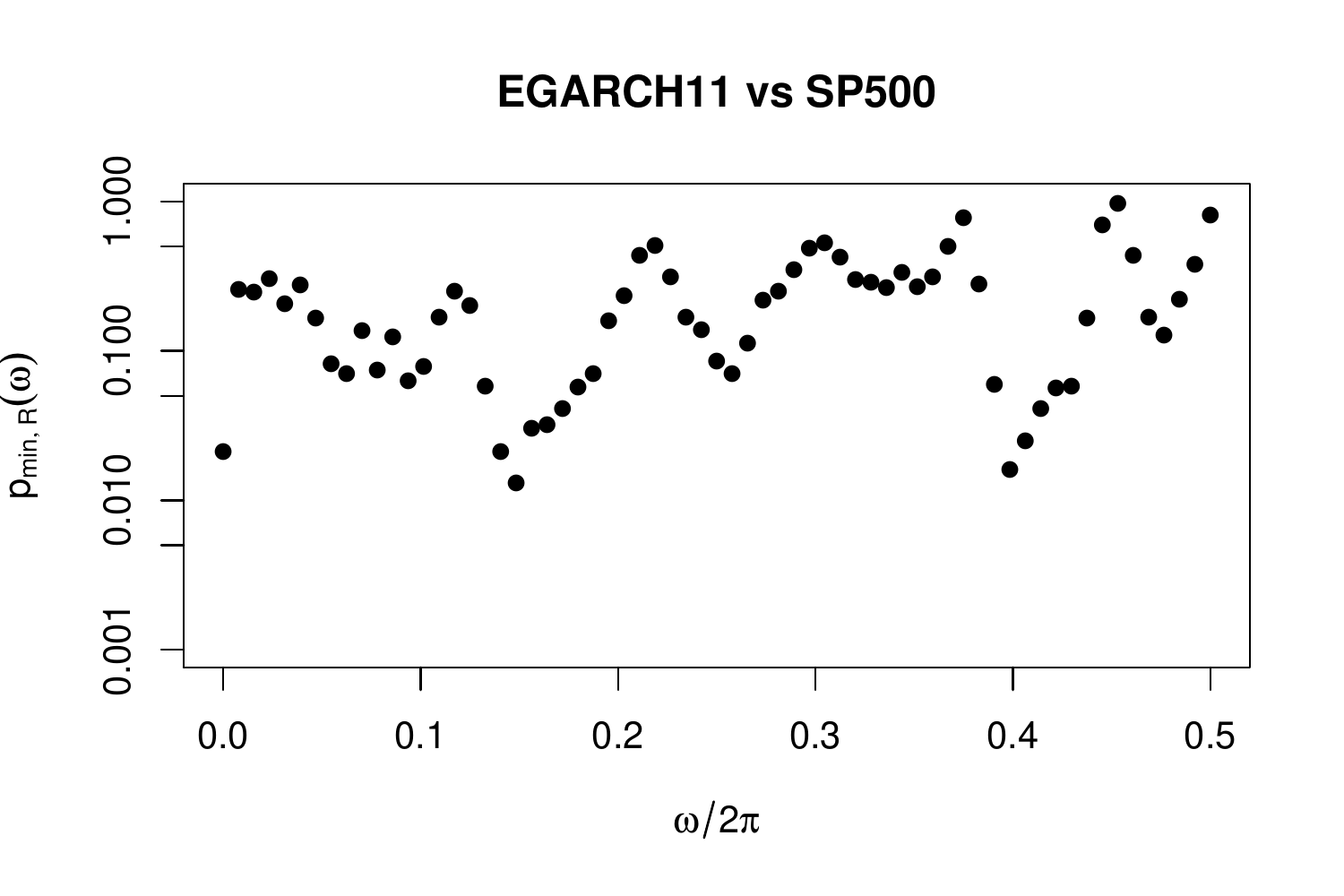}
\vspace*{-.8cm}
\end{center}
\caption{Summary plots produced by Algorithm~\ref{alg2} based on the daily log-returns of the S\&P~500 between 2000 and 2005. Figure from top left to bottom right correspond to the following candidate model classes (a) AR(3), (b) ARCH(1), (c) GARCH(1,1) and (d) EGARCH(1,1).} \label{fig.data_example2}

\end{figure}

\begin{figure}[h!]
\includegraphics[width = 0.5\textwidth]{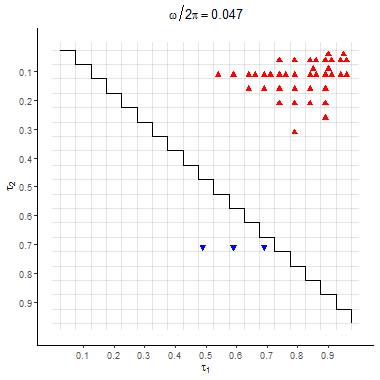}
\includegraphics[width = 0.5\textwidth]{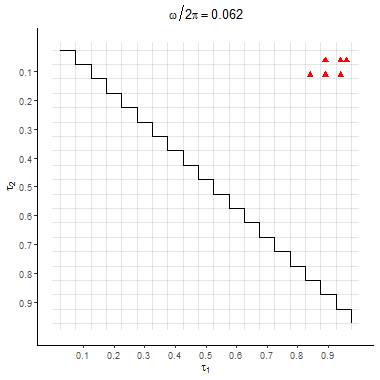}
\caption{Detailed plots produced by Algorithm~\ref{alg2} at two particular frequencies based on the daily log-returns of the S\&P~500 between 2000 and 2005 with GARCH(1,1) as candidate model class.} \label{fig.data_example3}

\end{figure}

We first consider the daily log-returns of the S\&P~500 between 03.01.2000 and 30.12.2005. Assuming for the moment that we have no clue about financial time series we first attempt to fit an AR(3) model. Algorithm~\ref{alg1} with this model class produces Figure~\ref{fig.data_example}(a). This clearly shows that an AR(3) manages to capture the ``median dependence'' but can not account for the strong dependencies observed at $\btau = (0.1,0.1),(0.9,0.9)$ and $(0.9,0.1)$. This is further confirmed in the output produced by Algorithm~\ref{alg2} (see Figure~\ref{fig.data_example2}(a)). The most basic model which has the potential to model such dependencies is an ARCH(1) model, which is our next candidate. Plot (b) in Figure~\ref{fig.data_example} indeed shows that an ARCH(1) model produces the peaks around frequency $\omega = 0,$ but those peaks are not high enough to match the data, this is again confirmed by the summary plot from Algorithm~\ref{alg2} provided in Figure~\ref{fig.data_example2}(b). Our next try is a GARCH model which was specifically designed to model the types of dependence observed in financial data. Figure~\ref{fig.data_example}(c) shows that this model is well suited to reproduce the peaks for $\btau = (0.1,0.1),(0.9,0.9)$ and $(0.9,0.1).$ However, the imaginary parts of the spectra still don't match the data as can be seen from the part of Figure~\ref{fig.data_example}(c) corresponding to $\btau = (0.1,0.9)$; the mismatch between model and data dynamics is confirmed in the summary plot from Figure~\ref{fig.data_example2}(c). A closer look at the corresponding detailed plot in the top row of Figure~\ref{fig.data_example3} sheds light on the specific combinations of quantile levels for which a significant mismatch occurs.
Based on the discussion in Section~\ref{sec:interpr} about asymmetric dependence a reasonable model to try is an EGARCH(1,1) model. The output of Algorithm~\ref{alg1} in Figure~\ref{fig.data_example}(d) indeed indicates that among all models considered this leads to the best performance, although we still detect slight deviations for some of the imaginary parts. The impression that this model still does not provide a perfect fit is further strengthened by the summary plot in Figure~\ref{fig.data_example2}(d) where we see a fairly high proportion of p-values below $5\%$ which is much higher than we would expect even after adjusting for multiple testing across frequencies. %Detailed plots corresponding to several frequencies with low p-values are provided in Figure~\ref{fig:add_GARCH_Sp} of the online supplement. A close look at those plots reveals that most of the deviations take place for the imaginary parts of copula spectra although the deviation is less pronounced compared to the GARCH model.
%A closer look at more detailed plots corresponding to two of the frequencies with low p-values in the bottom row of Figure~\ref{fig.data_example2} further reveals that most of the deviations take place for the imaginary parts of copula spectra although the deviation is less pronounced compared to the GARCH model. This is confirmed by looking at figures corresponding to additional frequencies which are not displayed here due to space limitations. \textbf{Add those plots to supplement?}

\begin{figure}[h!]
\hspace{0.22\textwidth}(a) \hspace{0.5\textwidth} (b)

\vspace*{-0.4cm}\hspace*{-0.8cm}\includegraphics[width = 0.55\textwidth]{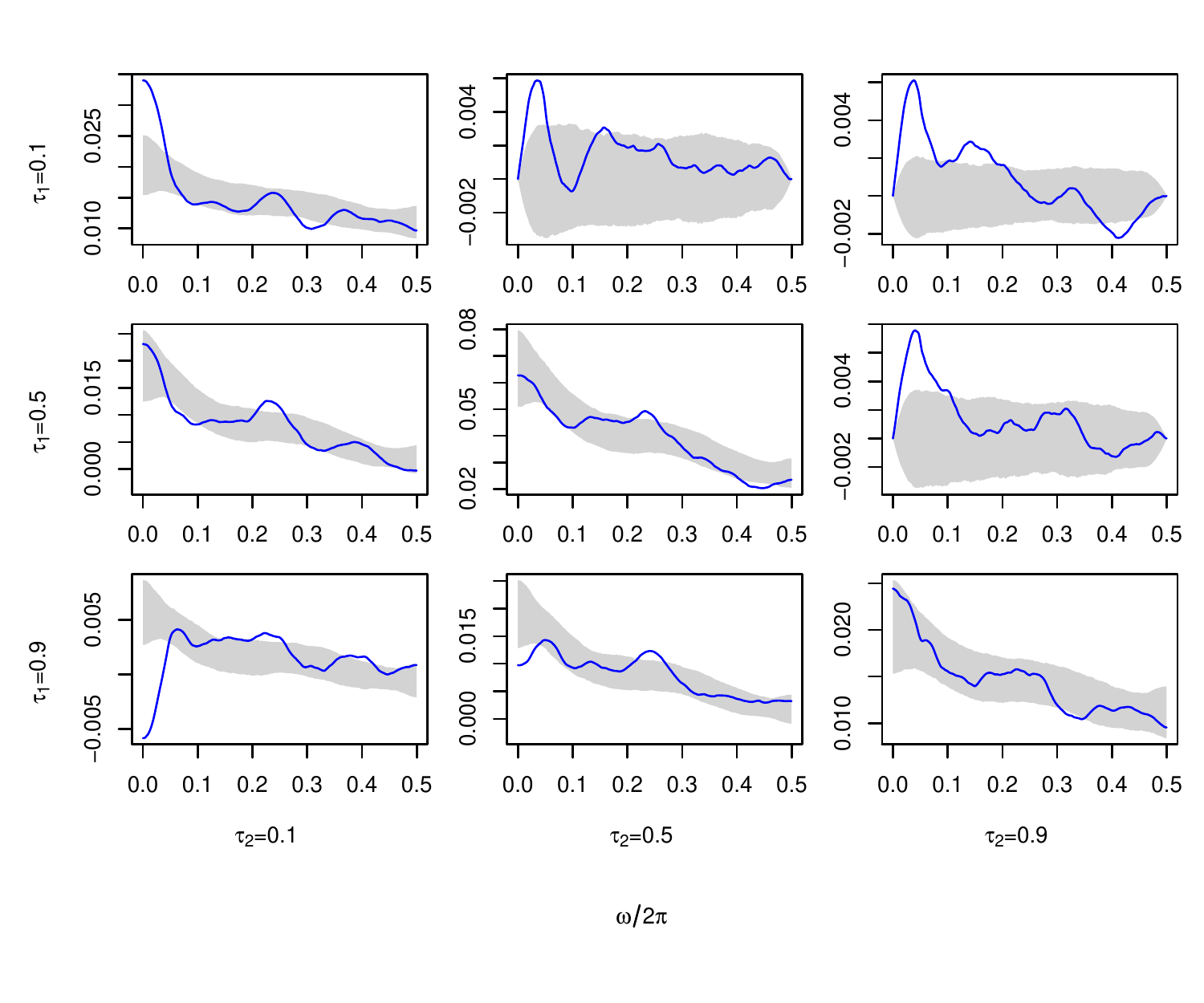}
\includegraphics[width = 0.55\textwidth]{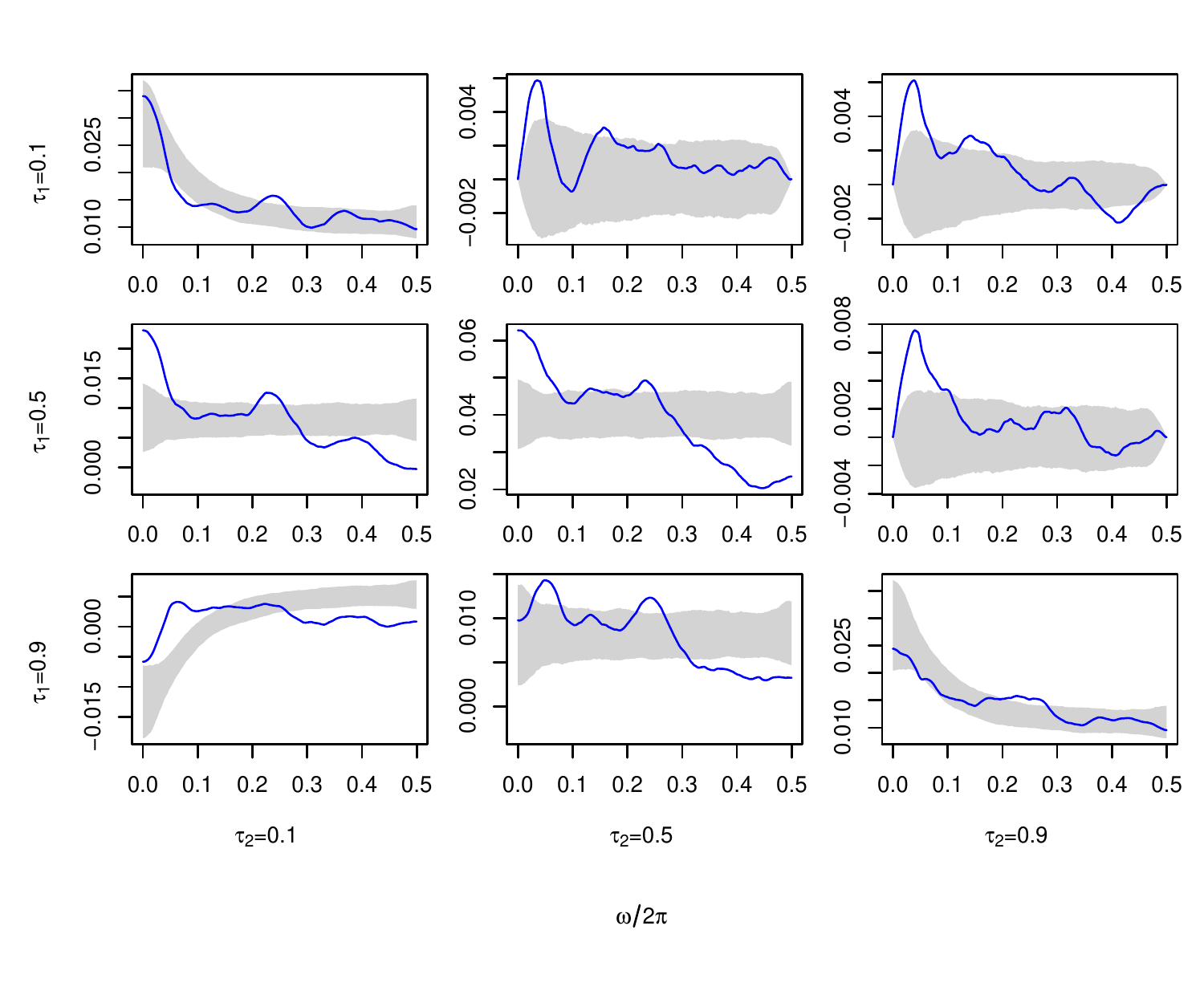}

\hspace{0.22\textwidth}(c) \hspace{0.5\textwidth} (d)

\vspace*{-0.4cm}\hspace*{-0.8cm}\includegraphics[width = 0.55\textwidth]{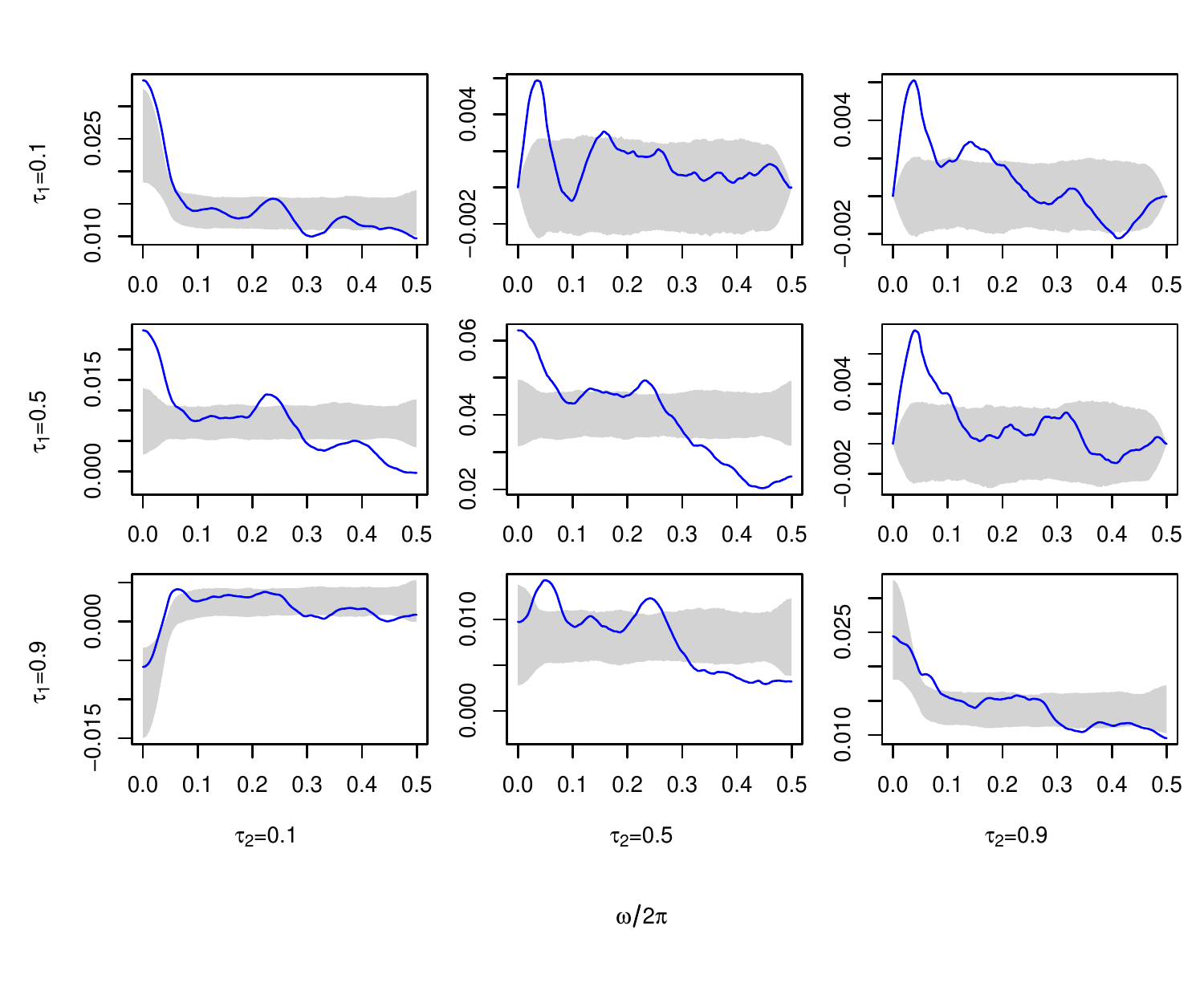}
\includegraphics[width = 0.55\textwidth]{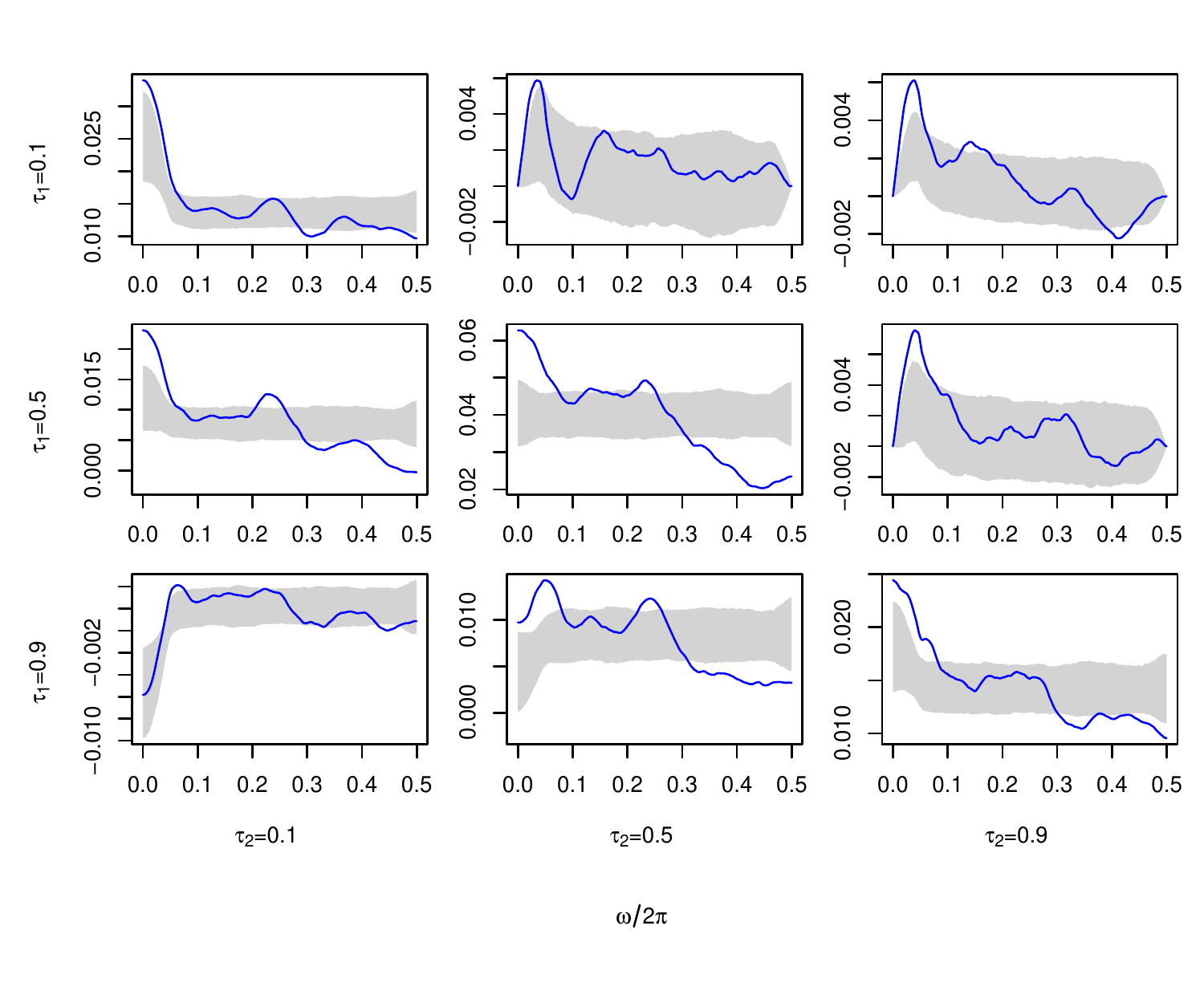}
\vspace*{-1cm}
\caption{Estimated copula spectral densities based on the daily log-returns of the S\&P~500 between 1966 and 1970. Figure displays the plots produced by Algorithm~\ref{alg1} for the following model classes (a) AR(3), (b) ARCH(1), (c) GARCH(1,1) and (d) EGARCH(1,1).} \label{fig.data_example_70}

\end{figure}

\begin{figure}[h!]
\begin{center}

\includegraphics[width = 0.45\textwidth]{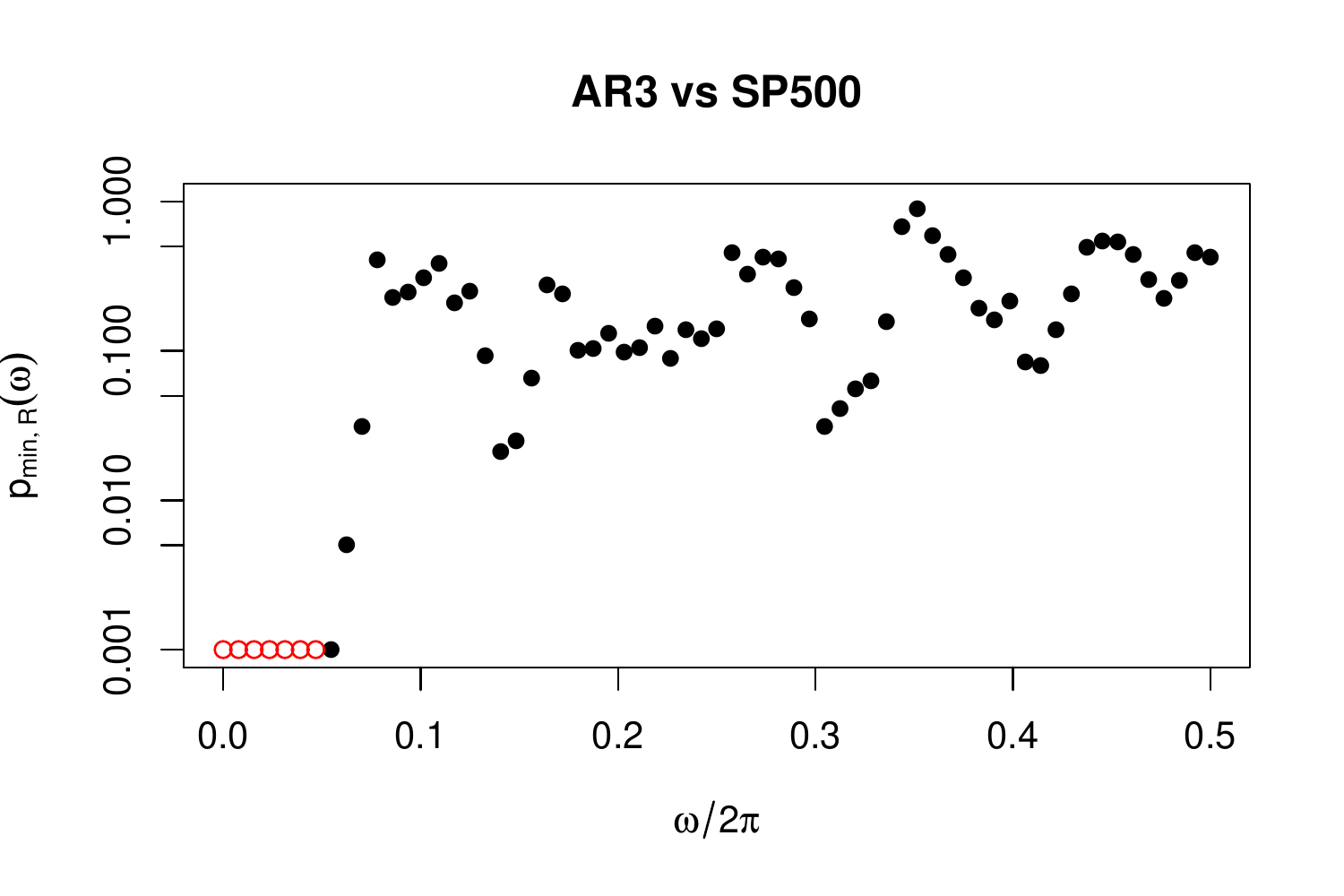}
\includegraphics[width = 0.45\textwidth]{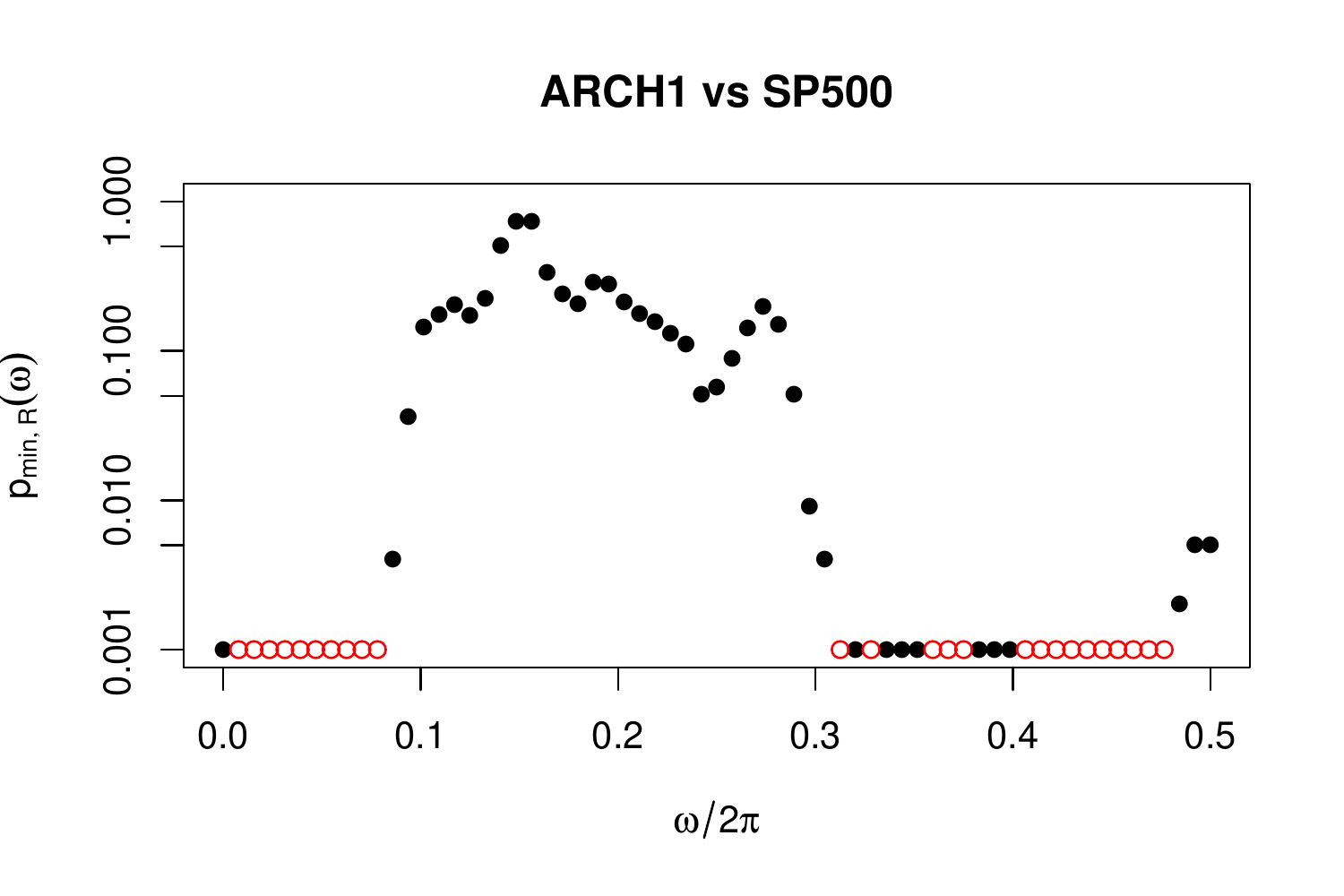}

\includegraphics[width = 0.45\textwidth]{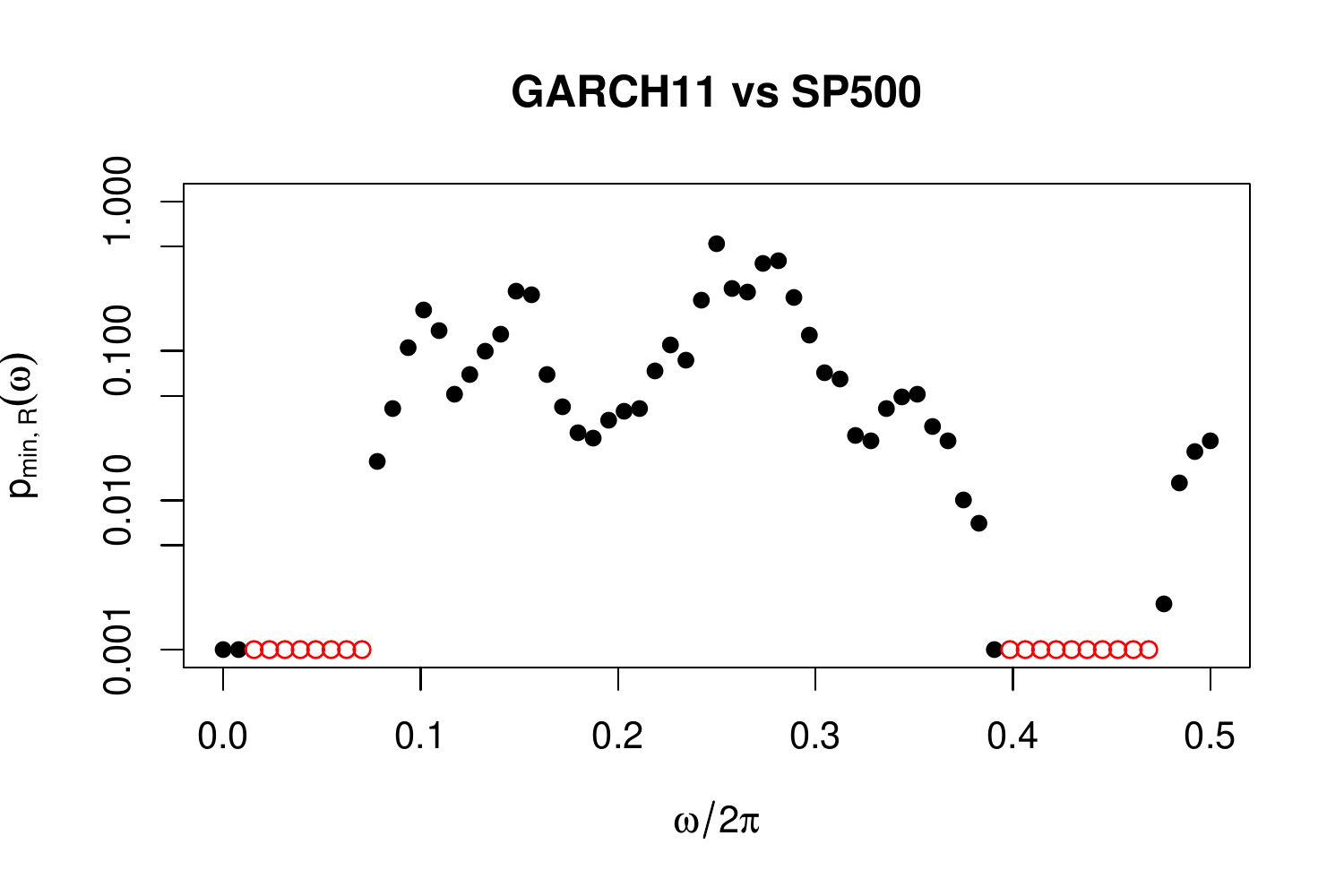}
\includegraphics[width = 0.45\textwidth]{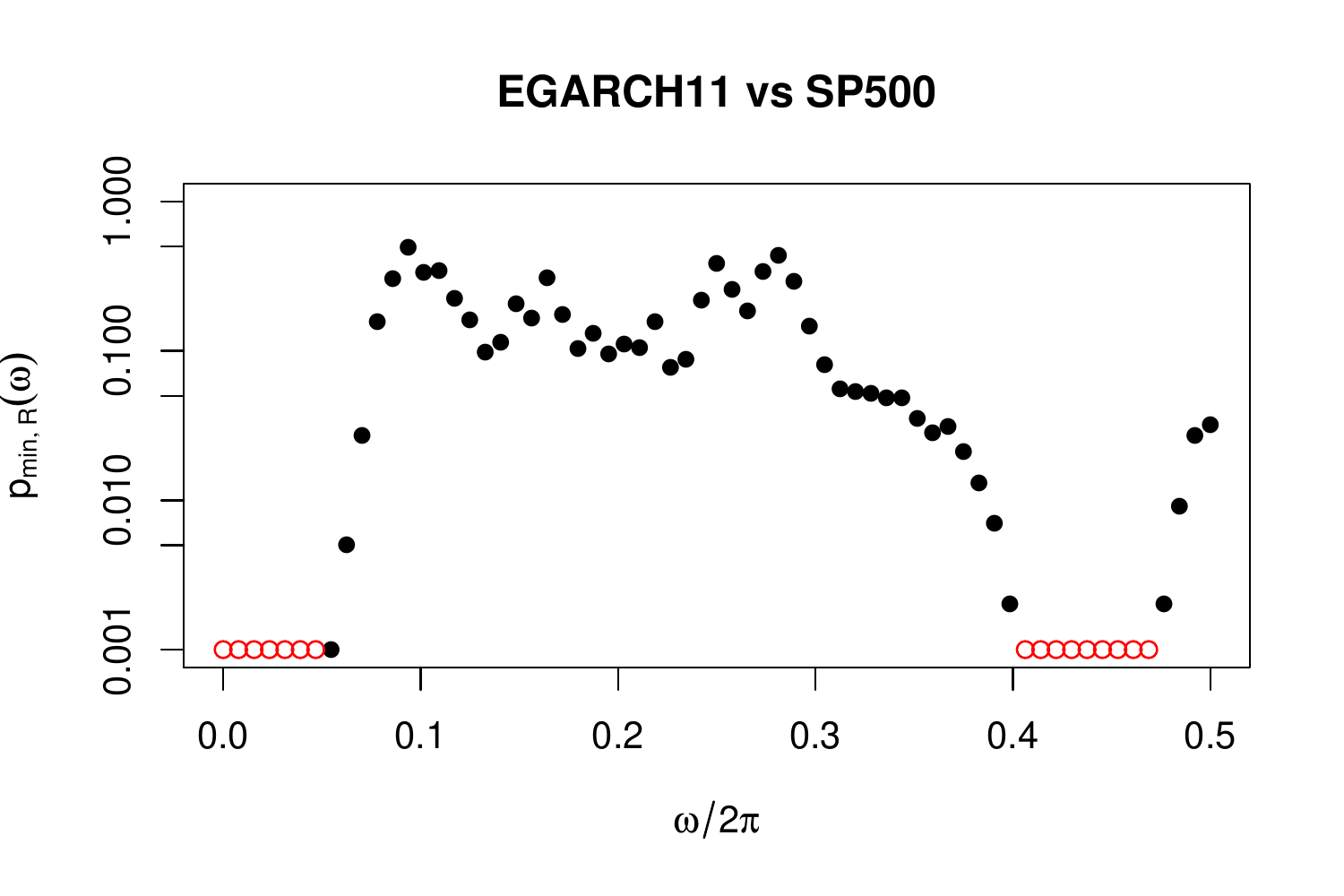}

\end{center}
\caption{Summary plots produced by Algorithm~\ref{alg2} based on the daily log-returns of the S\&P~500 between 1966 and 1970 with GARCH(1,1) as candidate model class.} \label{fig.data_example2_70}

\end{figure}

\begin{figure}[h!]
\begin{center}
\includegraphics[width = 0.3\textwidth]{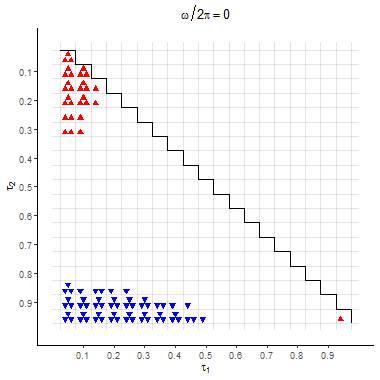}
\includegraphics[width = 0.3\textwidth]{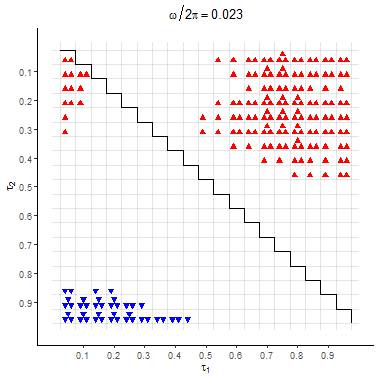}
\includegraphics[width = 0.3\textwidth]{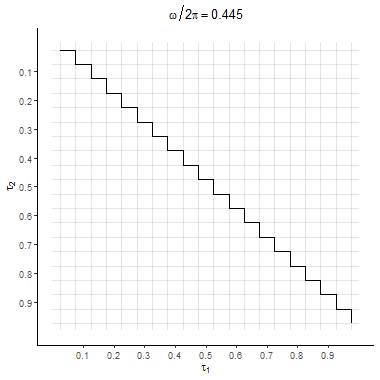}

\includegraphics[width = 0.3\textwidth]{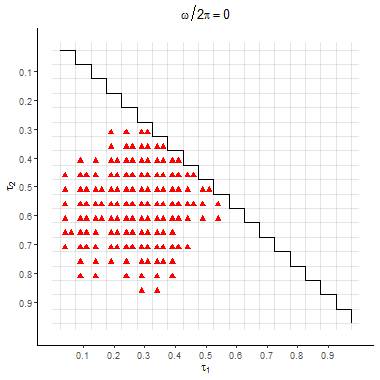}
\includegraphics[width = 0.3\textwidth]{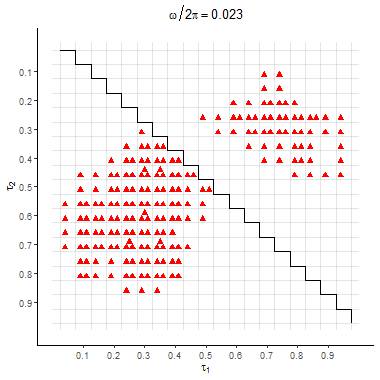}
\includegraphics[width = 0.3\textwidth]{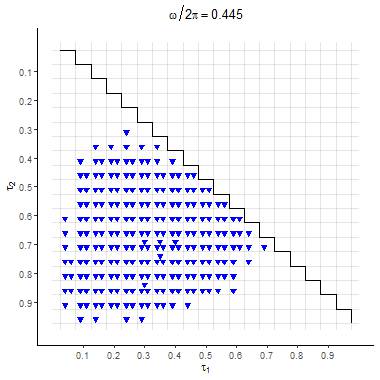}

\includegraphics[width = 0.3\textwidth]{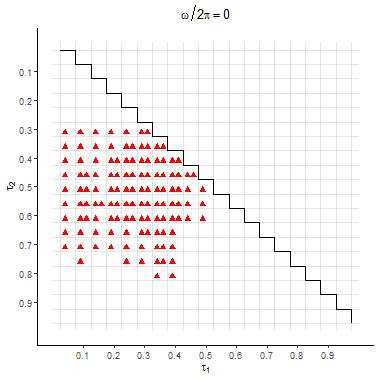}
\includegraphics[width = 0.3\textwidth]{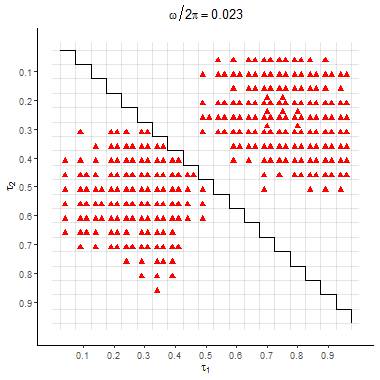}
\includegraphics[width = 0.3\textwidth]{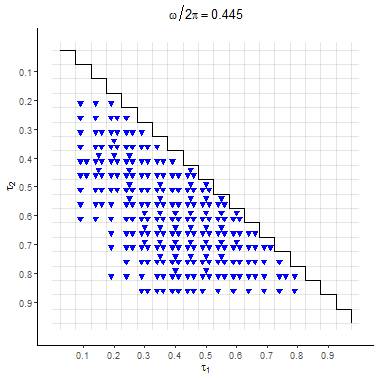}

\includegraphics[width = 0.3\textwidth]{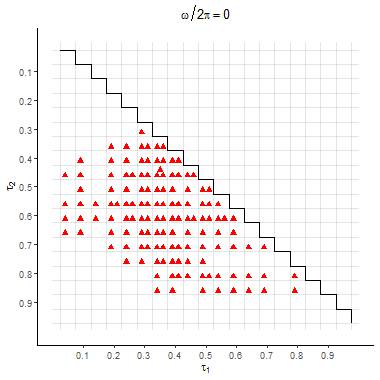}
\includegraphics[width = 0.3\textwidth]{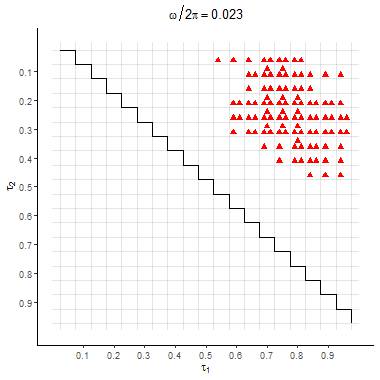}
\includegraphics[width = 0.3\textwidth]{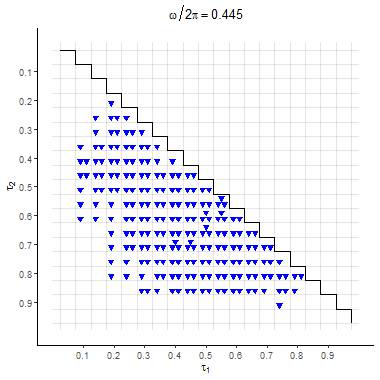}
\end{center}
\vspace*{-.8cm}

\caption{Detailed plots produced by Algorithm~\ref{alg2} at two particular frequencies based on the daily log-returns of the S\&P~500 between 1966 and 1970 with AR(3) (top row), ARCH(1) (second row), GARCH(1,1) (third row) and EGRACH(1,1) (bottom row) as candidate model classes.} \label{fig:add_Sp70}

\end{figure}

In the final part of this section, we consider the daily log-returns of the S\&P~500 between 1966 and 1970. The output of Algorithm~\ref{alg1} for the same four model classes as considered above is depicted in Figure~\ref{fig.data_example_70} (a)-(d). Interestingly, we find that none of the four model classes provide an adequate description of the dynamics observed in the data since the data contain both - linear type dynamics at the median level, but also strong GARCH-like tail dependencies and EGRACH-like imaginary parts (which are, however, appear to be too steep to be captured by an EGARCH(1,1) model) indicating a strongly asymmetric behaviour of the process going forward and backward in time. The inability of all considered models to capture the dynamics in the data is further confirmed by summary plots from Algorithm~\ref{alg2} as depicted in Figure~\ref{fig.data_example2_70}. Additional detailed plots from Algorithm~\ref{alg2} corresponding to specific frequencies are provided in Figure~\ref{fig:add_Sp70} in the online supplement. The middle column corresponding to $\omega = 4*\pi/64$ confirms that none of the considered models, including the EGARCH model, are able to produce a sufficiently sharp peak in the imaginary part which is observed in the spectrum of the data. The ARCH, GARCH and EGARCH models further struggle to produce the right amount of dependence at central quantile values while the AR(3) process does not have the right kind of dependence in low quantiles.

\newpage

\subsection{Simulation study} %\label{chap:simandex}

In this section we illustrate the finite sample properties of Algorithm~\ref{alg1} and Algorithm~\ref{alg2} with simulated data. First, we show that in settings where the data are generated from a model that is contained in the candidate parametric class, the simulated `typical regions' from Algorithm~\ref{alg1} contain the estimated spectral densities with probability $1-\alpha$ across a range of models, sample sizes and bandwidth parameters (note that this is counted pointwise in $\tau,\omega$). To this end, we consider the following three data generating processes.
\begin{align*}
(a_0)&\quad  X_t = 0.1X_{t-1} + 0.8Z_{t-1} +Z_t
\\
(b_0)& \quad X_t = 0.2X_{t-1} -0.4X_{t-2} + 0.2X_{t-3} + Z_t
\\
(c_0)& \quad X_t = \sigma_t Z_t, \textrm{ where } \sigma_t^2 = 0.01 + 0.4X_{t-1}^2 + 0.5\sigma_{t-1}^2
\end{align*}

In each case we simulate time series of length $n = 256,512,1024$ and consider the fixed bandwidth parameters $b_n = 0.01,0.02,0.05,0.1,0.4$. For each possible combination we simulate $1000$ repetitions of our algorithm with $\alpha = 0.05$ and the following candidate classes of parametric models (here, $\theta_j$ denote unknown parameters of the models)
\begin{align*}
(P_{a})&\quad  X_t = \theta_1 X_{t-1} + \theta_2 Z_{t-1} + Z_t, \quad Z_t \sim \mathcal{N}(0,1)
\\
(P_{b})& \quad X_t = \theta_1 X_{t-1} + \theta_2 X_{t-2} + \theta_3 X_{t-3} + Z_t, \quad Z_t \sim \mathcal{N}(0,1) 
\\
(P_{c})& \quad X_t = \sigma_t Z_t, \textrm{ where } \sigma_t^2 = \theta_0 + \theta_1 X_{t-1}^2 + \theta_2\sigma_{t-1}^2, \quad Z_t \sim \mathcal{N}(0,1) 
\end{align*}

\begin{figure}
\hspace*{-0.8cm} \includegraphics[width = 0.55\textwidth]{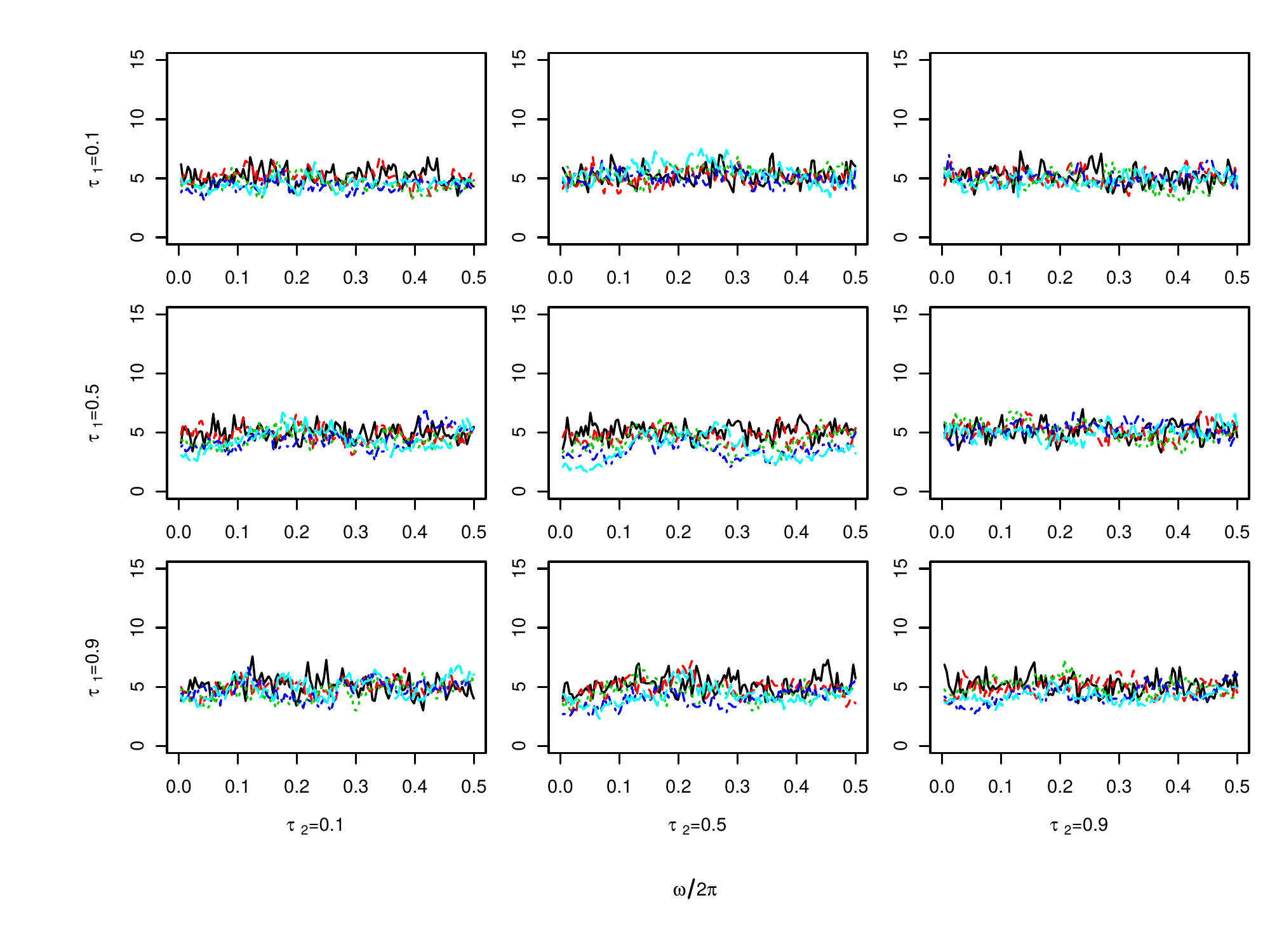}
\includegraphics[width = 0.55\textwidth]{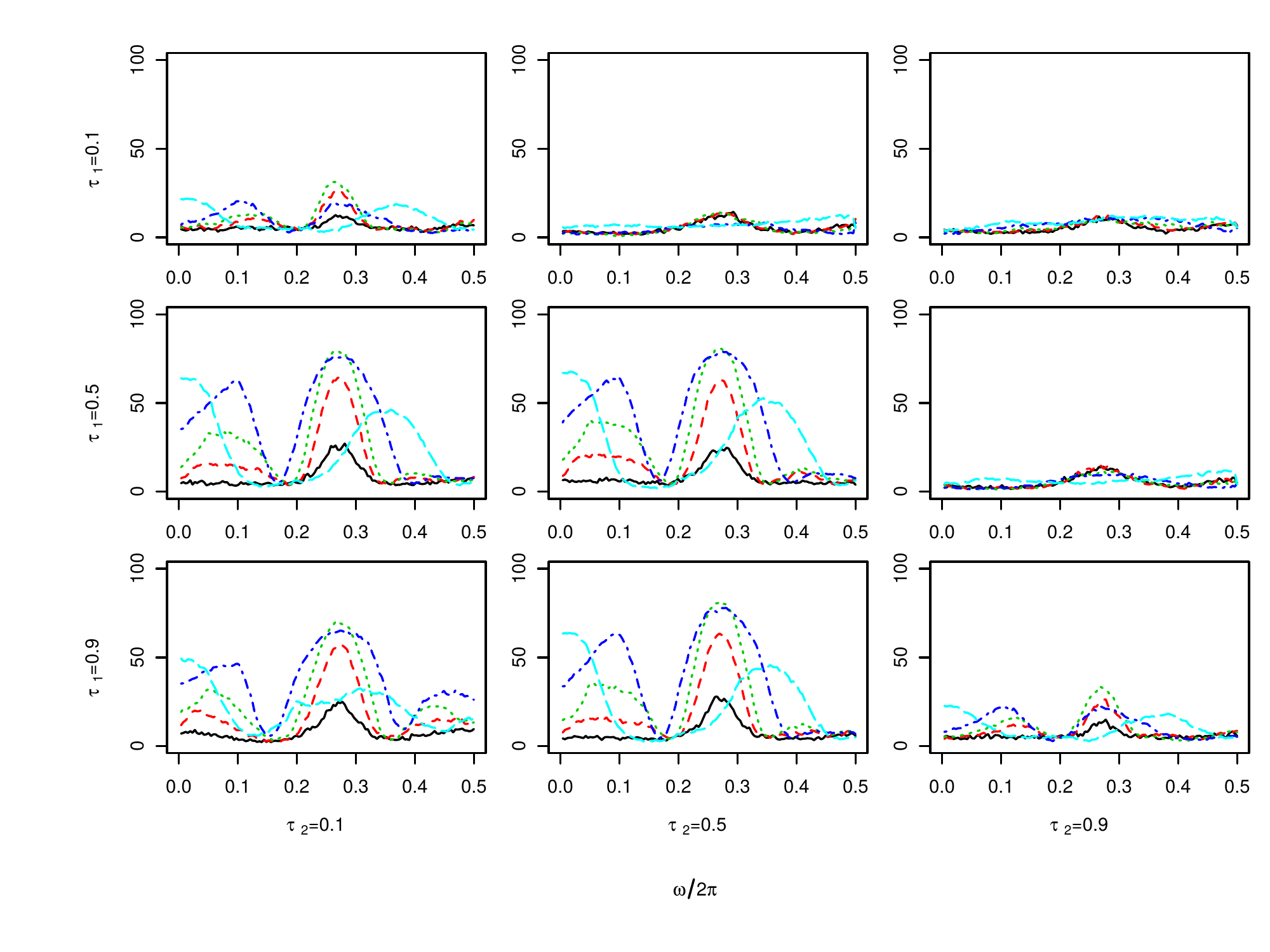}
\hspace*{-0.8cm} \includegraphics[width = 0.55\textwidth]{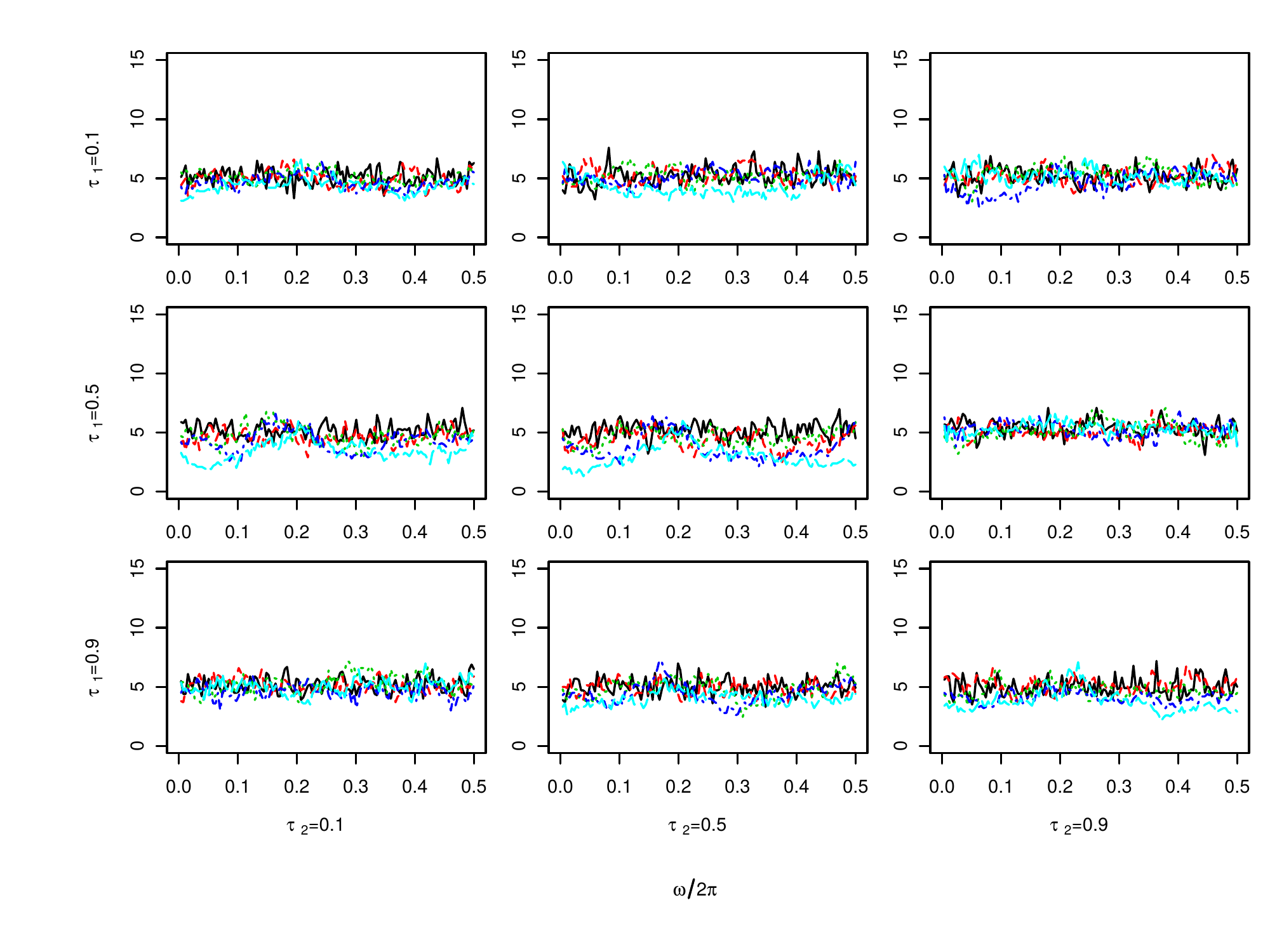}
\includegraphics[width = 0.55\textwidth]{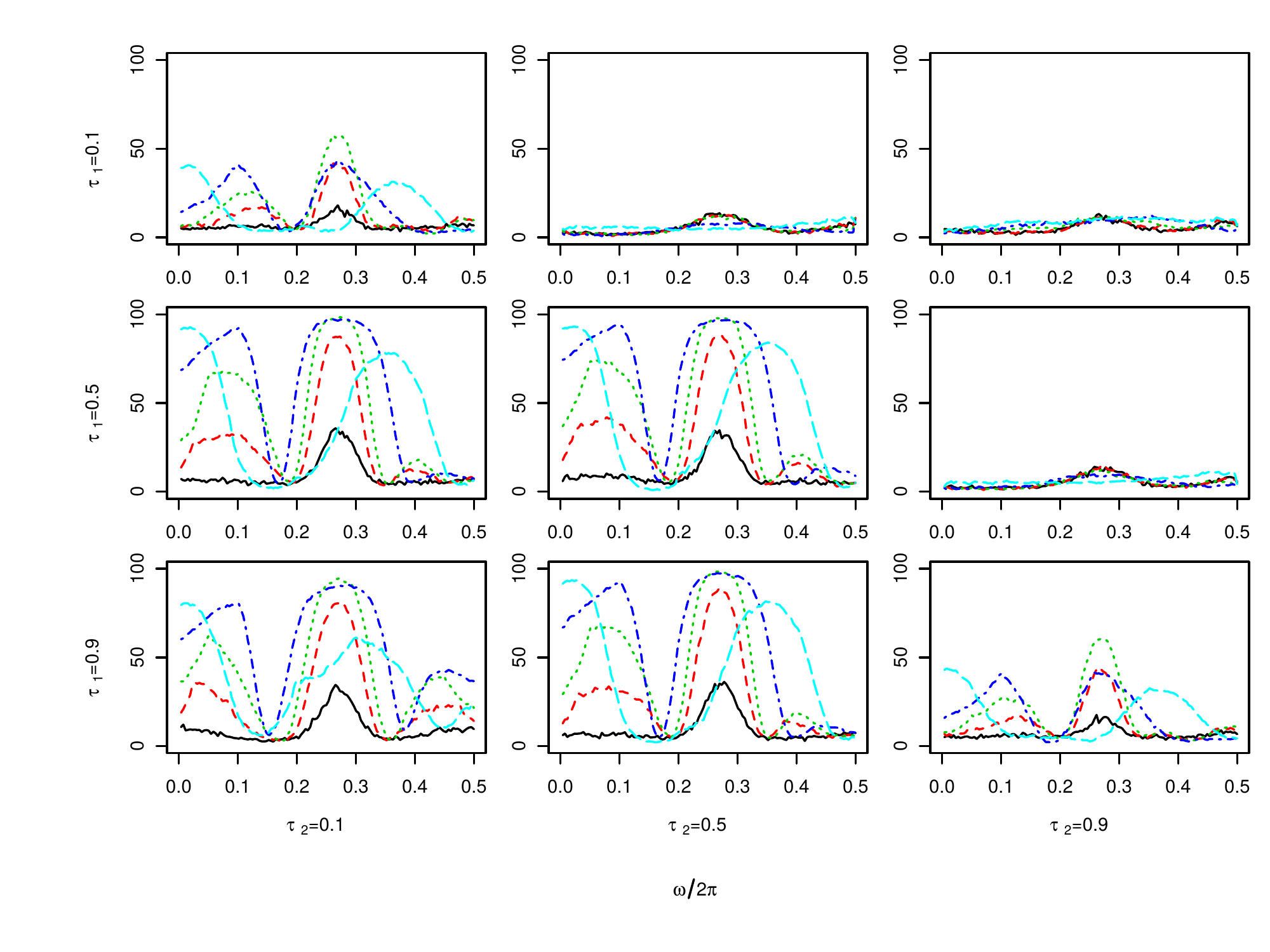}
 \hspace*{-0.8cm} \includegraphics[width = 0.55\textwidth]{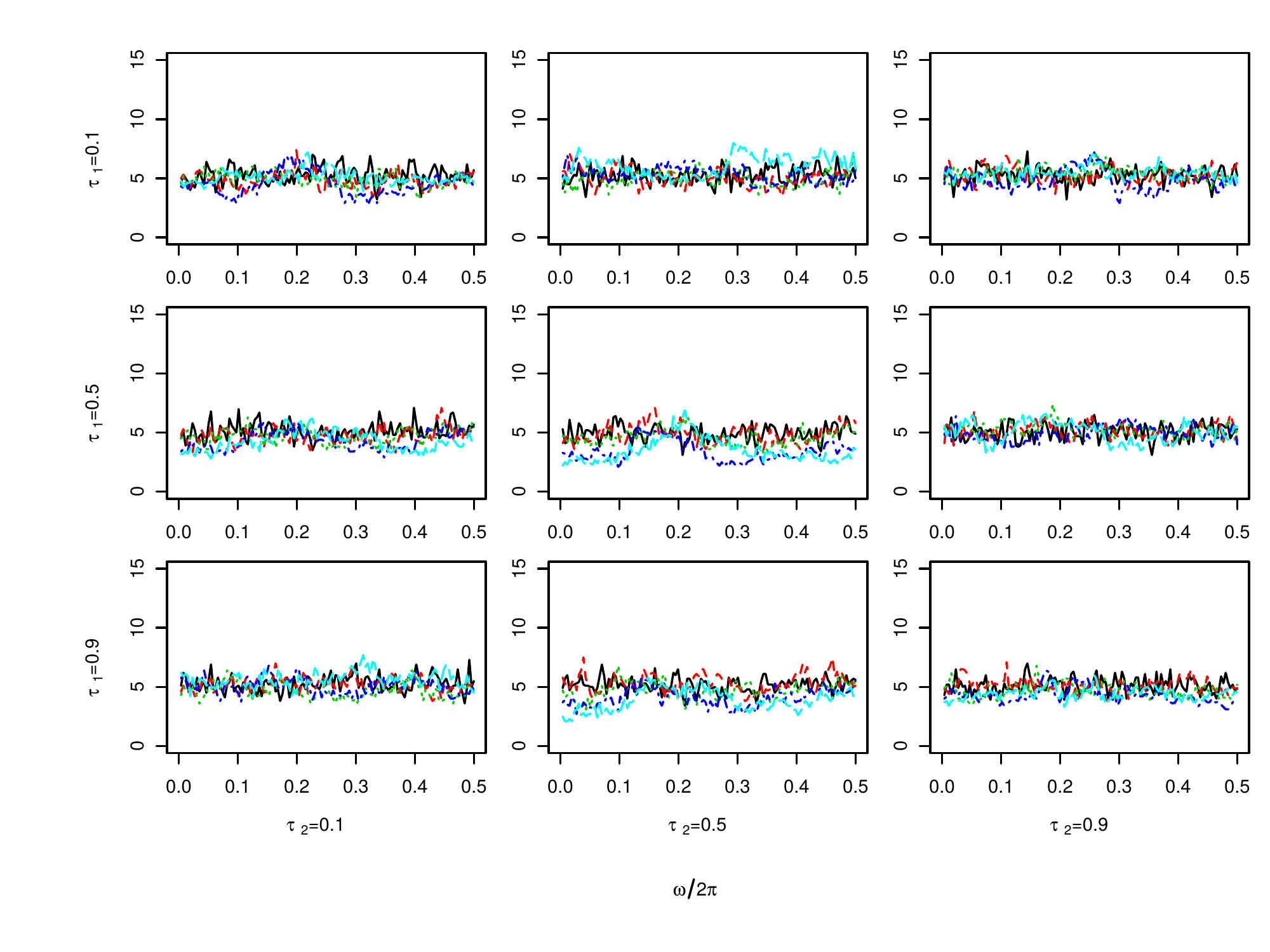}
\includegraphics[width = 0.55\textwidth]{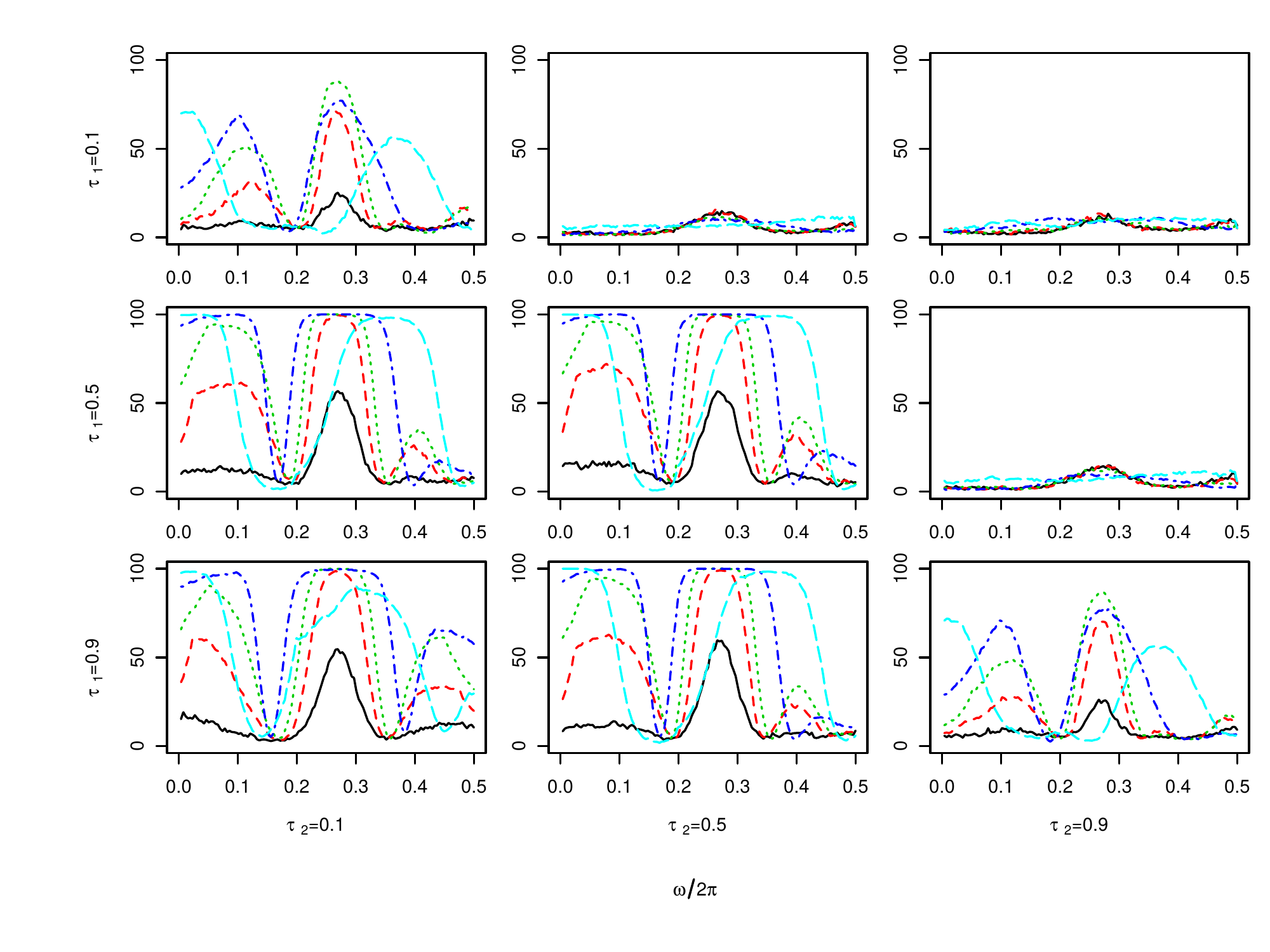}
\caption{Coverage of the estimator $\hat f$ by the critical regions obtained by Algorithm~\ref{alg1}. Model class used for the critical regions: $P_a$ (ARMA(1,1)). Data generated according to $(a_0)$ (ARMA(1,1), left panel) or $(a_1)$ (AR(3), right panel). We use $n = 256, 512$ and $1024$ observations in the first, second and third row respectively. Different bandwidth choices are shown using different colors and line types. The solid line (black), the lines with short (red), medium (green), alternating-length (blue) and long (cyan) dashes correspond to $b_n = 0.01,0.02,0.05,0.1,0.4$, respectively.} \label{fig.sim.1}
\end{figure}

\begin{figure}
\hspace*{-1cm} \includegraphics[width = 0.55\textwidth]{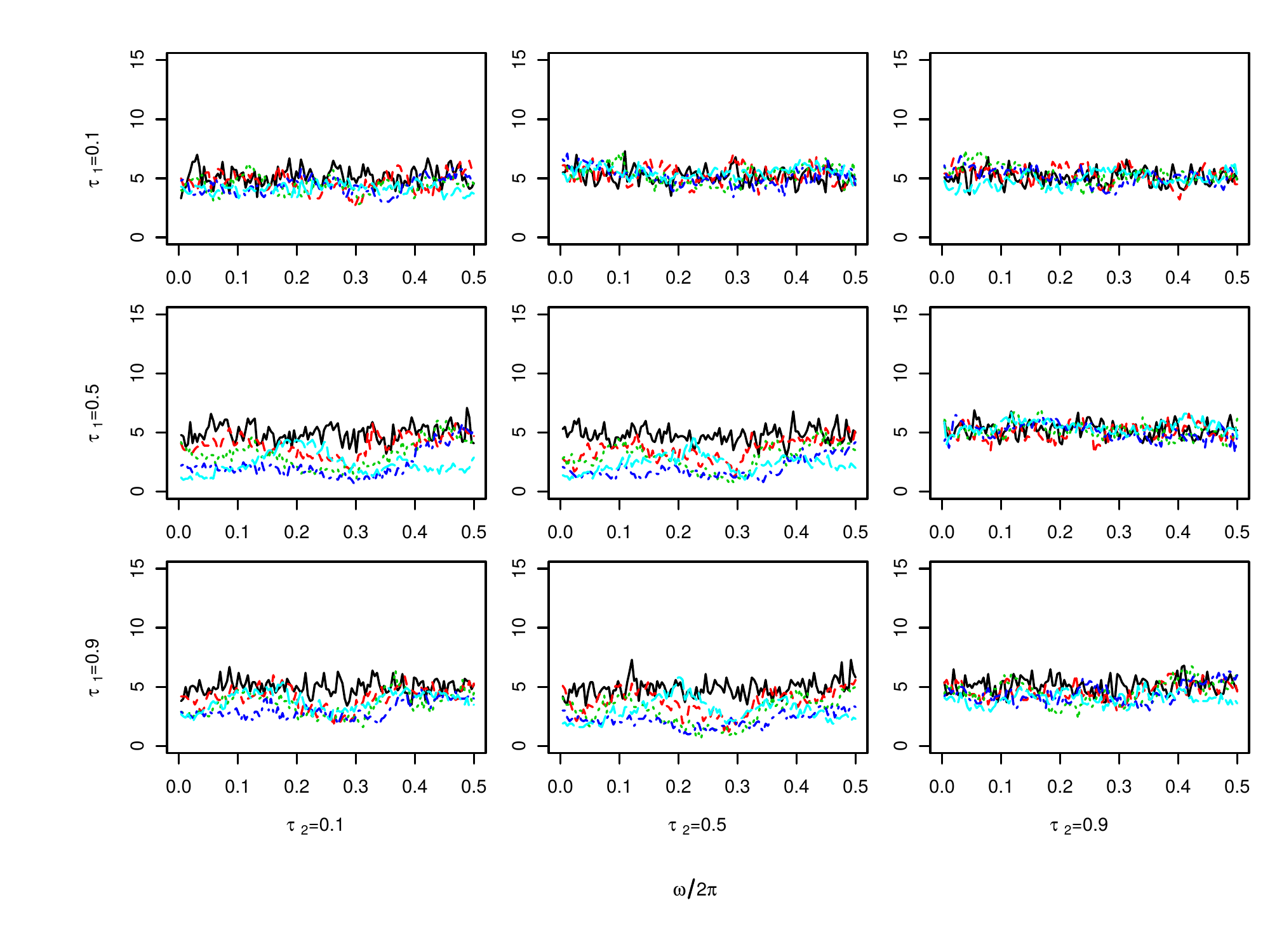}
\includegraphics[width = 0.55\textwidth]{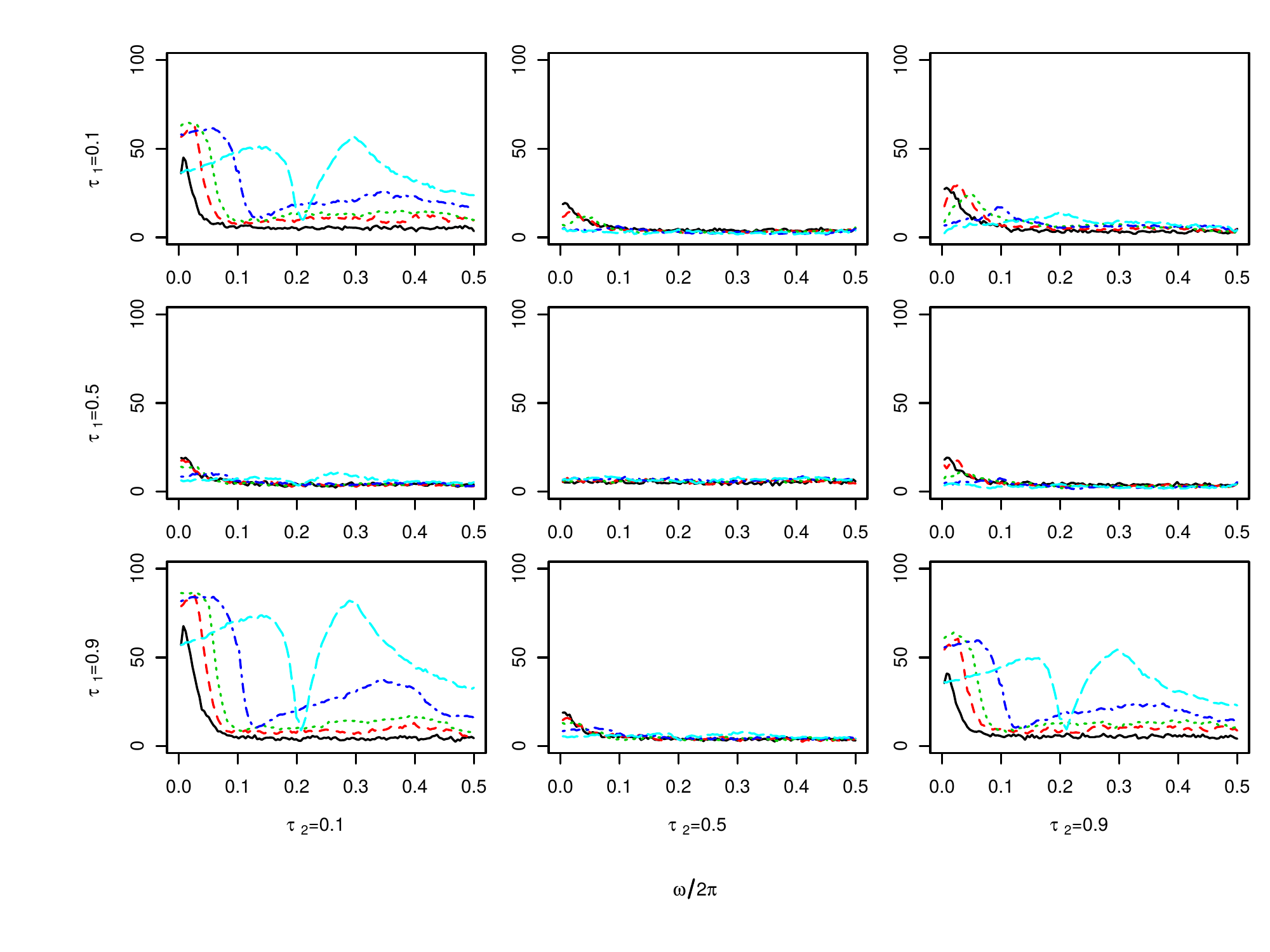}
\hspace*{-1cm} \includegraphics[width = 0.55\textwidth]{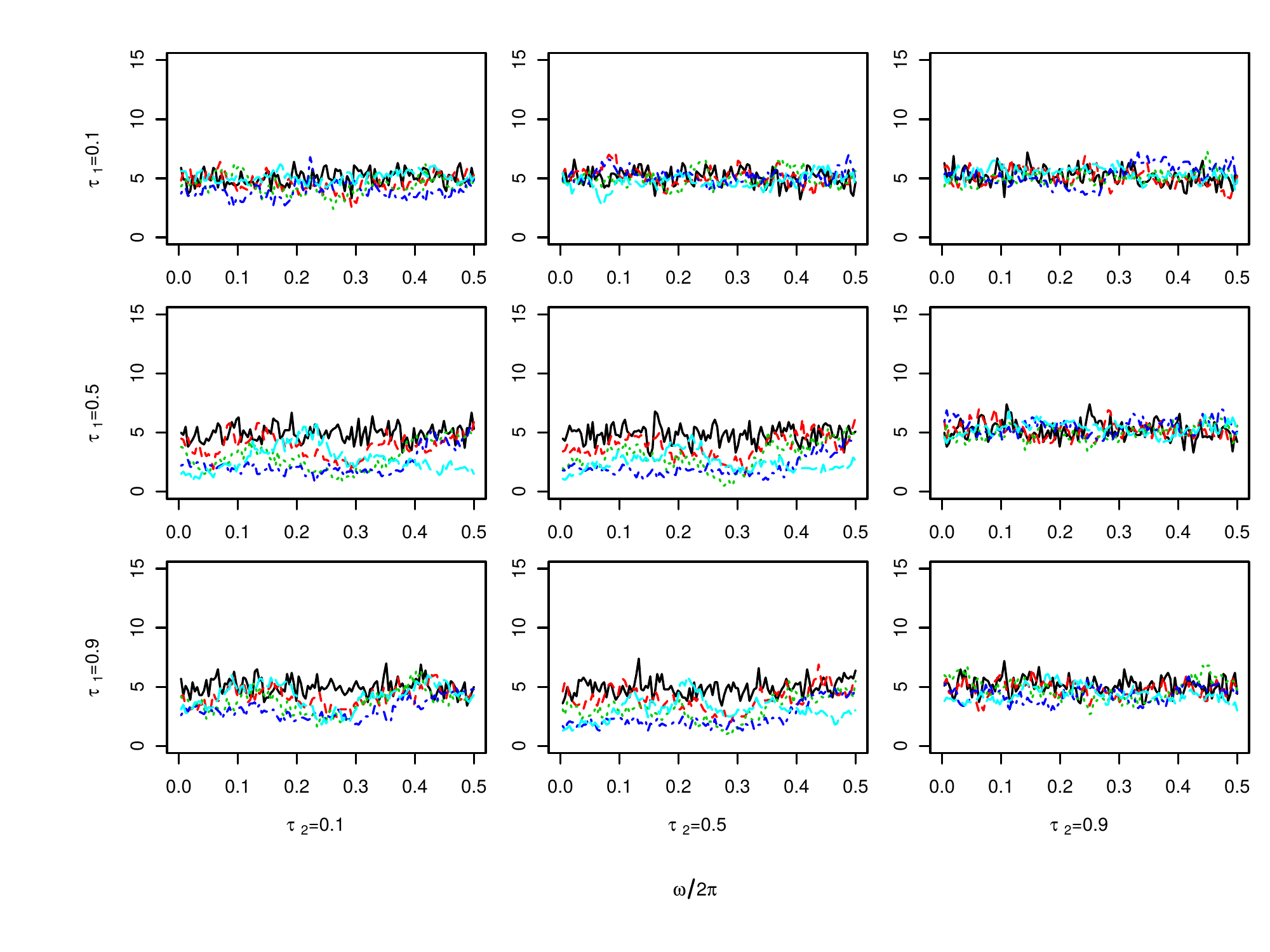}
\includegraphics[width = 0.55\textwidth]{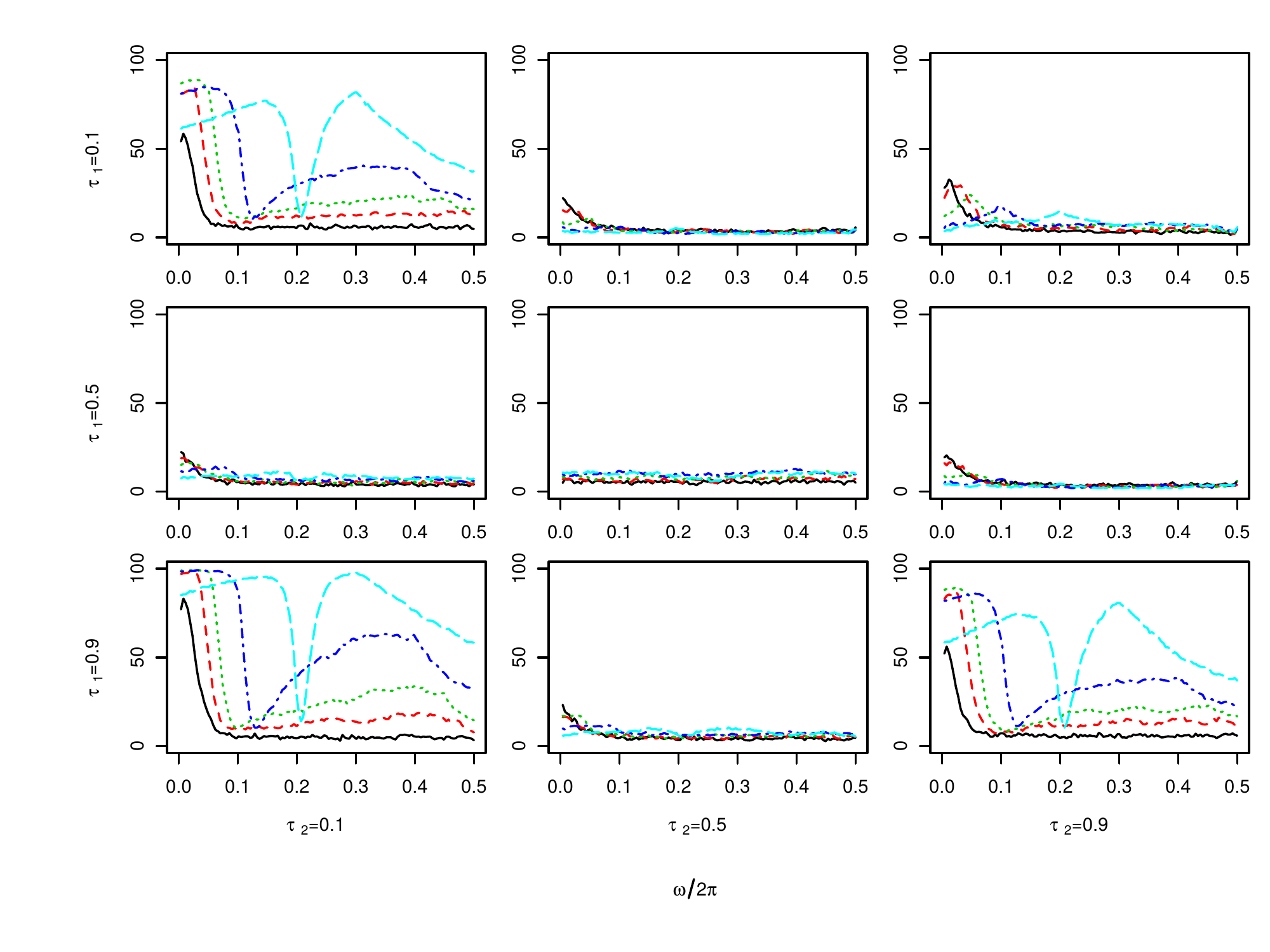}
\hspace*{-1cm}\includegraphics[width = 0.55\textwidth]{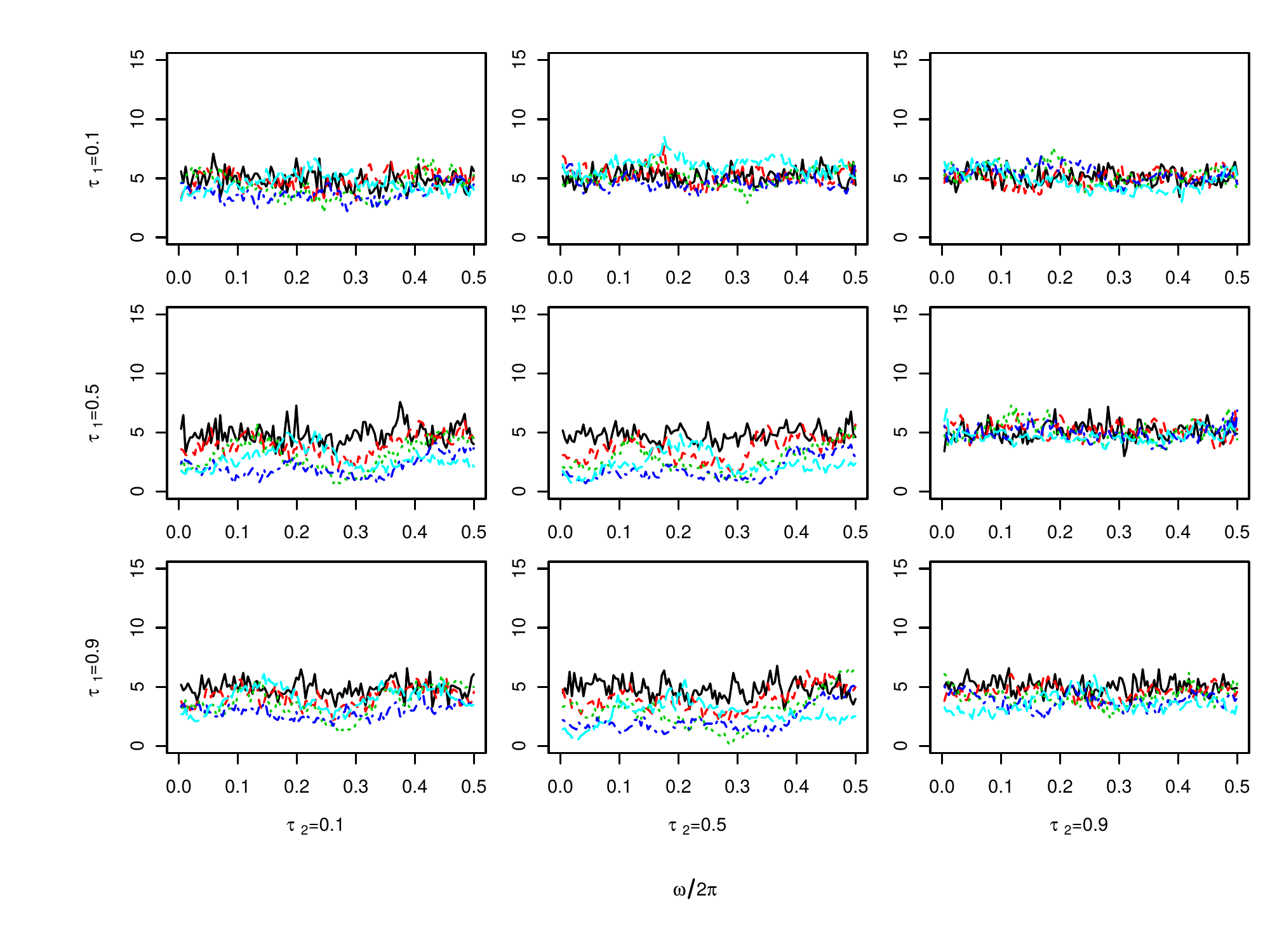}
\includegraphics[width = 0.55\textwidth]{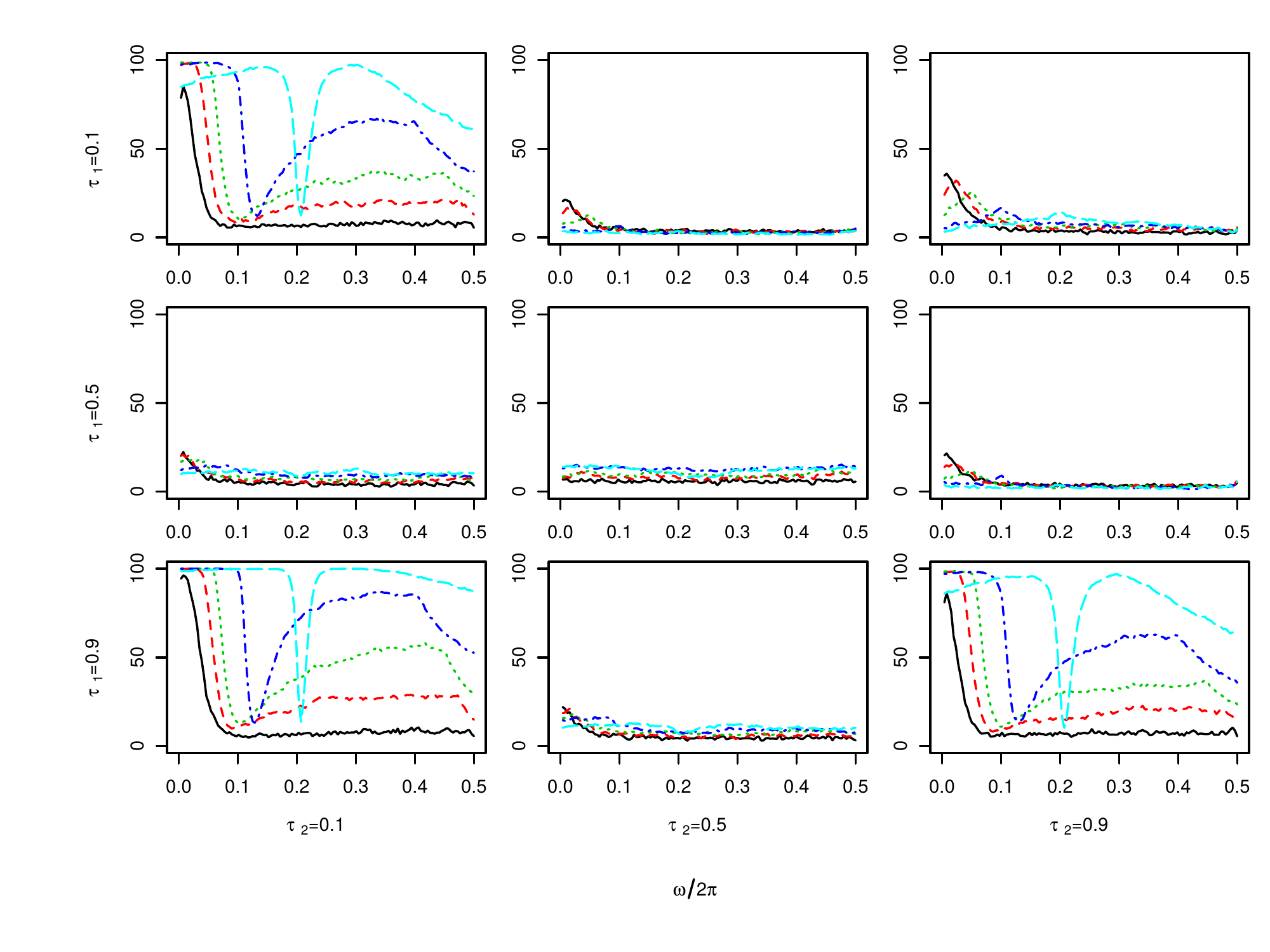}
\caption{
Coverage of the estimator $\hat f$ by the critical regions obtained by Algorithm~\ref{alg1}. Model class used for the critical regions: $P_b$ (AR(3)). Data generated according to $(b_0)$ (AR(3), left panel) and $(b_1)$ (GARCH(1,1), right panel). We use $n = 256, 512$ and $1024$ observations in the first, second and third row respectively. Different bandwidth choices are shown using different colors and line types. The solid line (black), the lines with short (red), medium (green), alternating-length (blue) and long (cyan) dashes correspond to $b_n = 0.01,0.02,0.05,0.1,0.4$, respectively.} \label{fig.sim.2}
\end{figure}

\begin{figure}
\hspace*{-1cm}\includegraphics[width = 0.55\textwidth]{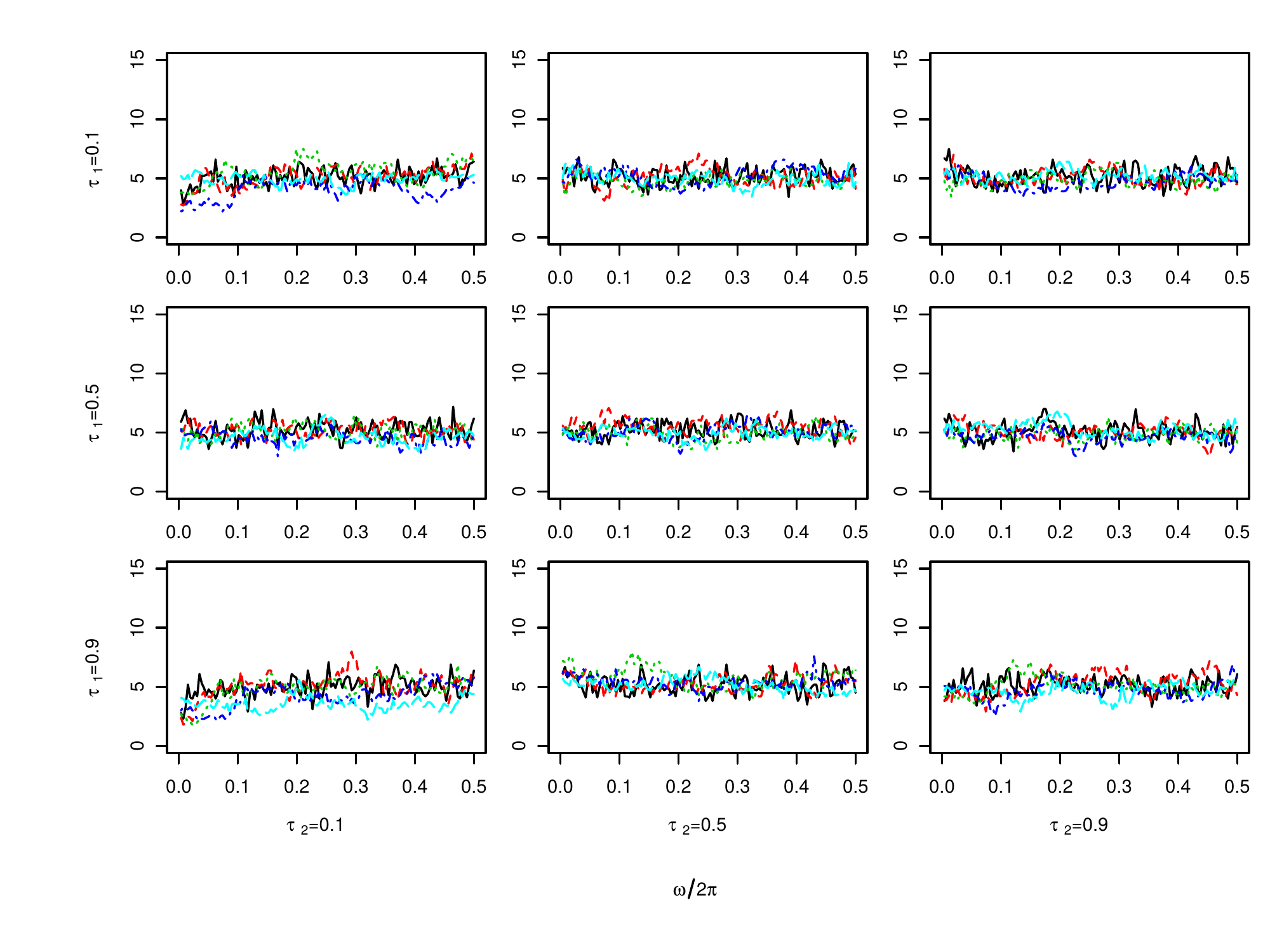}
\includegraphics[width = 0.55\textwidth]{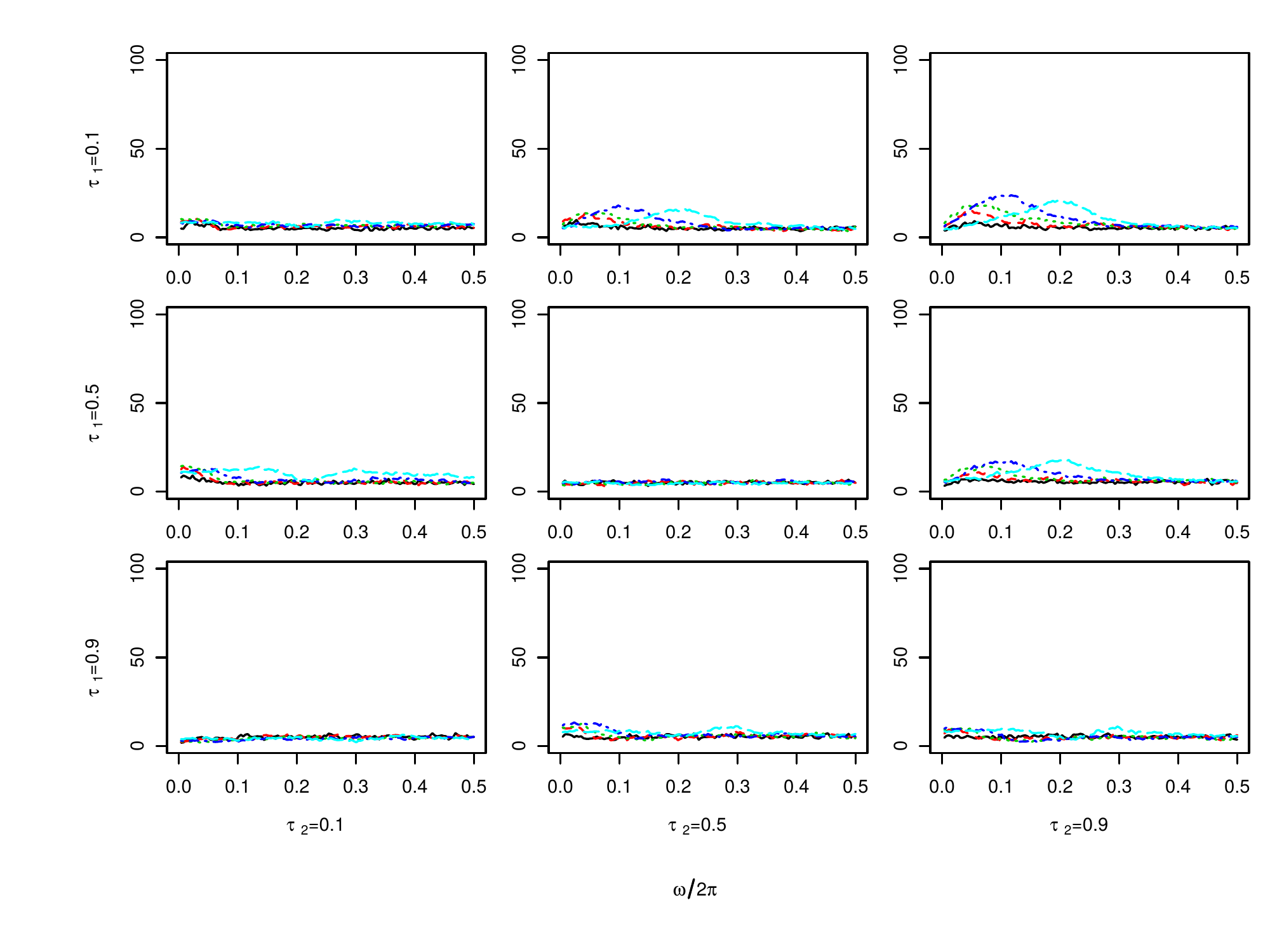}
\hspace*{-1cm}\includegraphics[width = 0.55\textwidth]{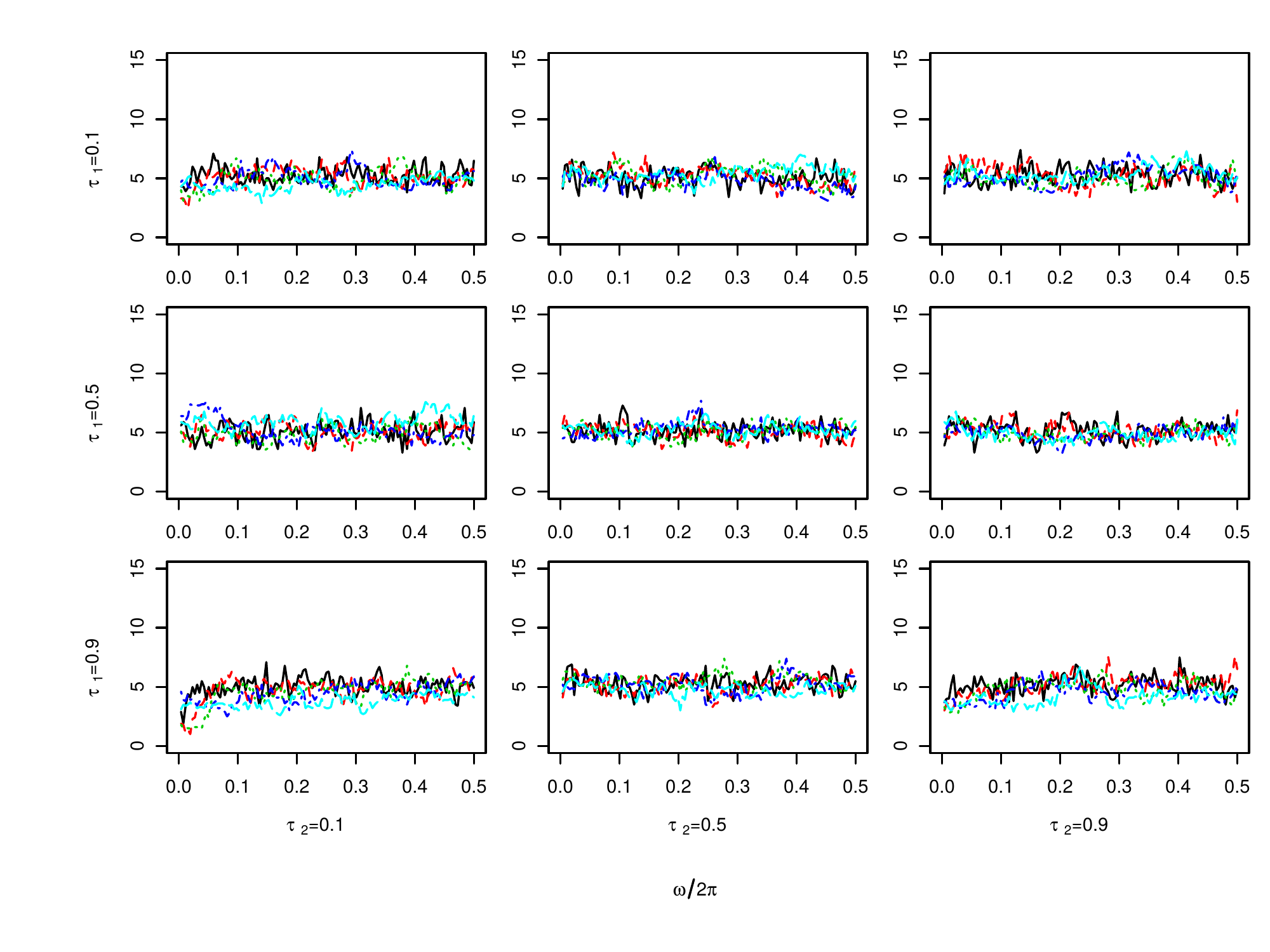}
\includegraphics[width = 0.55\textwidth]{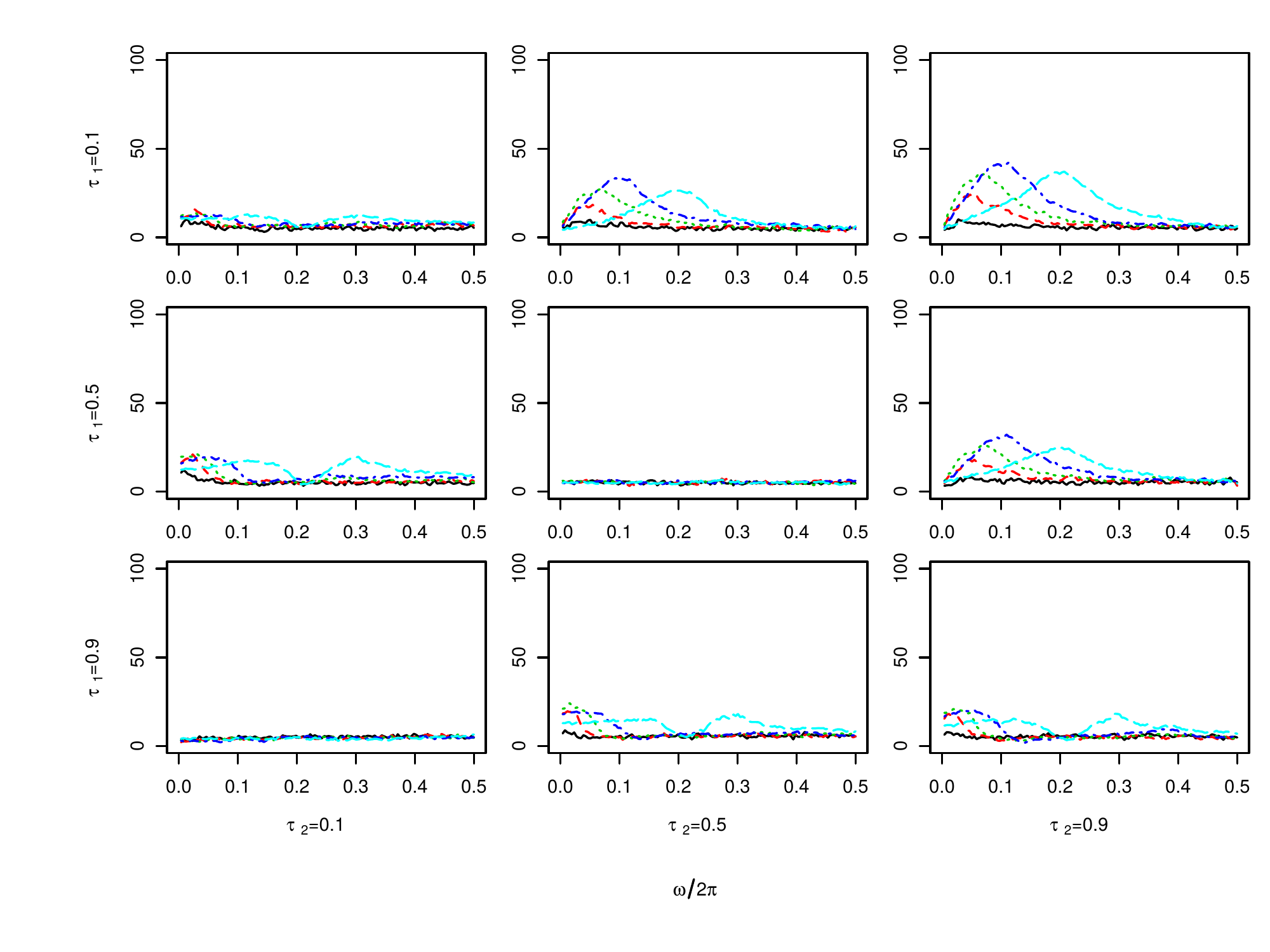}
\hspace*{-1cm}\includegraphics[width = 0.55\textwidth]{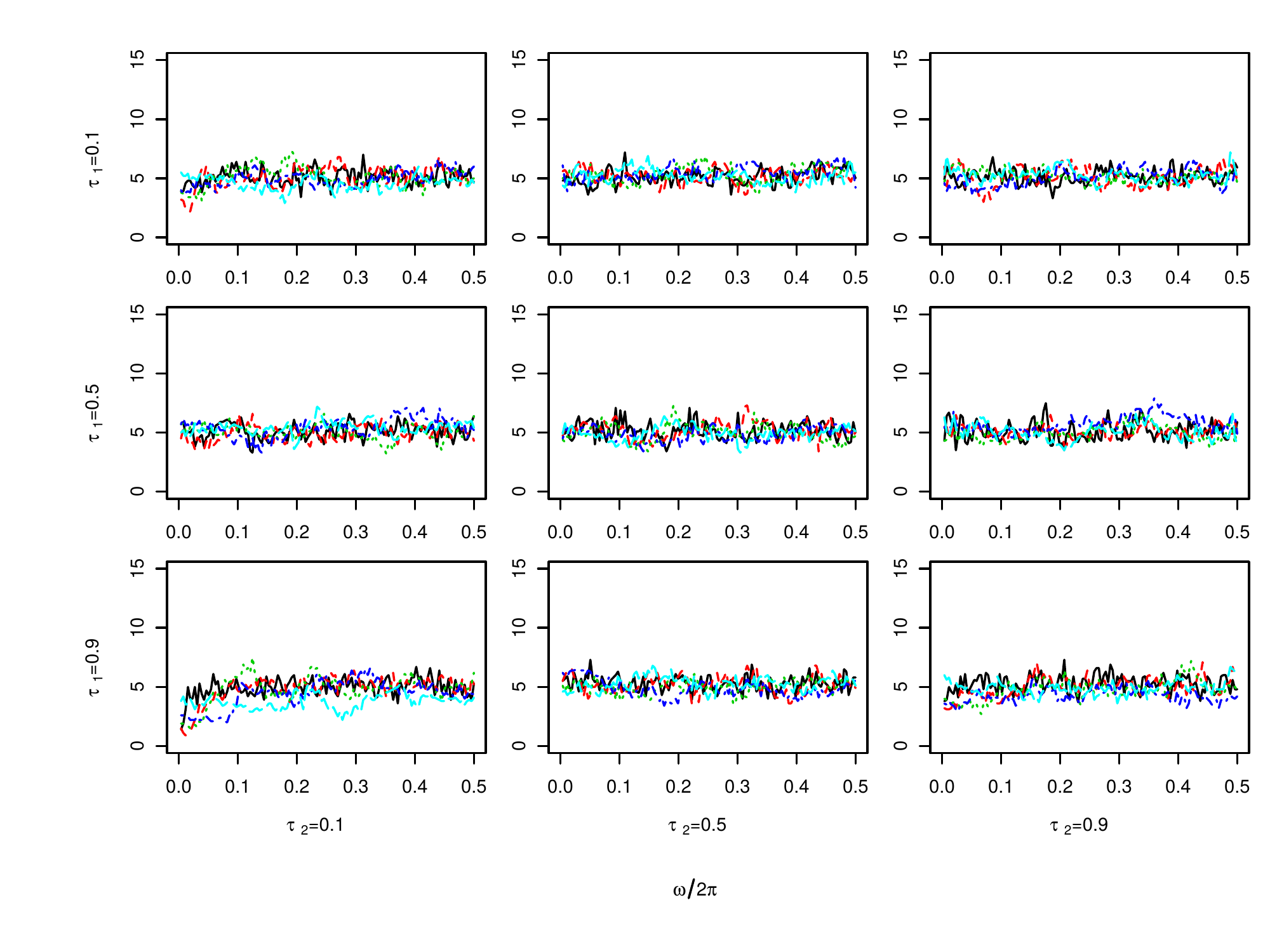}
\includegraphics[width = 0.55\textwidth]{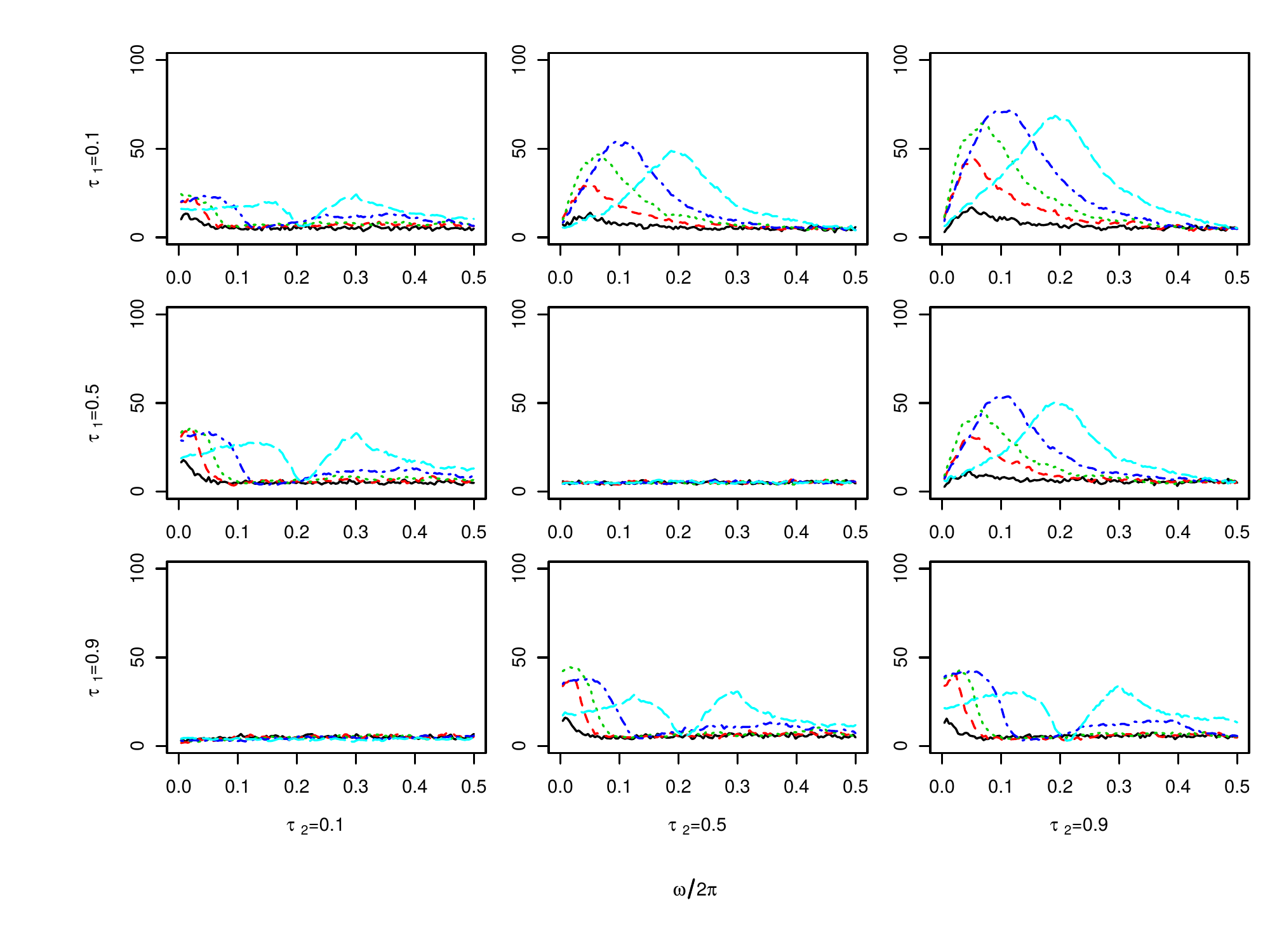}
\caption{
Coverage of the estimator $\hat f$ by the critical regions obtained by Algorithm~\ref{alg1}. Model class used for the critical regions: $P_c$ (GARCH(1,1)). Data generated according to $(c_0)$ (GARCH(1,1), left panel) and $(c_1)$ (EGARCH(1,1), right panel). We use $n = 256, 512$ and $1024$ observations in the first, second and third row respectively. Different bandwidth choices are shown using different colors and line types. The solid line (black), the lines with short (red), medium (green), alternating-length (blue) and long (cyan) dashes correspond to $b_n = 0.01,0.02,0.05,0.1,0.4$, respectively.} \label{fig.sim.3}
\end{figure}

We use the R packages \textbf{QPBoot} \citep{QPBoot} which contains useful functions for parametric bootstrap procedures for quantile spectra, \textbf{quantspec} \citep{kley2016} to compute the estimators for the copula spectral densities, and \textbf{rugarch} \citep{rugarch} to estimate and simulate the GARCH-type models. For each frequency $\omega$ we count the number of times the estimated spectral density $\hat{f}_{\btau}(\omega)$ does not lie in the interval $(l_{\tau,R}(\omega),u_{\tau,R}(\omega))$ (separately for real and imaginary parts). The resulting counts are shown (numbers normalized by $1000$) in the left panel of Figures \ref{fig.sim.1}, \ref{fig.sim.2} and \ref{fig.sim.3}, respectively. We can see that the simulated `typical regions' contain the estimator $\hat f_\btau$ with prescribed probability across a wide range of scenarios. 

\begin{figure}
\begin{center}
\rotatebox{90}{\hspace*{0.7cm} \small $P_a$ versus $(a_0)$} ~~
\includegraphics[width = 0.9\textwidth]{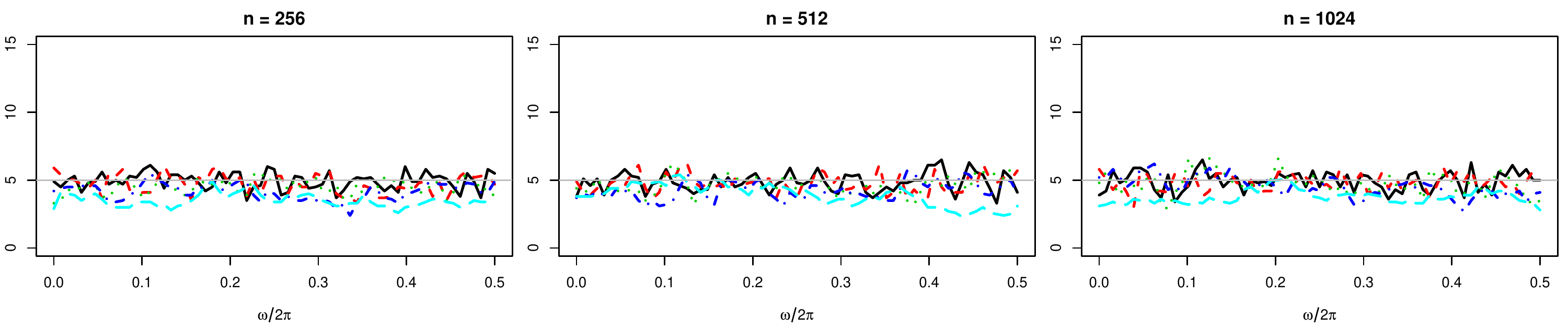}\\
\rotatebox{90}{\hspace*{0.7cm} \small $P_b$ versus $(b_0)$} ~~
\includegraphics[width = 0.9\textwidth]{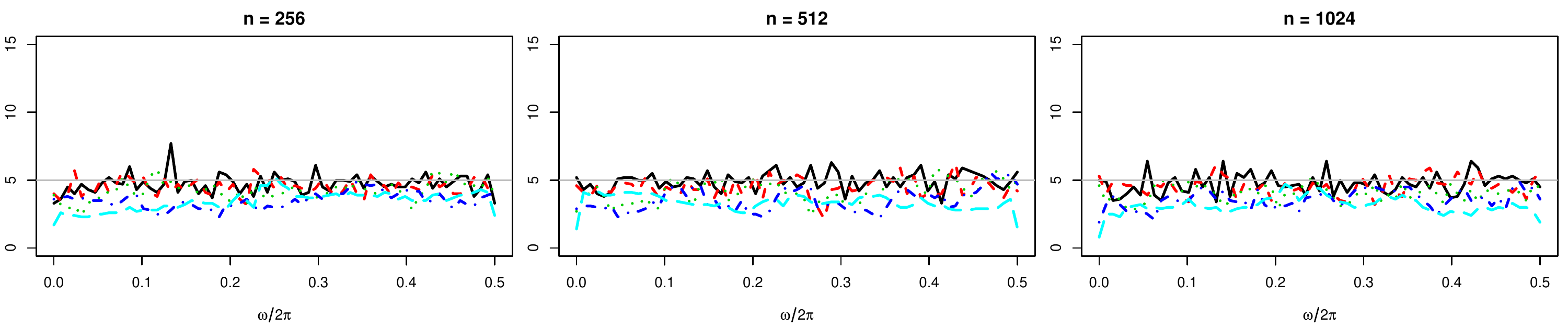}\\
\rotatebox{90}{\hspace*{0.7cm} \small $P_c$ versus $(c_0)$} ~~
\includegraphics[width = 0.9\textwidth]{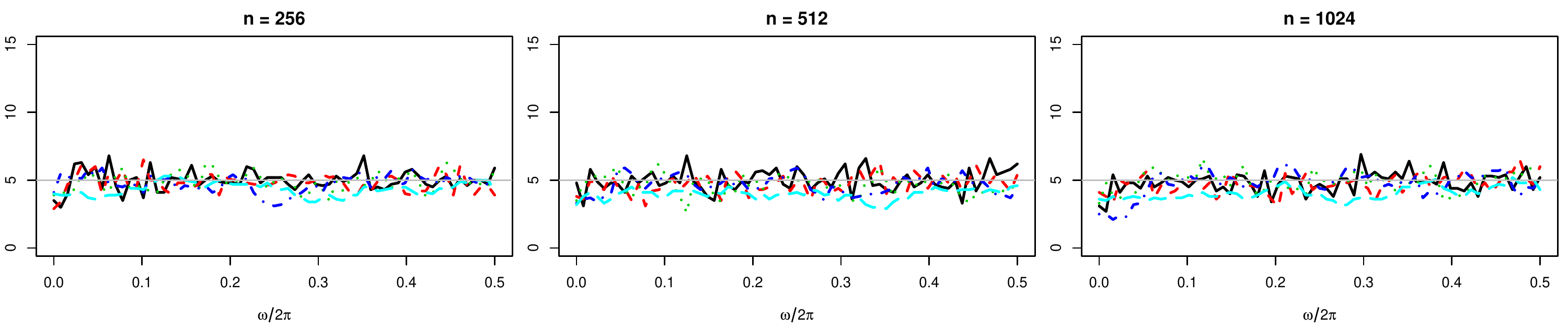}\\
\rotatebox{90}{\hspace*{0.7cm} \small $P_a$ versus $(a_1)$} ~~
\includegraphics[width = 0.9\textwidth]{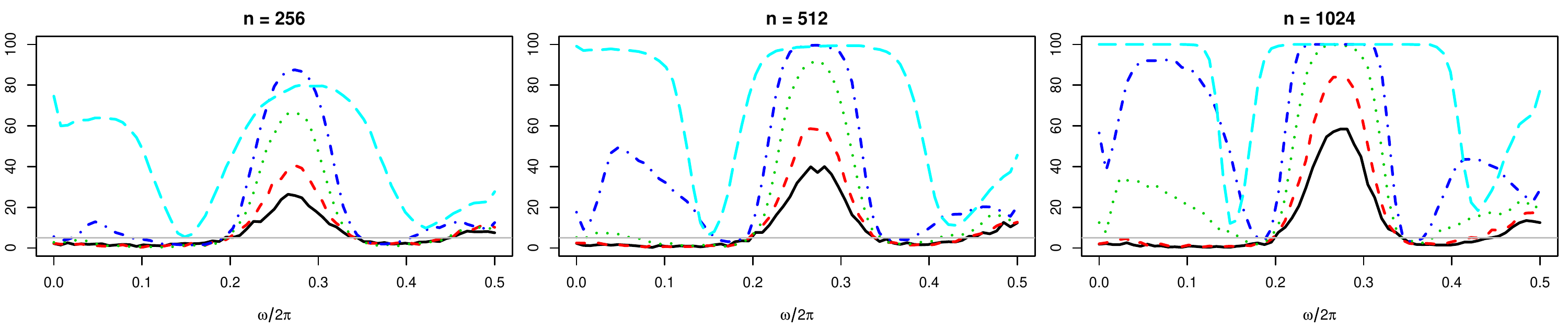}\\
\rotatebox{90}{\hspace*{0.7cm} \small $P_b$ versus $(b_1)$} ~~
\includegraphics[width = 0.9\textwidth]{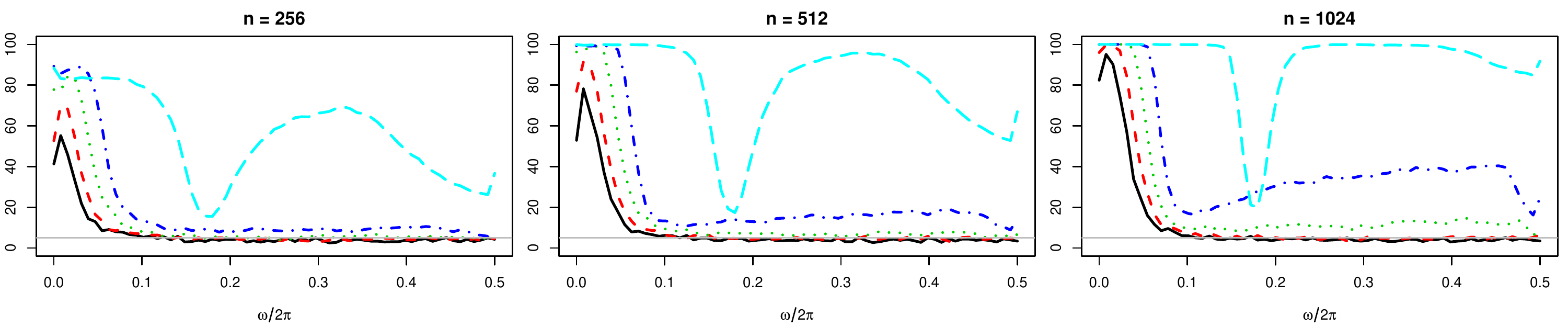}\\
\rotatebox{90}{\hspace*{0.7cm} \small $P_c$ versus $(c_1)$} ~~
\includegraphics[width = 0.9\textwidth]{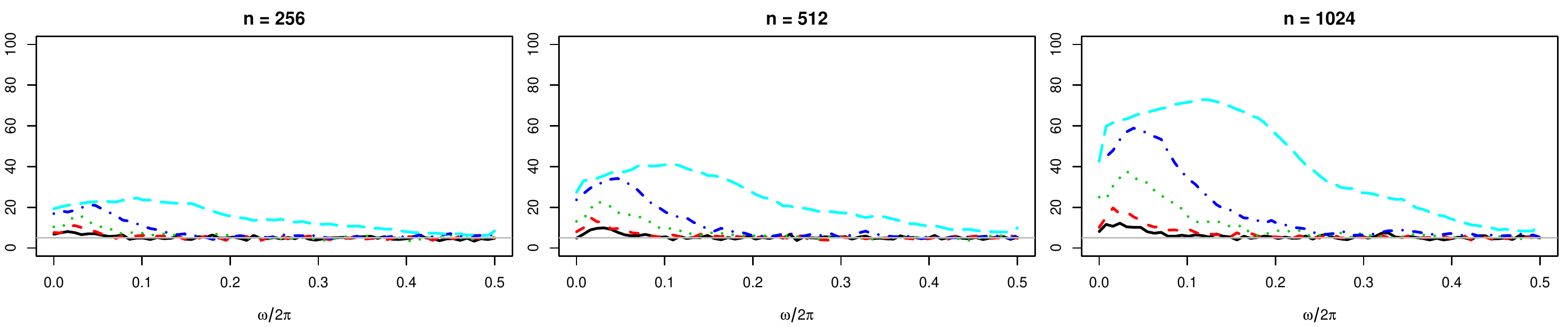}\\
\end{center}
\caption{
Proportion of cases, per frequency, where a p-values obtained from Algorithm~\ref{alg2} are below $\alpha = 0.05$, for at least on quantile level. First, second and third row show $P_a$ versus $(a_0)$ (ARMA(1,1)), $P_b$ versus $(b_0)$ (AR(3)), and $P_c$ versus $(c_0)$ (GARCH(1,1)), respectively. Fourth, fifth and sixth row show $P_a$ versus $(a_1)$ (ARMA(1,1) and AR(3)), $P_b$ versus $(b_1)$ (AR(3) and GARCH(1,1)), and $P_c$ versus $(c_1)$ (GARCH(1,1) versus EGARCH(1,1)), respectively. We use $n = 256, 512$ and $1024$ observations in the first, second and third column, respectively. Different bandwidth choices are shown using different colors and line types. The solid line (black), the lines with short (red), medium (green), alternating-length (blue) and long (cyan) dashes correspond to $b_n = 0.01,0.02,0.05,0.1,0.4$, respectively.} \label{fig.sim.4}
\end{figure}

Next, we show that the aggregated p-values obtained from Algorithm~\ref{alg2} are calibrated properly. To this end we consider the same models and bandwidth parameters as described above and use $1000$ simulation replications to approximate the probabilities
\begin{equation} \label{eq:punif}
P\Big(\min_{\btau \in M} \min\{p_{\btau,R}^\Re(\omega), p_{\btau,R}^\Im(\omega) \}\leq \alpha \Big)
\end{equation}
where we use $M = \{0.05,...,0.95\}^2$ and $R=1000$. The results corresponding to model $(a_0)-(c_0)$ with candidate model classes $(P_{a})-(P_{c})$ are depicted in the top three rows of Figure~\ref{fig.sim.4} with frequencies on the x-axis and simulated values for the probabilities in~\eqref{eq:punif}, with $\alpha = 0.05$, on the y-axis. The plots suggest that the p-values perform as specified in all settings considered. 

Next, consider the case when the observations are created by the following models,
\begin{align*}
(a_1)& \quad X_t = 0.2X_{t-1} -0.4X_{t-2} + 0.2X_{t-3} + Z_t\\
(b_1)& \quad X_t = \sigma_t Z_t, \textrm{ where } \sigma_t^2 = 0.01 + 0.4X_{t-1}^2 + 0.5\sigma_{t-1}^2\\
(c_1)& \quad X_t =\sigma_t Z_t , \textrm{ where } \ln(\sigma_t^2) = 0.1 + 0.21(|X_{t-1}| - \E|X_{t-1}|) - 0.2X_{t-1} + 0.8\ln(\sigma_{t-1}^2),
\end{align*}
while the candidate parametric model classes are still $(P_{a}), (P_{b}), (P_{c})$ and thus are misspecified. The results for Algorithm~\ref{alg1} are depicted in the right panels of Figures \ref{fig.sim.1}, \ref{fig.sim.2} and \ref{fig.sim.3}, respectively. The plots corresponding to Algorithm~\ref{alg2} are in rows 4--6 of Figure~\ref{fig.sim.4}.

The results in Figure \ref{fig.sim.1} and row four of Figure~\ref{fig.sim.4} show that copula spectral densities are informative for distinguishing different types of linear dynamics (although in this setting any of the classical tests that are tailored to linear models would also be applicable and have excellent power properties). Figure \ref{fig.sim.1} indicates that in this setting copula spectral densities corresponding to $\btau = (0.5,0.5), (0.1,0.5), (0.5,0.9)$ are most informative. This is not surprising since linear dynamics act similarly in all real parts of copula spectral densities and copula spectral densities corresponding to the quantile values mentioned above are easier to estimate (note that for more extreme quantiles only a smaller proportion of the data carry relevant information). Since linear Gaussian processes are time-reversible, the imaginary parts of copula spectra carry no relevant information in this case. Finally, we remark that for this particular data generation process intermediate bandwidth values lead to the most informative results in Figure \ref{fig.sim.1}. Row four of Figure~\ref{fig.sim.4} additionally shows that aggregating over different frequencies does not lead to a loss in power (despite the uniformity over $\btau$) and in fact improves this probability for the largest bandwidth $b_n = 0.4$.

Figure \ref{fig.sim.2} and row five of Figure~\ref{fig.sim.4} show what happens if data are generated by a GARCH model but we attempt to fit their dependence structure by an AR(3) process. In this case the AR(3) model tries to capture the serial correlation of the data, which is zero (so the AR(3) model essentially results in iid data without any serial dependence). This does capture the median dynamics corresponding to $\btau = (0.5,0.5)$, but completely fails to account for dependence in the more extreme quantiles. This is clearly visible for the real parts of the copula spectral densities corresponding to $\btau = (0.1,0.1), (0.1,0.9), (0.9,0.9)$ with $\btau = (0.1,0.9)$ leading to the clearest distinction. It is also interesting to observe how different bandwidth values pick up different aspects of the deviation between data and model dynamics. While smaller bandwidth values mainly pick up the sharp peak near zero frequencies, larger bandwidth values also find differences for intermediate frequency values.

The most complicated case that we investigate in our study is to differentiate between a GARCH and an EGARCH process. Results for this are shown in the right panel of Figure~\ref{fig.sim.3} and in the bottom row of Figure~\ref{fig.sim.4}. Both processes have a very similar serial dependence structure as they are uncorrelated but dependent in the extreme quantiles. The difference is that the EGARCH process is asymmetric in the sense, that the dependence is higher in the lower quantiles due to the negative leverage parameter of $-0.2.$ This difference is subtle and only present in the dependence at large quantiles and hence difficult to pick up and large sample sizes are needed to reliably pick up this distinction. It also turns out that the imaginary part corresponding to $\btau = (0.1,0.9)$ carries the most information here, with larger bandwidth parameters leading to higher probabilities of detecting relevant differences. The results in the bottom row of Figure~\ref{fig.sim.4} additionally show that by aggregating over different quantile levels we are likely to detect deviations between GARCH and an EGARCH processes across a wider range of frequencies. This is due to the fact that for different quantile levels the deviations between the two models are most pronounced at different frequencies.

\newpage

\newpage

%\section*{References}
\bibliographystyle{elsarticle-num} 
\bibliography{LocStatBib_2016-07-11}

\newpage

\begin{center}
{\huge\textbf{Online supplement}}
\end{center}

\section{Proofs}\label{sec:proofs}

\subsection{Proof of Example~\ref{lem:arma}}\label{sec:proofex}
Note that under the assumptions made $X_t^\theta$ has the representation
\[
X_t^{\theta} = \sum_{k = 0}^\infty \psi_k^\theta \epsilon_{t-k}
\]
where the coefficients are defined by
\[
\sum_{k=0}^\infty \psi_k^\theta z^k = \frac{Q^{\theta}(z)}{P^{\theta}(z)}, \qquad |z| \leq 1.
\]
By properties of the multivariate normal distribution it suffices to show that for some $L$
\[
\sup_{\ub \in \R^2} \Big|F_h^\theta(\ub) - F_h^{\theta_0}(\ub)\Big| \leq \|\theta - \theta_0\|.
\]
Applying the triangle inequality we find
\[
\sup_{\ub \in \R^2} \Big|F_h^\theta(\ub) - F_h^{\theta_0}(\ub)\Big| \leq 2\sup_{x \in \mathbb{R}} \E\Big[|\Ind{X_{0}^{\theta} \leq x} - \Ind{X_{0}^{\theta_0} \leq x}|\Big]
\]
and hence it suffices to show that
\[
\sup_{x \in \mathbb{R}} \E\Big[|\Ind{X_{0}^{\theta} \leq x} - \Ind{X_{0}^{\theta_0} \leq x}| \Big] \leq L\|\theta - \theta_0\|.
\]
Denote by~$\mathcal{A}_t$ the~$\sigma-$field generated by~$\{\epsilon_s|s < t\},$ and by $F_\epsilon$ the distribution function of $\epsilon.$ 
This yields
\begin{equation*}
\begin{split}
& \sup_{x \in \mathbb{R}} \E(|\Ind{X_{0}^{\theta} \leq x} - \Ind{X_{0}^{\theta_0}\leq x}|)
= \sup_{x \in \mathbb{R}} \E\Big[\E[| \Ind{X_{t}^\theta \leq x } - \Ind{X_{t}^{\theta_0 }\leq x }| \big| \mathcal{A}_t ] \Big]
\\
& \quad = \sup_{x \in \mathbb{R}} \E\Big[\E\Big[| \Ind{\epsilon_t \leq x - \sum_{j=1}^\infty \psi_j^\theta \epsilon_{t-j} } - \Ind{\epsilon_t \leq x - \sum_{j=1}^\infty \psi_j^{\theta_0}\epsilon_{t -j}}| \Big| \mathcal{A}_t \Big]\Big]
\\
& \quad =  \sup_{x \in \mathbb{R}} \E\Big[ \Big| F_{\epsilon}\Big(x - \sum_{j=1}^\infty \psi_j^\theta\epsilon_{t -j}\Big) - F_{\epsilon}\Big(x - \sum_{j=1}^\infty \psi_j^{\theta_0}\epsilon_{t -j}\Big) \Big| \Big]
\leq C_1 \E \Big|\sum_{j=1}^\infty \psi_j^\theta\epsilon_{t -j} - \sum_{j=1}^\infty \psi_j^{\theta_0}\epsilon_{t -j} \Big|
 \leq C_2 \sum_{j=1}^\infty |\psi_j^\theta - \psi_j^{\theta_0}|.
\end{split}
\end{equation*}
Finally, we bound the last term above. To shorten notation we write $p_\theta(z) = Q_\theta(z)/P_\theta(z).$ As $P_{\theta_0}(z)$ has no roots on the unit circle, there exist $\eta,\delta>0$ such that for all $\| \theta - \theta_0\|\leq \eta$
\[
P_{\theta_n}(z) \neq 0 \quad \forall z \in \mathbb{C}: |z| < 1 + 2\delta.
\]
(Otherwise we could derive a contradiction by using the fact that on $\mathbb{C}$ the locations of roots of a polynomial are a continuous function of the coefficients.) Therefore $p_{\theta}$ is a holomorphic function on $\{z \in \mathbb{C}: |z| \leq 1+2\delta\}$ and we can expand $p_{\theta}(z) = \sum_{j=0}^\infty \psi_j^\theta z^j$ and $p_{\theta_0}(z) = \sum_{j=0}^\infty \psi_j^{\theta_0} z^j$ with
\[ 
\psi_j^\theta = \frac{1}{2 \pi i} \oint_{|\zeta| = 1 + \delta} \frac{p_\theta(\zeta)}{\zeta^{j+1}} d\zeta, \quad j\in \mathbb{N}_0 
\]
by Cauchy's differentiation formula. This implies
\[ 
|\psi_j^\theta - \psi_j^{\theta_0}| =  \frac{1}{2 \pi} \Big|\oint_{|\zeta| = 1 + \delta} \frac{p_{\theta}(\zeta) - p_{\theta_0}(\zeta)}{\zeta^{j+1}} d\zeta \Big|.  
\]
And with $p_\theta(z) = Q_\theta(z)/P_\theta(z)$ we have that
\[ 
\sup_{|z| = 1+\delta} \Big|\frac{p_{\theta}(z) - p_{\theta_0}(z)}{z^{j+1}} \Big| \leq \frac{||\theta - \theta_0||}{(1+\delta)^{j+1}}, 
\]
which leads to
\[
\sum_{j=1}^\infty |\psi_j^\theta - \psi_j^{\theta}| \leq C_3||\theta - \theta_0|| \sum_{j=1}^\infty \frac{1}{(1+\delta)^{j}} =: L ||\theta - \theta_0||.
\]
\hfill $\Box$

%\newpage

\subsection{Proof of Proposition~\ref{prop:just}}

We begin by stating a useful Lemma.

\begin{lemma} \label{lem:help2}
Consider a sequence $r_n = o(1)$ and collection of distribution functions $F_{n,\xi}$ indexed by $\xi \in \Xi, n \in \N$ such that for any deterministic sequence $\xi_n$ in $\Xi$ with $\xi_n = \xi_0 + O(r_n)$ we have $F_{n,\xi_n} \leadsto F$ for some distribution function $F$. Then, for any sequence of random variables $\hat \xi_n$ in $\Xi$ with $\hat \xi_n = \xi_0 + O_P(r_n)$ we have: if $Y_1,..,Y_{m_n}$ are i.i.d. $F_{n,\hat\xi_n}$ conditional on $\hat\xi_n$, $m_n \to \infty$ and $q_n$ denotes the $\alpha$th sample quantile of $Y_1,..,Y_{m_n}$ then $q_n = F^{-1}(\alpha) + o_P(1)$ for any continuity point $\alpha$ of $F^{-1}$. 
\end{lemma}

\textbf{Proof of Lemma~\ref{lem:help2}} Let $\hat F_n$ denote the empirical cdf of $Y_1,..,Y_{m_n}$. For any fixed $t\in \R$ we have by the conditional Chebycheff inequality 
\[
P\big(|\hat F_n(t) - F_{n,\hat\xi_n}(t)| \geq \eps \big| \hat \xi_n \big) \leq (4m_n \eps^2)^{-1} \quad a.s.
\]
Taking the expectation with respect to $\hat \xi_n $ shows that $\hat F_n(t) - F_{n,\hat\xi_n}(t) = o_P(1)$ since by assumption $m_n \to \infty$. Next note that for arbitrary $C>0$ 
\[
P\big(|F(t) - F_{n,\hat\xi_n}(t)| \geq \eps \big) \leq \mathbb{I}\Big\{\sup_{|\xi-\xi_0|\leq Cr_n}|F(t) - F_{n,\xi}(t)| \geq \eps \Big\} + P(|\hat\xi_n - \xi_0| \geq Cr_n)
\]
We shall first show that the first term on the right-hand side converges to zero (for $n\to\infty$) for arbitrary $0<C<\infty$. Suppose this was not true. Then there exists $\delta > 0$, a subsequence $(n_k)_{k \in \N}$, and $\xi_{n_k} \in \Xi$ with $|\xi_{n_k}-\xi_0|\leq Cr_{n_k}$ and $|F(t) - F_{n_k,\xi_{n_k}}(t)| \geq \delta$ for all~$k\in\N$. However, by construction $\xi_{n_k} = \xi_0 + O(r_n)$ (for $k \to \infty$) which contradicts the assumption. Thus for all $C>0$
\[
\limsup_{n \to \infty} P\big(|F(t) - F_{n,\hat\xi_n}(t)| \geq \eps \big) \leq \limsup_{n \to \infty} P(|\hat\xi_n - \xi_0| \geq Cr_n).
\]
The right-hand side can be made arbitrarily small by choosing $C$ large since $\hat\xi_n = \xi_0 + O_P(r_n)$. Thus we have proved $\hat F_n(t) = F(t) + o_P(1)$ for all $t \in \R$. 

To complete the proof, observe that $\hat F_n$ is a sequence of distribution functions and $F$ is a distribution function. Thus a standard argument implies that $\sup_{t\in\R} |\hat F_n(t) - F(t)| = o_P(1)$. This implies $\hat F_n^{-1}(\alpha) = F^{-1}(\alpha)$ for all $\alpha$ that are continuity points of $F^{-1}$; the latter statement follows by the characterization of convergence in probability in term of a.s. convergence along subsequences and Lemma 21.2 in \cite{van1998asymptotic}. \hfill $\Box$

\medskip

With the preparations above we are ready to prove Proposition~\ref{prop:just}. 

We begin with the proof of~\eqref{eq:main}. Recall the setting and notation introduced in the beginning of Section~\ref{chap:asymptotics}. Let $q(\alpha,\theta_0)$ denote the $\alpha$-quantile of the distribution of $\Re H_0(\btau;\omega)$ (where $H_0$ denotes the weak limit in Theorem~\ref{thm:main}). Define 
\[
g_n := \Re\Big(f_{\btau}^{\theta_0}(\omega) + B_n^{(k)}(\btau,\omega)\Big)
\]
for $ B_n^{(k)}$ from Theorem~\ref{thm:main} and let 
\[
Z_n := \sqrt{nb_n}\Big(\Re\hat f_{\btau}(\omega) - g_n \Big). 
\]
By Theorem~\ref{thm:main} applied to the sequence $\theta_n \equiv \theta_0$, $Z_n \leadsto \Re H_0(\btau;\omega)$ with the limit being a centered normal random variable with non-zero variance.

Now consider the setting of Lemma~\ref{lem:help2} with $m_n = R_n, r_n = n^{-1/2}, \hat \xi_n = \hat\theta$, $F$ the cdf of $\Re H_0(\btau;\omega)$, $F_{n,\theta}$ the cdf of $\sqrt{nb_n}(\Re\hat f_{\btau}^\theta(\omega) - g_n)$ and $Y_i = \sqrt{nb_n}(\Re\hat f_{\btau}^{\hat\theta,i}(\omega) - g_n), i=1,...,R_n$. Note that $\hat\theta_n = \theta_0 + \Op(n^{-1/2})$ by assumption and $F_{n,\theta_n} \leadsto F$ for any sequence $\theta_n = \theta_0 + O(n^{-1/2})$ by Theorem~\ref{thm:main}. Hence, all conditions of Lemma~\ref{lem:help2} are satisfied and we obtain 
\begin{align*}
\sqrt{nb_n}\Big( l_{\btau,R_n}^\Re(\omega) - g_n \Big) &= q(\alpha/2,\theta_0) + o_P(1).
% \\
% \sqrt{nb_n}\Big( \Re u_{\tau,R_n}(\omega) - g_n \Big) &= q(1-\alpha/2,\theta_0) + o_P(1).
\end{align*}
Similarly
\[
\sqrt{nb_n}\Big( u_{\btau,R_n}^\Re(\omega) - g_n \Big) = q(1-\alpha/2,\theta_0) + o_P(1).
\]
By Slutzky's Lemma 
\[
Z_n - \sqrt{nb_n}\Big( u_{\btau,R_n}^\Re(\omega) - g_n \Big)\Dkonv \Re H_0(\btau;\omega) - q(1-\alpha/2,\theta_0),
\]
and since the distribution of the limit is continuous
\[
P(\Re\hat f_{\btau}(\omega) \leq u_{\btau,R_n}^\Re(\omega)) = P\Big(Z_n \leq \sqrt{nb_n}( u_{\btau,R_n}^\Re(\omega) - g_n )\Big) \to 1-\alpha/2.
\]
Similarly,
\[
P(\Re\hat f_{\btau}(\omega) < l_{\btau,R_n}^\Re(\omega)) = P\Big(Z_n < \sqrt{nb_n}( l_{\btau,R_n}^\Re(\omega) - g_n )\Big) \to \alpha/2.
\]
This completes the proof of~\eqref{eq:main}.\\
\\
Next let us prove~\eqref{eq:main:unif}.
Begin by observing that 
\[
x \mapsto 1 - \hat F_R(x-) = \frac{1}{R} \sum_{r=1}^R I\Big\{ x \leq \max\{A_{r}^\Re(\omega), A_{r}^\Im(\omega)\} \Big\}
\] 
is non-increasing, so 
\[
\min_{\btau \in M} \min\Big\{p_{\btau,R}^\Re(\omega),p_{\btau,R}^\Im(\omega)\Big\}
= \frac{1}{R} \sum_{r=1}^R I\Big\{ \max_{\btau \in M} \max\{E_{\btau}^\Re(\omega), E_{\btau}^\Im(\omega)\} \leq \max\{A_{r}^\Re(\omega), A_{r}^\Im(\omega)\} \Big\}.
\]
%\[
%\inf_{\btau \in M} \frac{1}{R} \sum_{r=1}^R I\Big\{  \Re E_{\btau}(\omega) \leq \Re A_{r}(\omega) \Big\} = \frac{1}{R} \sum_{r=1}^R I\Big\{ \sup_{\btau \in M} \Re E_{\btau}(\omega) \leq \Re A_{r}(\omega) \Big\}.
%\]
Define 
%\[
%Z_{n}(\btau,\omega) := \sqrt{nb_n}\Big( \hat{f}_{\btau}(\omega) - f_\btau^{\theta_0}(\omega) - B_n^{(k)}(\btau,\omega) \Big)
%\]
%and
\[
Z_{r,n}(\btau,\omega) := \sqrt{nb_n}\Big(\hat{f}^{\hat\theta,r}_{\btau}(\omega) - f_\btau^{\theta_0}(\omega) - B_n^{(k)}(\btau,\omega) \Big)
\]
where $B_n^{(k)}$ is defined in Theorem~\ref{thm:main}. Define 
\begin{align*}
\widetilde l_{\btau,R}^\Re(\omega) &= \beta/2-\text{quantile}(\Re Z_{r,n}(\btau,\omega),\dots,\Re Z_{r,n}(\btau,\omega)),\\
\widetilde u_{\btau,R}^\Re(\omega) &= (1-\beta/2)-\text{quantile}(\Re Z_{r,n}(\btau,\omega),\dots,\Re Z_{r,n}(\btau,\omega))
\end{align*} 
and similar for imaginary parts. Let $\widetilde \Delta^\Re_{\btau,R}, \widetilde c^\Re_{\btau,R}, \widetilde A_r^\Re(\omega)$ denote the corresponding versions of $\Delta^\Re_{\btau,R}, c^\Re_{\btau,R},  A^\Re_r(\omega)$ with $\widetilde u_{\btau,R}^\Re, \widetilde l_{\btau,R}^\Re, \Re Z_{r,n}(\btau,\omega)$ instead of $u_{\btau,R}^\Re,l_{\btau,R}^\Re,\hat{f}^{\hat\theta,r}_{\btau}(\omega)$ and note that by equivariance of quantiles under the given transformations we have 
\[
\widetilde u_{\btau,R}^\Re = \sqrt{nb_n}\Big(u_{\btau,R}^\Re - \Re f_\btau^{\theta_0}(\omega) - \Re B_n^{(k)}(\btau,\omega) \Big), \quad \widetilde l_{\btau,R}^\Re = \sqrt{nb_n}\Big(l_{\btau,R}^\Re - \Re f_\btau^{\theta_0}(\omega) - \Re B_n^{(k)}(\btau,\omega) \Big)
\] 
which implies $\widetilde A_r^\Re(\omega) \equiv A_r^\Re(\omega)$ after some simple algebra. From Lemma~\ref{lem:help2} we obtain by similar arguments as above (noting that $M$ is finite) that $\widetilde l^\Re_{\btau,R}(\omega)$ converges to the $\beta/2$-quantile of the distribution of $\Re H_0(\btau,\omega)$ and $\widetilde u^\Re _{\btau,R}(\omega)$ converges to the $1-\beta/2$-quantile of the distribution of $\Re H_0(\btau,\omega)$ (both convergences are in probability). Since $\Re H_0(\btau,\omega)$ follows a normal distribution with non-zero variance this implies
\[
\max_{\btau \in M} |\widetilde \Delta_{\btau,R}^\Re(\omega) - \Delta_{\btau}^\Re(\omega)| = o_P(1), \quad \max_{\btau \in M} |\widetilde c_{\btau,R}^\Re(\omega)| = o_P(1)
\]
where
\[
\Delta_{\btau}^\Re(\omega) := \sigma_{\btau}^\Re(\omega)\Big\{\frac{1}{2} - \Phi^{-1}(\beta/2)\Big\} 
\]
and $\sigma_{\btau}^\Re(\omega)$ denotes the standard deviation of $\Re H_0(\btau,\omega)$. Similar results hold for the imaginary parts. By a combination of Slutzky's Lemma and the continuous mapping theorem we now obtain from Theorem~\ref{thm:main} that 
\begin{equation} \label{eq:weakmax}
\max\{\widetilde A_{1}^\Re(\omega), \widetilde A_{1}^\Im(\omega)\} \Dkonv \frac{1}{\frac{1}{2} - \Phi^{-1}(\beta/2)}\max_{\btau \in M} \max\Big\{\frac{\Re H_0(\btau;\omega)}{\sigma_{\btau}^\Re(\omega)}, \frac{\Im H_0(\btau;\omega)}{\sigma_{\btau}^\Im(\omega)}\Big\}.
\end{equation}
By similar arguments it follows that 
\begin{equation} \label{eq:weakmax2}
%\Big( \max_{\btau \in M} \widetilde E_{\btau}^\Re(\omega), \max_{\btau \in M} \widetilde E_{\btau}^\Im(\omega) \Big)\Dkonv \frac{1}{\frac{1}{2} - \Phi^{-1}(\beta/2)} \Big( \max_{\btau \in M} \frac{\Re H_0(\btau;\omega)}{\sigma_{\btau}^\Re(\omega)}, \max_{\btau \in M} \frac{\Im H_0(\btau;\omega)}{\sigma_{\btau}^\Im(\omega)} \Big).
\max_{\btau \in M} \max\{E_{\btau}^\Re(\omega), E_{\btau}^\Im(\omega)\} \Dkonv \max_{\btau \in M} \max\Big\{\frac{\Re H_0(\btau;\omega)}{\sigma_{\btau}^\Re(\omega)}, \frac{\Im H_0(\btau;\omega)}{\sigma_{\btau}^\Im(\omega)}\Big\} \sim F.
\end{equation}
%Next we observe that 
%\[
%\frac{1}{R} \sum_{r=1}^R I\Big\{ x \leq \max\{\Re A_{r}(\omega), \Im A_{r}(\omega)\} \Big\} = 1 - \hat F_R(x-)
%\] 
%where $\hat F_R(\cdot)$ is empirical cdf of the 'sample' $\max\{\Re \widetilde A_1(\omega), \Im \widetilde A_1(\omega)\},...,\max\{\Re \widetilde A_R(\omega), \Im \widetilde A_R(\omega)\}$. 
Denoting by $F_R$ the cdf of the random variable $\max\{\widetilde A_1^\Re(\omega), \widetilde A_1^\Im(\omega)\}$, the uniform Glivenko-Cantelli Theorem (see Theorem 2.8.1 in~\cite{vanderwell1996}) implies that $\sup_{x \in \R} |\hat F_R(x) - F_R(x)| = o_P(1)$. Together with~\eqref{eq:weakmax} and continuity of the cdf, say $F$, of the random variable $\max_{\btau \in M} \max\Big\{\frac{\Re H_0(\btau;\omega)}{\sigma_{\btau}^\Re(\omega)}, \frac{\Im H_0(\btau;\omega)}{\sigma_{\btau}^\Im(\omega)}\Big\}$ (note that the latter is a maximum over a finite number of (dependent) standard normal random variables, hence has a continuous distribution), it follows that $\sup_{x \in \R} |F_R(x) - F(x)| = o(1)$, and hence 
\[
\frac{1}{R} \sum_{r=1}^R I\Big\{ \max_{\btau \in M} \max\{E_{\btau}^\Re(\omega), E_{\btau}^\Im(\omega)\} \leq \max\{A_{r}^\Re(\omega), A_{r}^\Im(\omega)\} \Big\} = 1 - F\Big(\max_{\btau \in M} \max\{E_{\btau}^\Re(\omega), E_{\btau}^\Re(\omega)\}\Big) + o_P(1).
\]   
Now, by~\eqref{eq:weakmax2} and by continuity of $F$ combined with the continuous mapping Theorem and Slutzky's Lemma we finally obtain 
\[
\min_{\btau \in M} \min\Big\{p_{\btau,R}^\Re(\omega),p_{\btau,R}^\Im(\omega)\Big\} = \frac{1}{R} \sum_{r=1}^R I\Big\{ \max_{\btau \in M} \max\{E_{\btau}^\Re(\omega), E_{\btau}^\Im(\omega)\} \leq \max\{A_{r}^\Re(\omega), A_{r}^\Im(\omega)\} \Big\} \Dkonv 1- U[0,1].
\]
This completes the proof of~\eqref{eq:main:unif}. \hfill $\Box$

\subsection{Proof of Theorem~\ref{thm:main}}

We will make use of the following notation: $U_t^{\theta} := F^{\theta}(X_t^{\theta})$
\begin{align*}
d^{U,\theta}_{\tau,n}(\omega) &:= \sum_{t=0}^{n-1} \Ind{U_t^{\theta} \leq \tau} e^{-i\omega t},
\\
I^{U,\theta}_{(\tau_1,\tau_2),n}(\omega) &:= \frac{1}{2\pi n} d^{U,\theta}_{\tau_1,n}(\omega)d^{U,\theta}_{\tau_2,n}(-\omega),
\\
\hat f^{U,\theta}_{\btau,n}(\omega) &:= \frac{1}{2\pi n}\sum_{s=1}^{n-1} W_n(\omega - 2\pi s/n) I^{U,\theta}_{\btau,n}(2\pi s /n).
\end{align*}

Theorem~\ref{thm:main} follows from the following four statements.

\begin{description}
\item[(i)] for any fixed $\omega \in \R$ and an arbitrary sequence $\theta_n$ in $\Theta$ with $\theta_n = \theta_0 + o(1)$ we have for $n \rightarrow \infty$
\begin{equation*}
\sqrt{nb_n}(\hat f^{U,\theta_n}_{\btau,n}(\omega) - \E\hat f^{U,\theta_n}_{\btau,n}(\omega) )_{\tau \in \mathcal{T}} \leadsto H_0(\cdot;\omega) \quad in~~\ell^\infty(\mathcal{T})
\end{equation*}
\item[(ii)] for $n \rightarrow \infty$ we obtain the following result for the expectation
\begin{equation*}
\sup_{\substack{\btau \in [0,1]^2 \\ \omega \in \Rb}} \big| \E\hat f^{U,\theta_0}_{\btau,n}(\omega) - f_{\btau}^{\theta_0}(\omega) - B_n^{(k)}(\btau,\omega)  \big| = O((nb_n)^{-1}) + o(b_n^k),
\end{equation*}
\item[(iii)] For any fixed $\omega$
\begin{equation*}
\sup_{\btau \in [0,1]^2} |\hat f_{\btau}^{\theta_n}(\omega) - \hat f^{U,\theta_n}_{\btau,n}(\omega)| = o_P((nb_n)^{-1/2} + b_n^k),
\end{equation*}
\item[(iv)] for any sequence $\theta_n$ in $\Theta$ with $\theta_n = \theta_0 + O(n^{-1/2})$ we have for $n \rightarrow \infty$
\begin{equation*}
\sqrt{nb_n} \sup_{\btau \in \mathcal{T}, \omega \in \R} \Big|\E\hat f^{U,\theta_n}_{\btau,n}(\omega) - \E\hat f^{U,\theta_0}_{\btau,n}(\omega)\Big| = o(1).
\end{equation*}
\end{description}

Note that (ii) is proved in Theorem~3.6(ii) in \cite{kvdh2014} so that it remains to prove (i), (iii), (iv).

\subsubsection{Convergence as a process and the proofs of $(i)$} \label{sec:pf1}

Throughout this section, let $\Delta_n(\omega) := \sum_{t=0}^{n-1} e^{i \omega t}$ and $\mathcal{F}_n := \{2\pi j/n: j = 1,...,n-1 \}$. For intervals $A \subset [0,1]$ and $\omega \in \R$ define
\[
d_{A,n}^{U,\theta_n}(\omega) = \sum_{t=0}^{n-1} \Ind{U_{t}^{\theta_n} \in A}e^{-it\omega}.
\]
Let
\[
H_n(\btau,\omega) := \sqrt{nb_n} (\hat f^{U,\theta_n}_{\btau,n}(\omega) - \E\hat f^{U,\theta_n}_{\btau,n}(\omega)) 
\]
and denote by $\hat{F}_n^{U,\theta_n}$ the empirical cdf of $U_{1}^{\theta_n},...,U_{n}^{\theta_n}$.

We begin by stating several generalizations of results in \cite{kvdh2014}. The proofs of those results are very similar to the corresponding proofs in \cite{kvdh2014} and are omitted for the sake of brevity. 

The following statement can be proved similarly to Lemma A.4 in \cite{kvdh2014}: for arbitrary intervals $A_1,\dots,A_p \subset [0,1]$ define $\eps :=  \min_{1 \leq j \leq p} \lambda(A_j).$ Then there exist constants $C$ and $d$ that depend only on $p$ and $K_i, \rho_i, i=2,...,p$ from assumption (C) such that for all $\omega_1,\dots,\omega_p \in \R$  
\begin{equation} \label{eq:lem72}
\Big| \cum(d_{A_1,n}^{U,\theta_n}(\omega_1),\dots,d_{A_p,n}^{U,\theta_n}(\omega_p))\Big| \leq C\Big(\Big|\Delta_n\Big(\sum_{i=1}^p \omega_i\Big)\Big| + 1\Big)\eps(|\log(\eps)| + 1)^d.  
\end{equation}
Utilizing this statement and following the proof of Lemma 1.6 in the online supplement of \cite{kvdh2014} we find that for any $k \in \Nb$ there exists a constant $d_k$ such that, as $\delta_n \rightarrow 0$,
\begin{equation} \label{eq:lem86}
\sup_{\substack{ x,y \in [0,1] \\ |x-y| \leq \delta_n }} \sqrt{n}|\hat{F}^{U,\theta_n}_n(x) - \hat{F}^{U,\theta_n}_n(y) -x + y| = \Op((n^2 \delta_n + n)^{1/2k}(\delta_n |\log \delta_n|^{d_k} + n^{-1})^{1/2} ).
\end{equation}
This equation combined with the arguments in the proof of Lemma A.6 in \cite{kvdh2014} shows that for any $k \in \N$
\begin{equation} \label{eq:lem76}
\sup_{\omega \in \mathcal{F}_n} \sup_{\tau \in [0,1]} |d_{\tau,n}^{U,\theta_n}(\omega)| = \Op(n^{1/2+1/k}).
\end{equation}
Now~\eqref{eq:lem72}, \eqref{eq:lem86} and \eqref{eq:lem76} can be used to replace Lemma A.2, Lemma 1.6 (online supplement), and Lemma~A.6 from \cite{kvdh2014} in the proof of Lemma~A.7 in \cite{kvdh2014} to show the following: if $\delta_n = O\big((nb_n)^{-1/\gamma}\big)$ for some $\gamma \in (0,1)$ then
\begin{equation} \label{eq:lem77} 
\sup_{\omega \in \Rb}   \sup_{\substack{ \ub,\vb \in [0,1]^2 \\ \|\ub-\vb\| \leq \delta_n }} |H_n(\ub,\omega) - H_n(\vb,\omega)| = \op(1).
\end{equation} 

We are now ready for the proof of (i). In view of Theorem 1.5.4 and 1.5.7 in \cite{vanderwell1996} it suffices to show

\begin{description}
\item[(i1)] convergence of the finite-dimensional distributions
\begin{equation}
(H_n(\btau_j,\omega_j))_{j=1,\dots,k} \rightarrow^d (H_0(\btau_j,\omega_j))_{j=1,\dots,k})
\end{equation}
for any $(\btau_j,\omega_j) \in \mathcal{T} \times \Rb, j=1,...,k$ and $k \in \Nb.$
\item[(i2)] stochastic equicontinuity: for any $x > 0$ and any $\omega \in \Rb$
\begin{equation}
\lim_{\delta \downarrow 0} \limsup_{n \rightarrow \infty} \p\Big(\sup_{\ub,\vb \in [0,1]^2, \|\ub-\vb\| < \delta} |H_n(\ub,\omega) - H_n(\vb,\omega)| > x\Big) = 0.
\end{equation}
\end{description}

\bigskip

\textbf{Proof of $(i2)$} Apply Lemma A.2 from \cite{kvdh2014} with $L=3$ to obtain
\[ 
\sup_{\omega \in \Rb} \sup_{\|\ub-\vb\|_1 \leq \epsilon} \E|H_n(\ub,\omega) - H_n(\vb,\omega)|^{6} \leq K \sum_{l=0}^2 \frac{g(\epsilon)^{3-l}}{(nb_n)^l}, 
\]
here condition (A.2) from Lemma A.2 in \cite{kvdh2014} is satisfied with $g(x) = x(|\log x| + 1)^d$ by~\eqref{eq:lem72}.
With $\Psi(x) := x^6$ the Orlicz norm $||X||_\Psi$ coincides with the $L_6$ norm $||X||_6 = (\E|X|^6)^{1/6}$ so that we have, for any $\kappa \in (0,1)$ and sufficiently small $||a-b||_1,$
\[ 
\|H_n(\ub,\omega) - H_n(\vb,\omega)\|_\Psi \leq C\Big(\frac{\|\ub-\vb\|_1^\kappa}{(nb_n)^2}+\frac{\|\ub-\vb\|_1^{2\kappa}}{(nb_n)^1}+\|\ub-\vb\|_1^{3\kappa}\Big).
\]

To complete the proof of (i2) follow the arguments in the proof of Theorem 3.6, step (i2), in \cite{kvdh2014}. Replace Lemma A.7 therein by \eqref{eq:lem77} to obtain for all $x,\mu > 0, 2/3 < \gamma < 1$ 
\[
\lim_{\delta \downarrow 0} \limsup_{n \rightarrow \infty} \p(\sup_{\ub, \vb \in [0,1]^2} \sup_{\|\ub-\vb\|_1 < \delta} |H_n(\ub,\omega) - H_n(\vb,\omega)| > x) \leq \Big[\frac{8C}{x}\int_0^\mu z^{-2/(3\gamma)} dz \Big]^6.
\]
$(i2)$ follows since the integral tends to zero for $\mu \rightarrow 0.$

\bigskip

\textbf{Proof of $(i1)$} we have to show, that for any $\btau_1,\dots,\btau_k \in \mathcal{T}, k \in \Nb$ where $\btau_i = (\tau_{i1},\tau_{i2})$ and $\omega_1,\dots,\omega_k \neq 0 \mod 2\pi$ all cumulants of $(H_n(\btau_j,\omega_j))_{j=1,\dots,k}$ converge to the corresponding cumulants of $(H_0(\btau_j,\omega_j))_{j=1,\dots,k}),$ which by Lemma P4.5 in \cite{brill} gives the desired result. 
By construction
\[
\E(H_n(\btau,\omega)) = 0
\]
and
\[
\Cov(H_n(\btau_1,\omega_1),H_n(\btau_2,\omega_2)) = nb_n\Cov(\hat f_{\btau_1,n}^{U,\theta_n}(\omega_1),\hat f_{\btau_2,n}^{U,\theta_n}(\omega_2)).
\]
Under assumption (C) the random processes $(\Ind{U_t^{\theta_n} \leq \tau_{11}},...,\Ind{U_t^{\theta_n} \leq \tau_{k2}})_{t \in \Z}$ satisfy a uniform version of Assumption 2.6.2(2) in \cite{brill} while the weight functions $W_n$ satisfy Assumption 5.6.1 in \cite{brill}. A close look at the proof of Theorem 7.4.3 and Corollary 7.4.3 in \cite{brill} shows that all proofs go through without change and leads to the representation
\begin{multline*}
nb_n \Cov(\hat f_{\btau_1,n}^{U,\theta_n}(\omega_1),\hat f_{\btau_2,n}^{U,\theta_n}(\omega_2))
= 2\pi \int_{-\pi}^\pi W^2(u) {\rm d}u \Big[ f_{(\tau_{11}, \tau_{21})}^{\theta_n}(\omega_1) f_{(\tau_{12}, \tau_{22})}^{\theta_n}(-\omega_1)\Ind{\omega_1 = \omega_2}
 \\
+ f_{(\tau_{11},\tau_{22})}^{\theta_n}(\omega_1)f_{(\tau_{12},\tau_{21})}^{\theta_n}(-\omega_1)\Ind{\omega_1 = 2\pi-\omega_2}\Big]
+ O(b_n) + O((n b_n)^{-1}). 
\end{multline*}
Next we note that
\begin{align*}
\sup_{\btau \in \mathcal{T}, \omega \in \R}\Big|f_{\btau}^{\theta_n}(\omega) - f_{\btau}^{\theta_0}(\omega)\Big| \leq \sum_{h \in \Z} \sup_{\btau \in \mathcal{T}}|C_h^{\theta_n}(\btau)-C_h^{\theta_0}(\btau)| = o(1)
\end{align*}
by dominated convergence. Hence
\begin{multline*}
nb_n \Cov(\hat f_{\btau_1,n}^{U, \theta_n}(\omega_1),\hat f_{\btau_2,n}^{U, \theta_n}(\omega_2))
\to 2\pi \int_{-\pi}^\pi W^2(u) {\rm d}u \Big[ f_{(\tau_{11}, \tau_{21})}^{\theta_0}(\omega_1) f_{(\tau_{12}, \tau_{22})}^{\theta_0}(-\omega_1) \Ind{\omega_1 = \omega_2}
 \\
+ f_{(\tau_{11},\tau_{22})}^{\theta_0}(\omega_1)f_{(\tau_{12},\tau_{21})}^{\theta_0}(-\omega_1)\Ind{\omega_1 = 2\pi-\omega_2}\Big].
\end{multline*}

To complete the proof it remains to show that the cumulants of order $K \geq 3$ vanish as $n$ tends to infinity. We have with $\btau_i = (\tau_{i1},\tau_{i2}), 1 \leq i \leq K$ that 
\begin{align*}
&\cum(H_n(\btau_1,\omega_1),\dots,H_n(\btau_K,\omega_K))
= (n b_n)^{K/2} \cum(\hat f_{\btau_1,n}^{U,\theta_n}(\omega_1),\dots,\hat f_{\btau_K,n}^{U,\theta_n}(\omega_K)) \\
&= (2\pi)^{2K} n^{-3K/2} (b_n)^{K/2} \sum_{s_1 = 1}^{n-1} \cdots \sum_{s_K = 1}^{n-1} W_n(\omega_1 - 2\pi s_1/n) \cdots W_n(\omega_k - 2\pi s_K/n)\\
&\cum(d^{U,\theta_n}_{\tau_{11},n}(2\pi s_1/n)d^{U,\theta_n}_{\tau_{12},n}(-2\pi s_1/n)),\dots,d^{U,\theta_n}_{\tau_{K1},n}(2\pi s_K/n)d^{U,\theta_n}_{\tau_{K2},n}(-2\pi s_K/n)).
\end{align*}
To see that these cumulants tend to zero we will need arguments similar to those used in Step 2 of the proof of Lemma $A.2$ in \cite{kvdh2014}. Applying the product Theorem $2.3.2$ in \cite{brill} to the last cumulant leads to
\begin{multline} \label{eq:cumproduct}
\cum(d^{U,\theta_n}_{\tau_{11},n}(2\pi s_1/n)d^{U,\theta_n}_{\tau_{12},n}(-2\pi s_1/n)),\dots, d^{U,\theta_n}_{\tau_{K1},n}(2\pi s_K/n)d^{U,\theta_n}_{\tau_{K2},n}(-2\pi s_K/n)) \\
= \sum_{(\nu_1,\dots,\nu_N)} \prod_{k=1}^N \cum(d^{U,\theta_n}_{\tau_{ij},n}( (-1)^{j+1} s_i 2\pi/n); (i,j) \in \nu_k),
\end{multline}
where the sum runs over all indecomposable partitions $(\nu_1,\dots,\nu_N)$ (see \cite{brill} p. 20) of
\[
\begin{matrix}
(1,1) & (1,2) \\
 \vdots & \vdots \\
 (K,1) & (K,2).
\end{matrix}
\]
Note that an indecomposable partition consists of at most $N \leq K + 1$ sets. Now by~\eqref{eq:lem72} the absolute values of those cumulants are bounded by
\begin{equation*}
(\ref{eq:cumproduct}) \leq C\sum_{(\nu_1,\dots,\nu_N)} \prod_{k=1}^N \Big[\Delta_n\Big(\frac{2\pi}{n}\sum_{ (i,j) \in \nu_k} (-1)^{j+1} s_i \Big) + 1\Big] = C\sum_{(\nu_1,\dots,\nu_N)} \sum_{I \subset \{1,\dots,N\} } \prod_{k \in I} \Delta_n\Big(\frac{2\pi}{n}\sum_{ (i,j) \in \nu_k} (-1)^{j+1} s_i \Big)
\end{equation*}
where $C$ is some constant that depends on $K$ and the constants $K_p, \rho_p, p=1,...,2K$ from assumption (C) only. Furthermore, since 
\[ 
\Delta_n\Big(\frac{2\pi}{n} \omega\Big) = 
\begin{cases}
 n & \omega \in n\Zb \\
 0 & \omega \notin n\Zb
\end{cases}
\]
we have that for each combination of $\nu = \{\nu_1,...,\nu_N\}$ and $I \subset \{1,...,N\}$
\[ 
\prod_{k \in I} \Delta_n\Big(\frac{2\pi}{n}\sum_{ (i,j) \in \nu_k} (-1)^{j+1} s_i \Big) = 0
\]
unless
\[ 
\sum_{ (i,j) \in \nu_k} (-1)^{j+1} s_i \in n\Zb \quad \forall \nu_k \in \nu: k \in I. 
\]
In the latter case
\[ 
\prod_{k \in I} \Delta_n\Big(\frac{2\pi}{n}\sum_{ (i,j) \in \nu_k} (-1)^{j+1} s_i \Big) =  n^{|I|}.
\]
Now we can restrict the sum over the indices $(s_1,\dots,s_K)$ to the set 
\begin{equation*}
S(\nu,I) := \Big\{ (s_1,\dots,s_K) \in \{1, \dots, n-1 \}^K :
\sum_{ (i,j) \in \nu_k} (-1)^{j+1} s_i \in n\Zb \quad \forall \nu_k \in \nu: k \in I \Big\}.
\end{equation*}
To complete the proof follow the arguments starting at the bottom of page 16 of the online supplemntary meaterial in the proof of Lemma A.2 in \cite{kvdh2014} (note that the supplement states this as \textit{proof of Lemma 7.2}). First, note that $S(\nu,I)$ is empty for $|I| = K + 1$ and
\begin{multline*}
\sum_{s_1,\dots,s_K = 1}^{n-1} \prod_{m = 1}^{K} \big|W_n(\omega_m - 2\pi s_m/n)\big|\prod_{k \in I} \Delta_n\Big(\frac{2\pi}{n}\sum_{ (i,j) \in \nu_k} (-1)^{j+1} s_i \Big)
= \sum_{(s_1,\dots,s_K) \in S(\nu,I)} \prod_{m = 1}^{K} \big|W_n(\omega_m - 2\pi s_m/n)\big| n^{|I|} \\
= O((b_n^{-1})^{|I|-\lfloor |I|/N \rfloor} n^{K-(|I|-\lfloor |I|/N\rfloor)} n^{|I|} ),
\end{multline*}
where the last equality follows from the arguments around equation (1.26) in the online supplement of \cite{kvdh2014} . Finally, the number of indecomposable partitions $N$ does not depend on $n$ so that $\cum(H_n(\btau_1,\omega_1),\dots,H_n(\btau_K,\omega_K))$ is of order
\[ 
n^{-3K/2} (b_n)^{K/2} \max_{N \leq K} \max_{|I| \leq N} (b_n^{-1})^{|I|-\lfloor |I|/N \rfloor} n^{K-(|I|-\lfloor |I|/N\rfloor)} n^{|I|} = O((nb_n)^{1-K/2}), 
\]
which tends to zero for $K \geq 3.$ \hfill $\Box$

%\newpage

\subsubsection{Proof of (iii)}

Following the proof of Lemma~A.3 and the arguments in the end of the proof of Lemma~A.4 in \cite{kvdh2014} and using (C) it is straightforward to prove that $\omega \mapsto f_{\btau}^\theta(\omega)$ is infinitely often continuously differentiable (for any $\btau \in (0,1)^2$ and $\theta \in U_\eps(\theta_0)$) and that there exist constants $C,d$ that are independent of $\btau_1,\btau_2, \theta$ with
\begin{equation}\label{eq:lem73}
\sup_{\omega \in \R, \theta \in U_\eps(\theta_0)} \Big|\frac{d^j}{d \omega^j} f_{\btau_1}^\theta(\omega) - \frac{d^j}{d \omega^j}f_{\btau_2}^\theta(\omega) \Big| \leq C \|\btau_1-\btau_2\|_1(1+\log\|\btau_1-\btau_2\|_1 )^d.
\end{equation}
Moreover, the proof of Lemma A.5 \cite{kvdh2014} can be modified to obtain (recall the definition of $\hat F_n^{U,\theta_n}$ in the beginning of Section~\ref{sec:pf1})
\begin{equation}\label{eq:lem75}
\sup_{\tau\in [0,1]} |(\hat F_n^{U,\theta_n})^{-1}(\tau) - \tau| = \Op(n^{-1/2}).
\end{equation}
As in \cite{kvdh2014} (A.4) it follows that for any $k \in \N$ we have
\begin{equation*}
\sup_{\omega \in \Rb} \sup_{\tau \in [0,1]} |d_{\tau,n}(\omega)-d_{(\hat F_n^{U,\theta_n})^{-1}(\tau),n}^U|
\leq n \sup_{\tau \in [0,1]} |\hat{F}_n^{U,\theta_n}(\tau)-\hat{F}^{U,\theta_n}_n(\tau-)| \leq \Op(n^{1/(2k)}),
\end{equation*}
where $\hat{F}_n^{U,\theta_n}(\taus-) := \lim_{\xi \uparrow 0}\hat{F}_n^{U,\theta_n}(\tau - \xi)$ and the last inequality follows from~\eqref{eq:lem86}. The remaining part of the proof is analogous to the arguments given in Section A.3 of \cite{kvdh2014} and details are omitted for the sake of brevity.  \hfill $\Box$

\subsubsection{Proof of (iv)}

Begin by observing that for some constant $C_W$ that depends on $W$ only we have for any $\btau \in [0,1]^2$
\begin{equation*}
\begin{split}
&\Big|\E\hat f^{U,\theta_n}_{\btau,n}(\omega) - \E\hat f^{U,\theta_0}_{\btau,n}(\omega)\Big|
\leq \frac{1}{2\pi n} \sum_{s=1}^{n-1} \Big|W_n(\omega - 2\pi s/n)\Big| \Big|\E[I^{U,\theta_n}_{\btau,n}(2\pi s /n)] - \E[I^{U,\theta_0}_{\btau,n}(2\pi s /n)]\Big|
\\
& \leq C_W \max_{\omega \in \mathcal{F}_n} \Big|\E[I^{U,\theta_n}_{\btau,n}(\omega)] - \E[I^{U,\theta_0}_{\btau,n}(\omega)]\Big|
= C_W \max_{\omega \in \mathcal{F}_n} \Big|\frac{1}{2\pi n}\sum_{t_1,t_2 = 0}^{n-1} e^{-i(t_1-t_2)\omega} (C_{t_2-t_1}^{\theta_n}(\btau) - C_{t_2-t_1}^{\theta_0}(\btau)) \Big|
\\
& \leq \frac{C_W}{2\pi n} \sum_{|t_1|\leq n} \sum_{k \in \Z} \Big|C_k^{\theta_n}(\btau) - C_k^{\theta_0}(\btau) \Big|
\leq C_W \sum_{k \in \Z} \Big|C_k^{\theta_n}(\btau) - C_k^{\theta_0}(\btau) \Big|.
\end{split}
\end{equation*}
Now under (C) we have $|C_k^{\theta_n}(\btau) - C_k^{\theta_0}(\btau)| \leq 2 K_2 \rho_2^{|k|}$. Hence, for any fixed $N \in \N$ we have by (LC)
\begin{equation*}
\sup_{\btau\in\mathcal{T}} \sum_{k \in \Z} \Big|C_k^{\theta_n}(\btau) - C_k^{\theta_0}(\btau) \Big|
\leq \sum_{|k| \leq N} \sup_{\btau\in\mathcal{T}} \Big|C_k^{\theta_n}(\btau) - C_k^{\theta_0}(\btau) \Big| + 4K_2 \sum_{k > N} \rho_2^{|k|}
\leq (2N+1)\|\theta_n - \theta_0\| + \frac{4K_2\rho_2^{N+2}}{1-\rho_2}. 
\end{equation*}
Now by assumption $\|\theta_n - \theta_0\| = O(n^{-1/2})$, so picking $N = C \log n$ for a constant $C > 0$ such that $4K_2\rho_2^{C\log n} = o(n^{-1})$ we obtain 
\[
\sup_{\btau\in\mathcal{T}}\Big|\E\hat f^{U,\theta_n}_{\btau,n}(\omega) - \E\hat f^{U,\theta_0}_{\btau,n}(\omega)\Big| = O(\log n) \|\theta_n - \theta_0\| = o(\sqrt{nb_n})
\]
since by assumption $nb_n = o(n^{1-1/(2k+1)})$ for some $k \in \N$. \hfill $\Box$

\newpage

\section{Additional plots}\label{sec:addplots}

%\begin{figure}[h!]
%\includegraphics[width = 0.3\textwidth]{Pictures/uni_fine_EGARCH11_vsSP500_14_}
%\includegraphics[width = 0.3\textwidth]{Pictures/uni_fine_EGARCH11_vsSP500_21_}
%\includegraphics[width = 0.3\textwidth]{Pictures/uni_fine_EGARCH11_vsSP500_63_}
%%\includegraphics[width = 0.5\textwidth]{Pictures/uni_fine_EGARCH11_vsSP500_14_}
%%\includegraphics[width = 0.5\textwidth]{Pictures/uni_fine_EGARCH11_vsSP500_21_}
%\caption{Detailed plots produced by Algorithm~\ref{alg2} at three particular frequencies based on the daily log-returns of the S\&P~500 between 2000 and 2005 with EGARCH(1,1) as candidate model class.} \label{fig:add_GARCH_Sp}
%
%\end{figure}

%\newpage

%\section*{References}
%\bibliographystyle{elsarticle-num} 
%\bibliography{LocStatBib_2016-07-11}




%\section*{References}


%\begin{thebibliography}{00}
%%% Text of bibliographic item
%
%\bibitem{}
%
%\end{thebibliography}
\end{document}